\documentclass[fleqn,usenatbib]{mnras}
\usepackage{newtxtext,newtxmath}

\usepackage[T1]{fontenc}

\DeclareRobustCommand{\VAN}[3]{#2}
\let\VANthebibliography\thebibliography
\def\thebibliography{\DeclareRobustCommand{\VAN}[3]{##3}\VANthebibliography}

\usepackage{graphicx}	
\usepackage{amsmath}	
\usepackage{pdflscape}

\newcommand \tmtsI {TMTS-I }

\title[Minute-cadence Observations of the TMTS II]{Minute-Cadence Observations of the LAMOST Fields with the TMTS:\\ II. Catalogues of Short-Period Variable Stars from the First Two-Year Surveys}

\author[Lin et al.]{
Jie Lin,$^{1}$\thanks{E-mail: linjie2019@mail.tsinghua.edu.cn}
Xiaofeng Wang,$^{1,2}$\thanks{E-mail: wang\_xf@mail. tsinghua.edu.cn}
Jun Mo,$^{1}$
Gaobo Xi,$^{1}$
Alexei V. Filippenko,$^{3}$
Shengyu Yan,$^{1}$ \newauthor
Thomas G. Brink,$^{3}$
Yi Yang,$^{3}$
Chengyuan Wu,$^{4}$
P\'eter N\'emeth,$^{5,6}$
Gaici Li,$^{1}$
Fangzhou Guo,$^{1}$
Jincheng Guo,$^{2}$\newauthor
Yongzhi Cai,$^{4,7,1}$
Heran Xiong,$^{8,4}$
WeiKang Zheng,$^{3}$
Qichun Liu,$^{1}$
Jicheng Zhang,$^{9}$
Xiaojun Jiang,$^{10,11}$\newauthor
Liyang Chen,$^{1}$
Qiqi Xia,$^{1}$
Haowei Peng,$^{1}$
Zhihao Chen,$^{1}$ 
Wenxiong Li,$^{12}$
Weili Lin,$^{1}$
Danfeng Xiang,$^{1}$ \newauthor
Xiaoran Ma $^{1}$ 
and Jialian Liu$^{1}$
\\
$^{1}$Physics Department and Tsinghua Center for Astrophysics, Tsinghua University, Beijing, 100084, People's Republic of China\\
$^{2}$Beijing Planetarium, Beijing Academy of Sciences and Technology, Beijing, 100044, People’s Republic of China\\
$^{3}$Department of Astronomy, University of California, Berkeley, CA 94720-3411, USA\\
$^{4}$Yunnan Observatories, Chinese Academy of Sciences, Kunming, 650216, People’s Republic of China\\
$^{5}$Astronomical Institute of the Czech Academy of Sciences, Fri\v{c}ova 298, Ond\v{r}ejov, 25165, Czech Republic\\
$^{6}$Astroserver.org, F\H{o} t\'er 1, Malomsok, 8533, Hungary\\
$^{7}$Key Laboratory for the Structure and Evolution of Celestial Objects, Chinese Academy of Sciences, Kunming, 650216, People’s Republic of China\\
$^{8}$Research School of Astronomy \& Astrophysics, Australian National University, Canberra, 2611, Australia\\
$^{9}$Department of Astronomy, Beijing Normal University, Beijing, 100875, People's Republic of China\\
$^{10}$National Astronomical Observatories of China, Chinese Academy of Sciences, Beijing, 100012, People’s Republic of China\\
$^{11}$School of Astronomy and Space Science, University of Chinese Academy of Sciences, Beijing, 100049, People’s Republic of China\\
$^{12}$The School of Physics and Astronomy, Tel Aviv University, Tel Aviv 69978, Israel\\
}

\date{Accepted XXX. Received YYY; in original form ZZZ}

\pubyear{2023}

\begin{document}
\label{firstpage}
\pagerange{\pageref{firstpage}--\pageref{lastpage}}
\maketitle

\begin{abstract}
Over the past few years, wide-field time-domain surveys like ZTF and OGLE have led to discoveries of various types of interesting short-period stellar variables, such as ultracompact eclipsing binary white dwarfs, rapidly rotating magnetised white dwarfs (WDs), transitional cataclysmic variables between hydrogen-rich and helium accretion, and blue large-amplitude pulsators (BLAPs), which greatly enrich our understandings of stellar physics under some extreme conditions. In this paper, we report the first-two-year discoveries of short-period variables (i.e., P$<$2~hr) by the Tsinghua University--Ma Huateng Telescopes for Survey (TMTS). TMTS is a multi-tube telescope system with a field of view up to 18 deg$^2$, which started to monitor the LAMOST sky areas since 2020 and generated uninterrupted minute-cadence light curves for about ten million sources within 2 years. 
Adopting the Lomb-Scargle periodogram with period-dependent thresholds for the maximum powers, we identify over 1\,100 sources that exhibit a variation period shorter than 2~hr.
Compiling the light curves with the {\it Gaia} magnitudes and colours, LAMOST spectral parameters, VSX classifications, and archived observations from other prevailing time-domain survey missions, we identified 1\,076 as $\delta$~Scuti stars, which allows us study their populations and physical properties in the short-period regime. The other 31 sources include BLAPs, subdwarf B variables (sdBVs), pulsating WDs, ultracompact/short-period eclipsing/ellipsoidal binaries, cataclysmic variables below the period gap, etc., which are highly interesting and worthy of follow-up investigations. 
\end{abstract}

\begin{keywords}
surveys -- stars: oscillations (including pulsations) -- (stars:) binaries (including multiple): close -- (stars:) novae, cataclysmic variables
\end{keywords}



\section{Introduction}
\begin{figure*}
\centering
    \includegraphics[width=0.9\textwidth]{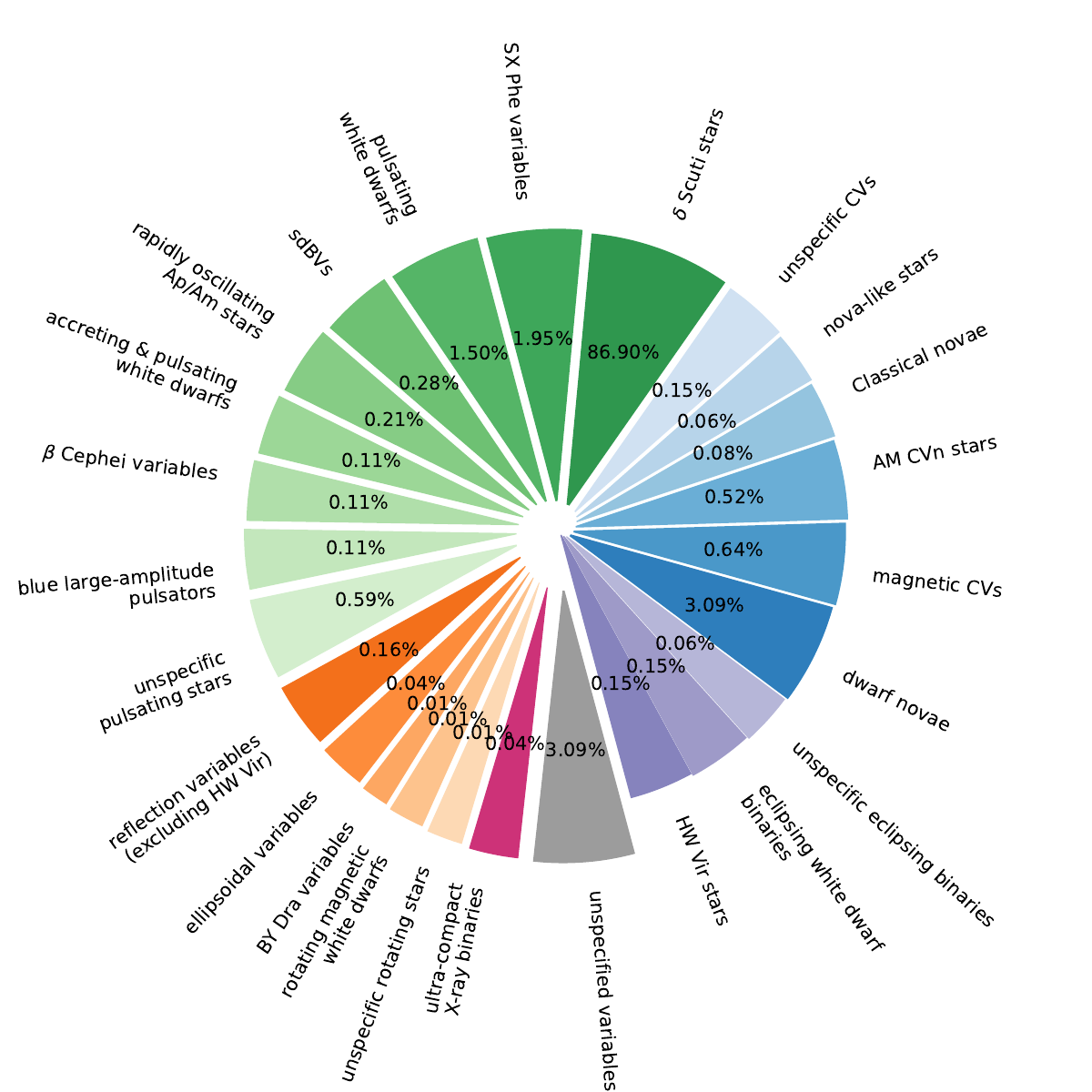}
    \caption{
Fractions of various classes of short-period ($P<2$~hr) variable stars according to the VSX.
Eclipsing binaries, cataclysmic variables, pulsating stars, rotating variables, and ultracompact X-ray binaries are colour-coded with purple, blue, green, orange, and violet red, respectively. The ``eclipsing white dwarf binaries'' include WD–MS binaries, WD-sdB binaries, and binary WDs, since these binaries were not well subdivided in the VSX.
The ``accreting \& pulsating white dwarfs'' are the cataclysmic binaries harboring a pulsating WD (e.g., GW~Lib variables).
The ``reflection variables (excluding HW~Vir)'' represent the reflection variables that have not been identified as HW~Vir stars.
The ``unspecified'' areas represent those variables lacking further identifications.
Note that $\delta$~Scuti stars constitute the vast majority of short-period variables.
     } 
    \label{fig:vsx_statistic}
\end{figure*}

Short-period variable stars are characterised by brightness variability with periods shorter than several hours or even a few tens of minutes \citep{Varadi+etal+2009+Gaia_spv,Macfarlane+etal+2015+omega_I,Rimoldini+etal+2022+GaiaDR3}. 
Currently known classes include the following.
\begin{itemize}
    \item \textbf{Eclipsing binaries}, e.g.,
    HW Vir stars (an eclipsing system that consists of a hot subdwarf and a cool companion; \citealt{Soszynski+etal+2015+USPbinary_OGLE,Schaffenroth+etal+2019+Short_binary,Keller+etal+2022+eclipsingWD});
    eclipsing binary white dwarfs (WDs) \citep{Brown+etal+2011+12min_EWD,Brown+etal+2017+40minWD,Brown+etal+2020+ELMVIII,Brown+etal+2022+ELMIX,Kilic+etal+2014+EWD,Soszynski+etal+2015+USPbinary_OGLE,Schaffenroth+etal+2019+Short_binary,Burdge+etal+2019+Nature+7minWD,Burdge+etal+2020+systematic,Burdge+etal+2020+9minute,Keller+etal+2022+eclipsingWD};
    eclipsing WD–main-sequence (MS) binaries \citep{Gomez-Moran+2011+post_common_binary,Keller+etal+2022+eclipsingWD};
eclipsing hot subdwarf binaries (other than HW~Vir) \citep{Vennes+etal+2012+sd-WD,Burdge+etal+2020+systematic,Ramsay+etal+2022+BLAP_sdbWD,Krzesinski+etal+2022+hst_binary};
and eclipsing M-dwarf binaries \citep{Nefs+etal+2012+Mdwarfs}.
    \item \textbf{Cataclysmic variables (CVs)}, e.g., 
AM~CVn stars (interacting binary WDs; \citealt{Burdge+etal+2020+systematic,Keller+etal+2022+eclipsingWD,Roestel+etal+2022+AMCVn});
AM~Her/DQ~Her variables (polars/intermediate polars; \citealt{Downes+etal+2001+ATLAS_CV,Muno+etal+2003+Chandra_sources,Drake+etal+2014+Catalina_CV,Motch+etal+2010+XMM_sources});
U~Gem stars (dwarf novae; \citealt{Drake+etal+2009+catalina+realtime,Drake+etal+2014+Catalina_CV,Soszynski+etal+2015+USPbinary_OGLE,Coppejans+etal+2016+DN,Kato+etal+2013+DN,Kato+etal+2016+SUUMa});
UX~UMa stars (nova-like stars; \citealt{Downes+etal+1993+CVcatalog,Downes+etal+2001+ATLAS_CV,Drake+etal+2009+catalina+realtime});
and classical novae \citep{Woudt+Warner+2002+novae,Schaefer+2010+novae,Schaefer+2022+Novae,Walter+etal+2012+novae}.
    \item \textbf{Pulsating stars}, e.g., $\delta$~Scuti stars \citep{Smalley+etal+2011+Superroam,Chen+etal+2020+ZTFvariable,Jayasinghe+etal+2020+deltascuti,Pietrukowicz+etal+2020+deltascuti,Soszynski+etal+2021+24000deltascuti};
    SX~Phe variables \citep{Kaluzny+Krzeminski+1993+SXPHE,Ramsay+etal+2011+SXPHE,Darragh+Murphy+2012+SXPHE};
rapidly oscillating Ap/Am stars \citep{Handler+Paunzen+1999+roAp,Smalley+etal+2011+Superroam, Holdsworth+etal+2014+roapam};
$\beta$~Cep variables \citep{Jakate+1979+bceps,Arentoft+etal+2001+beta_ceph,Pigulski+Pojmanski+2008+beta_cep,Chang+etal+2015+beta_ceph};
V361~Hya/V1093~Her/DW~Lyn variables (subdwarf B variables, sdBVs; \citealt{Kilkenny+etal+1997+361_discovery,Ostensen+etal+2010+pmodesdBVs,Reed+etal+2010+gmodesdBVs,Kupfer+etal+2021+ZTFhighcadence,Uzundag+etal+2021+TESS_gmode_sdBVs});
ZZ ceti/V777 Her /GW Vir/GW Lib/ELMV/pre-ELMV variables (pulsating WDs; \citealt{Corsico+etal+2019+book+pulsatingWD,Castanheira+etal+2006+newZZ,Castanheira+etal+2010+newZZ,Hermes+etal+2012+ELMV,Grootel+etal+2013+newZZ,Maxted+etal+2013+preELMV});
blue large-amplitude pulsators (BLAPs; \citealt{Pietrukowicz+etal+2017+BLAPs,Kupfer+2019+high-g_BLAPs,McWhirter+Lam+2022+blap_candidates,Ramsay+etal+2022+BLAP_sdbWD,Pigulski+Kolaczek-Szymanski+2022+TESS_BLAP,lin+etal+2022+NatAs}).
    \item \textbf{Rotating variables}, e.g., 
    reflection variables (other than HW~Vir stars) \citep{Parsons+etal+2013+R_ell_binary,Steele+etal+2013+shortest_Rbinary,Burdge+etal+2022+Nature+widow};
    ellipsoidal variables \citep{Gomez-Moran+2011+post_common_binary,Hermes+etal+2012+ell_WD,Parsons+etal+2013+R_ell_binary,Soszynski+etal+2016+ell,Burdge+etal+2020+systematic,Rowan+etal+2021+Ell_ASASSN,Pelisoli+etal+2021_NatAst_ell,El-Badry+etal+2021+proto-ELM};
    BY~Dra variables \citep{Burggraaff+etal+2018+bright_variables,Chahal+etal+2022+BYDra_statistic};
    rotating magnetic WDs \citep{Ferrario+etal+1997+MWD,Pshirkov+etal+2020+WD1832,Caiazzo+etal+2021+moon,Williams+etal+2022+MWD}.
    \item \textbf{Ultracompact X-ray binaries} (UCXBs; \citealt{Heinke+etal+2001+XB1832,Wang+Chakrabarty+2004+4u1543,Shahbaz+etal+2008+0614,Zhong+Wang+2011+0918,Lin+Yu+2018+UCXB,Wang+etal+2021+UCXB}).
\end{itemize}

The International Variable Star Index (VSX; \citealt{Watson+etal+2006+VSX}) has provided a comprehensive catalogue for almost all known variable stars. 
Fig.~\ref{fig:vsx_statistic} shows the percentages of various classes for a total number of 14,254 short-period ($P<2$~hr) variable stars recorded in VSX (the version updated on May 30, 2022). 
More than 90\% of known short-period variable stars are pulsating stars, among which $\delta$~Scuti stars are the dominant subclass. 
Several rare and peculiar subclasses of pulsating stars can be distinguished from $\delta$~Scuti stars in terms of oscillation modes, temperatures, and luminosities, e.g., ZZ~Ceti stars \citep{Corsico+etal+2019+book+pulsatingWD}, sdBVs \citep{Kilkenny+etal+1997+361_discovery, Heber+2009+araa}, and blue large-amplitude pulsators \citep{Pietrukowicz+etal+2017+BLAPs,McWhirter+Lam+2022+blap_candidates}. 
Identification of rapidly oscillating Ap/Am stars and SX~Phe variables usually requires additional information, such as chemical abundances \citep{Holdsworth+etal+2014+roapam} and Galactic latitudes. 

Owing to the short-period cutoff at $P_{\rm orb}\sim 0.22$~day in contact eclipsing binaries \citep{Rucinski+1992+cutoff,Li+etal+2020+cutoff,Yang+etal+2020+LAMOST_binary}, short-period eclipsing binaries are moderately rare in current survey missions. 
The period for eclipsing binaries in the VSX database generally refers to the orbital period ($P_{\rm orb}$ hereafter) rather than the dominant photometric period ($P_{\rm pho}$ hereafter). 
The latter can be directly obtained from period-finding algorithms (e.g., Lomb-Scargle periodogram) and is usually half of $P_{\rm orb}$. 
Since we selected short-period variable stars (see Section~\ref{sec:data} in detail) using $P_{\rm pho}<2$~hr, our binary samples tend to include some binary systems with an orbital period between 2 and 4~hr, such as subdwarf binaries and M-dwarf binaries. 
Only a few dozen short-period binaries have been reported thus far \citep{Nefs+etal+2012+Mdwarfs,El-Badry+etal+2021+proto-ELM,Krzesinski+etal+2022+hst_binary}.
Furthermore, the light curves of some short-period eclipsing binaries show close resemblance to those of ellipsoidal variables \citep{Rowan+etal+2021+Ell_ASASSN,Krzesinski+etal+2022+hst_binary}. 
Because the eclipsing and ellipsoidal features are generally hard to distinguish for some short-period binary systems, it is also common to classify eclipsing and ellipsoidal binaries into a unified category \citep{Soszynski+etal+2015+USPbinary_OGLE}.

Cataclysmic variables can be divided into two groups according to the 2-to-3-hour period gap \citep{Spruit+Ritter+1983+period_gap}, and CVs with an orbital period below the gap account for a fraction up to $\sim 80$\% \citep{Drake+etal+2014+Catalina_CV}.
Since a period threshold of 2~hr helps us exclude most of eclipsing binaries (e.g., W~UMa binaries; \citealt{Xia+etal+2018+EW,Xia+etal+2021+contact}) and rotating stars (BY~Dra and RS~CVn stars; see \citealt{Chen+etal+2020+ZTFvariable}), the samples of CVs identified from short-period variables can thus be significantly less contaminated by other classes of variable stars.
According to the cases reported to the VSX, CVs account for $\sim 5$\% of short-period variable stars, and their number is second only to pulsating stars. 
The classes of short-period CVs include both magnetic and nonmagnetic CVs, in which the mechanisms of periodic light variations are not completely consistent \citep{Warner+1995+book_CV}.

\begin{figure*}
\centering
    \includegraphics[width=0.9\textwidth]{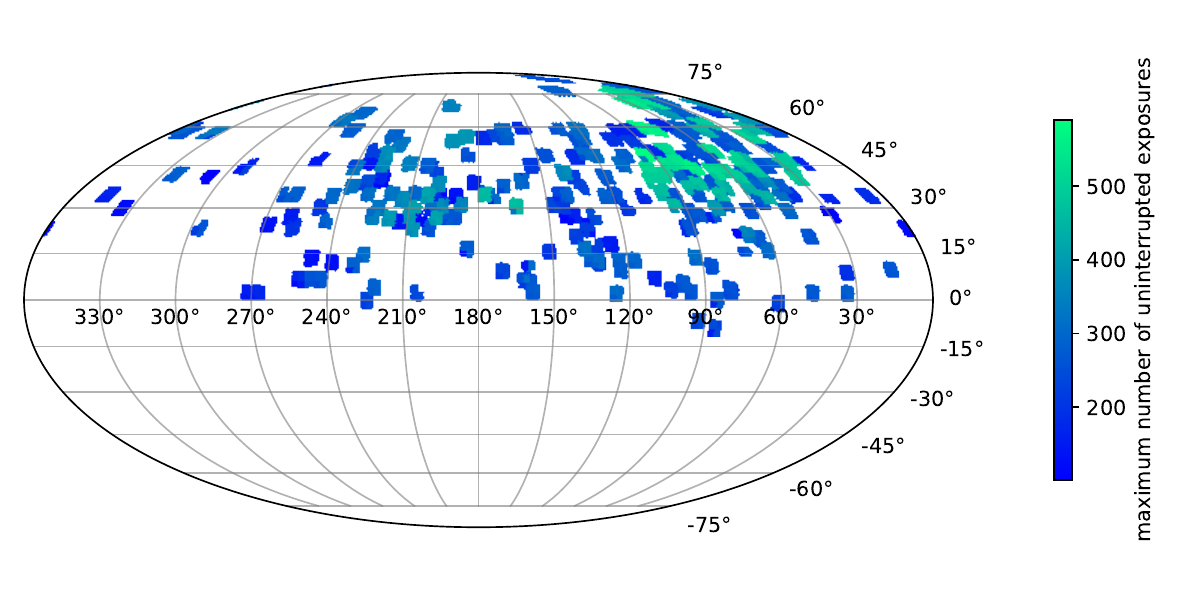}
    \caption{Sky areas of TMTS uninterrupted observations from 2020 to 2021 in equatorial coordinates. The colour represents the maximum number of uninterrupted exposures within an individual night.
    We used the {\sc HEALPIX} package (\url{ http://healpix.sourceforge.net}) with NSIDE=128 to plot this sky map \citep{Gorski+etal+2005}.
     } 
    \label{fig:skymap}
\end{figure*}

The short-period variations imply extraordinary physical conditions in pulsation, rotation, and orbital motion among these variable stars.
Hence, detections and observations of short-period variables enable studies of stellar physics under extreme conditions. 
Several frontier topics on short-period variable stars have emerged over the past decade, including the following.
\begin{itemize}
    \item \textbf{Verification binaries for gravitational waves (GWs)}. The verification binaries, which have an orbital period shorter than a few hours, play a key role in the functional tests of space GW observatories \citep{Shah+etal+2012+GW_aid}.
    The electromagnetic observations of verification binaries can help constrain independently the GW parameters such as frequency and inclination. A dozen verification binaries have been discovered by the Zwicky Transient Facility (ZTF) using the ultra-short periodicity signals of light curves \citep{Burdge+etal+2019+Nature+7minWD,Burdge+etal+2020+systematic,Burdge+etal+2020+9minute}. The ultracompact binaries (UCBs) with extremely short orbital periods could introduce new challenges for the theory of binary evolution \citep{Burdge+etal+2019+Nature+7minWD}. 
     \item \textbf{Ultracompact black hole binaries (UCBHBs)}. UCBHBs are the close binary systems consisting of a black hole (BH) and a star, which are expected to be detected by both electromagnetic and gravitational waves \citep{Chen+etal+2020+UCXB_LISA,Wang+etal+2021+UCXB}, thus providing the most direct evidence for the existence of stellar-mass black holes.
    Although tens of ultracompact X-ray binaries (UCXBs) have been revealed by X-ray observations, only one UCXB (47~Tuc~X9) was suspected to harbor a BH accretor \citep{Bahramian+etal+2017+BH_UCXB,Tudor+etal+2018+BH_UCXB_HST}. Population-synthesis studies predict that the total number of UCBHBs in the Milky Way could be over 10,000 \citep{Yungelson+etal+2006+BHUCXB_theory},  while the missing detections of short-period BH X-ray binaries might be understood as a consequence of radiatively inefficient accretion \citep{Menou+etal+1999,Wu+etal+2010+LMXB_period,Knevitt+etal+2014+inefficient_accretion} and the long recurrence time of X-ray outbursts \citep{Lin+etal+2019+outbursts}.
    Several BH binaries were discovered through periodic photometric variability and radial velocities (RVs; \citealt{Thompson+etal+2019+Science+noninteractive,Liu+etal+2019+Nature+LB}), supporting the notion that the search for black hole binaries (BHBs) is not exclusive to X-ray observations. The noninteracting or quiescent BHBs that cannot be detected by current X-ray detectors are able to be dynamically discovered by current optical survey missions \citep{Yi+etal+2019+search_BH}, such as the Large Sky Area Multi-object Fiber Spectroscopic Telescope (LAMOST).
    Since the photometry survey observations provide an opportunity to detect the short-period variations of ellipsoidal binaries harboring an unseen degenerate star \citep{Pelisoli+etal+2021_NatAst_ell},
    the wide-field photometry surveys have also the potential to search for UCBHBs.
    \item \textbf{Surviving remnants from mergers of binary WDs}. As the most numerous stellar graveyard in the Milky Way, $\sim 2.5\times10^8$ WDs reside in Galactic binary systems \citep{Nelemans+etal+2001+pop_WD,Napiwotzki+2009+pop_WD}. In some binary WDs, the components are sufficiently close to undergo mass transfer (i.e., AM~CVn stars) and evolve toward a final merger \citep{Wu+etal+2022+CO+He,Wu+etal+2023+ONe+CO}. 
    Depending on the masses of binary WDs, some mergers result in WDs with high magnetic field and rapid rotation, owing to strong magnetic dynamos and conservation of angular momentum during the mergers. 
    Recently, several rapidly rotating magnetic WDs have been discovered through their periodic photometric variations induced by rotation and high magnetic field \citep{Pshirkov+etal+2020+WD1832,Caiazzo+etal+2021+moon,Williams+etal+2022+MWD}. 
    These ultramassive WD merger remnants provide a window to study elemental mixing processes and angular momentum transport during the merger, and help us better understand the final fates of binary degenerates \citep{Yoon+etal+2007+CO_merger,Dan+etal+2014+WD_remnants}.
    \item \textbf{New realms of pulsating stars}. In the past few years, a new and rare class of hot pulsating stars with unusually large amplitudes and short periods, namely blue large-amplitude pulsators (BLAPs), were discovered by the Optical Gravitational Lensing Experiment (OGLE; \citealt{Pietrukowicz+etal+2017+BLAPs}). The high amplitudes in very hot pulsating stars challenge the current theory of stellar oscillations \citep{Corsico+etal+2019+book+pulsatingWD}. The physical origins of BLAPs are believed to be either helium-core pre-WDs or core helium-burning (CHeB) subdwarfs \citep{Pietrukowicz+etal+2017+BLAPs,Wu+etal+2018+CHeB_BLAP,Kupfer+2019+high-g_BLAPs,Pigulski+Kolaczek-Szymanski+2022+TESS_BLAP,Byrne+etal+2018+BLAPs,Byrne+etal+2020+faint_BLAP,Byrne+etal+2021+population_BLAP}. However, we recently proved that an 18.9~min BLAP (TMTS-BLAP-1) is a hot subdwarf at the short-lived phase of shell-helium ignition, which is prior to the stable shell-helium burning phase \citep{lin+etal+2022+NatAs,Xiong+etal+2022+SHeB}. We suggested that known BLAPs may have diverse physical origins through their distribution on the $\dot{P}/P-P$ diagram. 
    With the operations of wide-field survey missions, a growing sample of BLAPs will provide further details regarding their nature.
    \item \textbf{Transitional CVs}. The transitional CVs characterised by accretion of helium-rich material provide key evidence to support the evolutionary pathway from hydrogen-rich CVs into helium CVs (also known as AM~CVn stars; \citealt{Burdge+etal+2022+TCV+nature}). These transitional CVs are expected to have an orbital period between those of AM~CVn stars and the shortest-period hydrogen-rich CVs, bridging the evolution of orbit, mass transfer, and accretion between them. However, the transitional CVs are very rare \citep{Green+etal+2020+UCABs,Burdge+etal+2022+TCV+nature}, leading to an ambiguous link between hydrogen-rich and helium CVs. Another interesting type of transitional CVs is the magnetic CVs between the DQ~Her and AM~Her stars, as the former are thought to be the progenitors of the latter \citep{Chanmugam+Ray+1984+AMDQ,Patterson+1994+DQ}.
    Almost all known DQ~Her stars are above the period gap while most AM~Her stars are below the gap \citep{Pretorius+etal+2013+mCV_density,Pretorius+etal+2014+IP_density}, which supports such a hypothesis; however, whether DQ~Her stars will evolve into AM~Her stars or short-period DQ~Her stars (e.g., EX~Hya) is still unclear \citep{Southworth+etal+2007+longspin_IP,Norton+etal+2008+EXHYA}.
\end{itemize}

In order to search for short-period variable stars, the Tsinghua University--Ma Huateng Telescopes for Survey (TMTS), a multitube telescope system consisting of four 40~cm optical telescopes \citep{Zhang+etal+2020+tmts_performance}, started to monitor the LAMOST sky areas in the white-light band (i.e., TMTS $L$ band) since 2020. 
Thanks to the large field of view (FoV) of about 18~deg$^2$ and uninterrupted observation mode during the whole night, TMTS has detected a series of variable stars with a period below $\sim 7.5$~hr (see \citealt{tmtsI+2022}, hereafter \tmtsI).
In this paper, we focus on the periodic variable stars at the short-period end, $P_{\rm pho}<2$~hr, and release catalogues of these targets based on the data collected during the first-two-year survey. 
Since the orbital period of eclipsing/ellipsoidal binaries is usually twice their photometric period, our eclipsing/ellipsoidal binaries presented in this study may have an orbital period longer than 2~hr but still shorter than 4~hr.
The criteria adopted to identify periodic variable stars are described in Section~\ref{sec:data}, and the characteristics of the catalogues are introduced in Section~\ref{sec:catalog}. 
Section~\ref{sec:stars} discusses the properties of individual interesting short-period variable stars. 
A brief summary of the study is given in Section~\ref{sec:summary}.

\section{Data and Methods} \label{sec:data}

\begin{figure*}
\centering
    \includegraphics[width=0.9\textwidth]{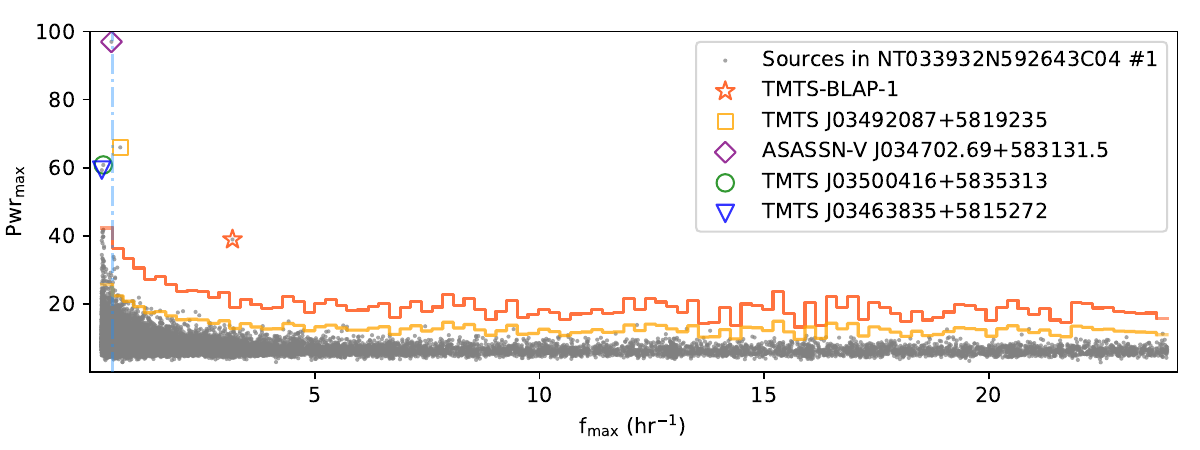}
    \caption{ 
    Pwr$_{\rm max}$--$f_{\rm max}$ diagram for sources captured by TMTS telescope \#1 in LAMOST plate NT033932N592643C04 on December 24, 2020. Pwr$_{\rm max}$ and $f_{\rm max}$ represent the power and frequency of the maximum peak in a Lomb–Scargle periodogram (LSP), respectively. The red and orange lines indicate the 10$\sigma$ and 5$\sigma$ thresholds, respectively. The vertical dot-dashed line indicates the frequency/period threshold (i.e., $P_{\rm pho}<2$~hr) for selection. 
    Periodic variable stars identified above the 10$\sigma$ threshold are highlighted by colour-coded symbols.
     } 
    \label{fig:fpmax_axample}
\end{figure*}

During the first-two-year survey, TMTS has monitored 251 LAMOST/TMTS plates\footnote{The counted plates here only include the observed plates with more than 100 visits, and thus these data can be adopted for our light-curve analysis.}, corresponding to a total sky coverage of 3\,857~deg$^2$, as shown in Fig.~\ref{fig:skymap}. 
These observations produced 10\,856\,233 uninterrupted light curves having at least 100 epochs for about ten million sources. 
We performed light-curve analysis and selected periodic variable sources with these uninterrupted photometric data.

\subsection{Selection}

We have introduced the period-finding algorithm using the Lomb–Scargle periodogram (LSP; \citealt{Lomb+1976,Scargle+1982}) and distribution of modified false-alarm probability (FAP) in \tmtsI. However, the distributions of FAPs are also dependent on the selected period ranges. 
An example for the distribution of the maximum powers in LSPs (hereafter Pwr$_{\rm max}$) computed with the TMTS light curves is shown in Figure~\ref{fig:fpmax_axample}; an increased Pwr$_{\rm max}$ toward lower frequencies is caused by red noise \citep{Vaughan+2005+rednoise,Vaughan+2010+rednoise}.
Obviously, a universal threshold across the entire period/frequency range tends to exclude candidates near the short-period end.
Hence, the selection threshold for each frequency range/bin was computed independently, as we are interested in short-period variable stars throughout this work.

Similar to the methods adopted in \tmtsI for variability detection at different magnitudes, we calculated the median and standard deviation (StD) of Pwr$_{\rm max}$ for each frequency bin, and thus determined the corresponding 5$\sigma$ and 10$\sigma$ thresholds.
Out of more than ten million light curves, 7\,319\,954 have a frequency of maximum LSP peak $f_{\rm max}$ above 0.5~hr$^{-1}$. Among them, 13\,311 light curves have values of Pwr$_{\rm max}$ above the 5$\sigma$ threshold while 1\,366 exceed the 10$\sigma$ threshold. By visually inspecting the 5$\sigma$ threshold light curves, we found a high false-positive rate. This is a common issue for high-cadence, single-night photometric observations, since some non-astrophysical variations can hardly be ruled out without resorting to multi-night observations \citep{Kupfer+etal+2021+ZTFhighcadence,tmtsI+2022}. Hence, we visually checked all light curves above the 10$\sigma$ threshold and selected those that exhibit signatures likely caused by astrophysically periodic variations. For light curves having a significance of periodic variations between 5$\sigma$ and 10$\sigma$, we also selected those interesting targets that appear to deviate from the main-sequence branch in the colour-magnitude diagram (CMD) according to {\it Gaia} measurements. The selection procedure described above results in a total of 1\,146 periodic light curves, corresponding to 1\,107 different stars. Periodic variable stars with relatively lower significance will be identified by machine-learning techniques in the future. Variable stars identified in this study at a high confidence level will be utilised to generate training and test datasets.

\subsection{Classification}
\begin{figure*}
\centering
    \includegraphics[width=0.9\textwidth]{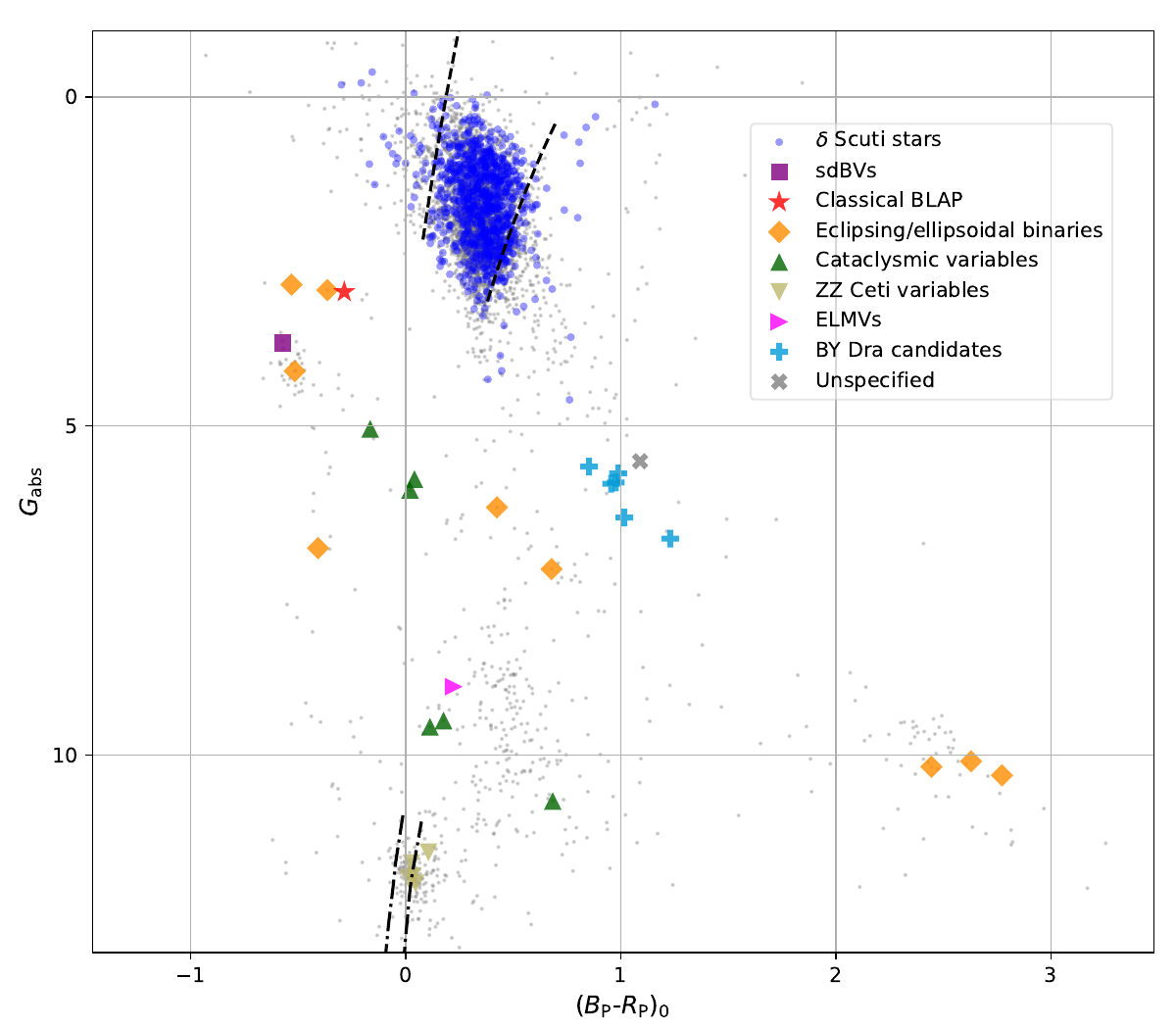}
    \caption{ 
    Distribution of short-period variable stars across the colour-magnitude diagram.
    Colour-coded symbols represent different types of short-period variable stars, among which the eclipsing/ellipsoidal binaries have an orbital period shorter than 4~hr and other variable stars have a period below 2~hr.
    The grey dots depict the short-period ($P<2$~hr) periodic variables reported to the VSX.
    The dashed lines and dot-dashed lines indicate the instability strip edges for $\delta$~Scuti \citep{Murphy+etal+2019+ds} and ZZ~Ceti \citep{Caiazzo+etal+2021+moon}, respectively.
     } 
    \label{fig:tmts_hrd}
\end{figure*}

\begin{figure*}
\centering
    \includegraphics[width=0.9\textwidth]{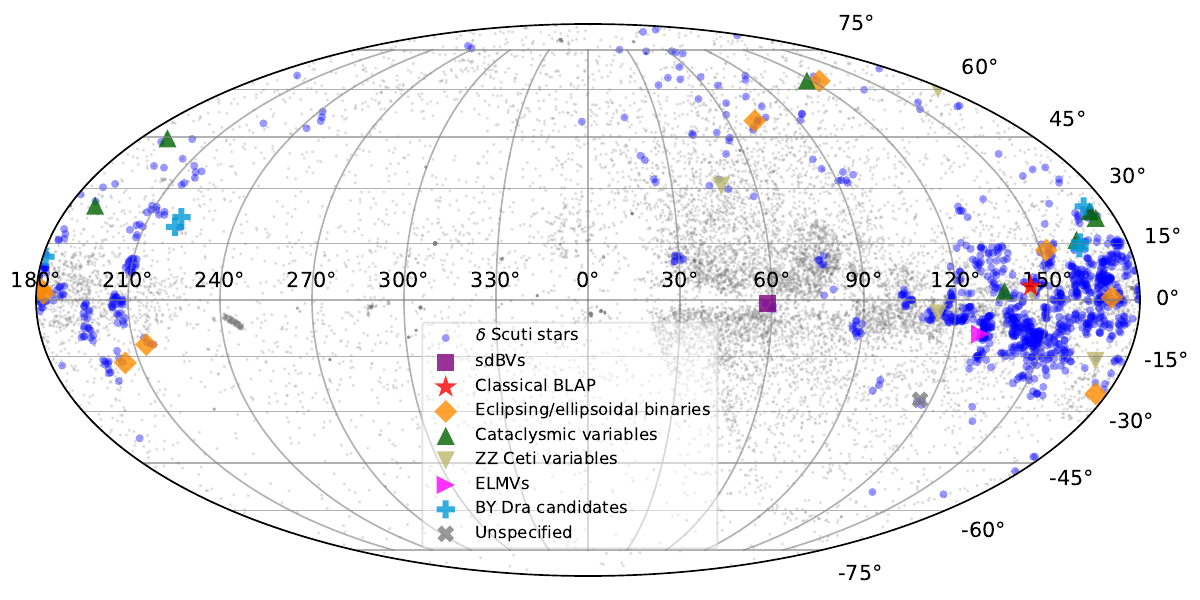}
    \caption{ 
    Distribution of short-period variable stars in Galactic coordinates.
    Colour-coded symbols represent different types of short-period variable stars selected from TMTS observations, and the grey dots depict the short-period ($P_{\rm pho}<2$~hr) periodic variables reported to the VSX. 
     } 
    \label{fig:tmts_Galaxy}
\end{figure*}

The colour-magnitude diagram (CMD) is a very powerful tool for preliminary classification of stars, especially for short-period variable stars, because variables of different categories are usually located in different regions of the CMD. Thanks to abundant data provided by {\it Gaia} \citep{Gaia_Collaboration+2016+performance,Gaia_collaboration+2018+data}, we can classify those selected periodic variable stars using corresponding extinction-corrected absolute magnitudes and colours in addition to the parameters derived from TMTS light curves. 
Among the selected targets, 954 stars have reliable parallax measurements (i.e., $\varpi/\sigma_\varpi \geq 5.0$). The absolute magnitudes of the stars are thus estimated as
\begin{equation}
G_{\rm abs}= G_{\rm obs} - 5\times \log_{10}{(\frac{{\rm 1~mas}}{\varpi } \times 100 )}- A_G \, ,
 \label{Eq:absG}
\end{equation}
where $G_{\rm obs}$ and $A_G$ represent the magnitude and interstellar extinction in the {\it Gaia} $G$ band, respectively. 
The interstellar reddening is obtained from the three-dimensional (3D) dust map \citep{Green+etal+2019+3dmap} using the \emph{DUSTMAPS Python} package\footnote{\url{ https://github.com/gregreen/dustmaps}} \citep{Green+2018+python}, and then corrected for the {\it Gaia} $G$ passband by assuming $A_G=0.835\times A_V$ \citep{GentileFusillo+etal+2019+gaia_wd}.
Similarly, the extinction in {\it Gaia} $B_P$ and $R_P$ passbands can be calculated as $A_{BP}=1.364\times A_G$ and $A_{RP}= 0.778\times A_G$. Thus, the dereddening colour can be computed as $(B-R)_0=G_{BP,{\rm obs}}-G_{RP,{\rm obs}}-0.586\times A_G$.

\begin{figure}
\centering
    \includegraphics[width=0.5\textwidth]{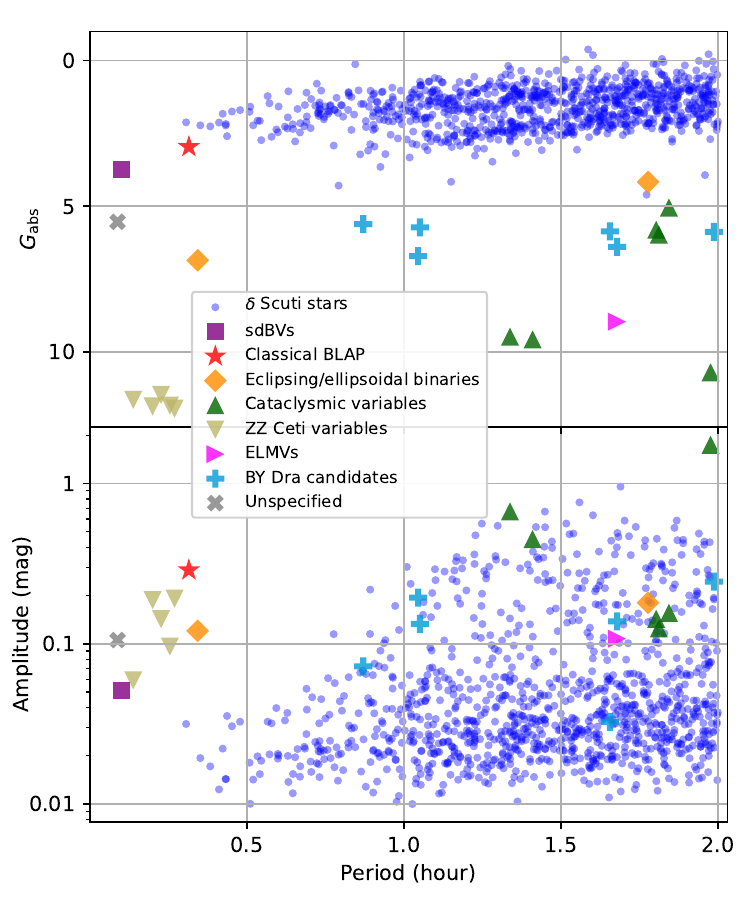}
    \caption{ 
    The Gaia $G$-band absolute magnitude (upper panel) and the peak-to-peak amplitude (lower panel) as a function of the pulsation/orbital period. A fraction of eclipsing/ellipsoidal binaries have an orbital period beyond 2~hr and are thus out of the period range of plot.
     } 
    \label{fig:tmts_period_relation}
\end{figure}

Fig.~\ref{fig:tmts_hrd} shows the short-period variable stars detected by the TMTS survey on the CMD. As identifications of rapidly oscillating Ap/Am stars and SX~Phe variables (i.e., old $\delta$~Scuti stars) require further information from spectra, we use ``$\delta$~Scuti stars'' to represent all short-period A-type and F-type main-sequence pulsating stars around the classical instability strip throughout this paper  \citep{Chen+etal+2020+ZTFvariable,Soszynski+etal+2021+24000deltascuti}.  
The distribution of $\delta$~Scuti stars spans slightly beyond the edges of the classical instability strip, which is likely owing to the uncertainties in extinction corrections or contributions from potential companion stars (see Section~\ref{sec:dsct_binary}). Furthermore, the boundaries of classical instability strip is expected to be further modified \citep{Bowman+Kurtz+2018+dsct_strip}. As shown in Fig.~\ref{fig:tmts_Galaxy}, some $\delta$~Scuti stars are found at high Galactic latitudes (e.g., $b>30^\circ$), which are very likely SX~Phe variables evolved from old stellar populations. 
About half of these $\delta$~Scuti stars have LAMOST (DR7) spectra, which can be used to differentiate chemically peculiar stars from normal stars without additional spectroscopic observations.
Fig.~\ref{fig:tmts_period_relation} also shows the presence of a few $\delta$~Scuti stars with dominant periods shorter than 30~min, which are similar to the roAp/Am stars \citep{Handler+Paunzen+1999+roAp,Holdsworth+etal+2014+roapam}. Hence, these ultra-short-period $\delta$~Scuti stars can be also taken as candidates of roAp/Am stars.

In principle, the eclipsing/ellipsoidal binaries can appear anywhere in the CMD. 
For short-period binaries with $P_{\rm orb} < 4$~hr, however, their components are unlikely to be early-type main-sequence stars because they are usually far larger than the inferred size of the Roche lobes. 
Moreover, all short-period binaries, except for M-dwarf binaries, should be located below the main-sequence branch in the CMD. 
Thus, most of the short-period eclipsing/ellipsoidal binaries selected from the TMTS data should consist of subdwarf binaries  \citep{Krzesinski+etal+2022+hst_binary,El-Badry+etal+2021+proto-ELM} and M-type dwarf binaries. 
These binaries span a wide colour range from $(G_{BP}-G_{RP})_0\approx -0.5$~mag to 2.5~mag (i.e., from type B to M). 
Owing to their compact orbits, these systems usually undergo strong tidal effects \citep{Soszynski+etal+2015+USPbinary_OGLE}.

Pulsating hot subdwarfs and BLAPs are located on the blue side of early-type main-sequence stars in the CMD, which is similar to the location of HW~Vir stars.
The pulsating hot subdwarfs cluster at regions of the extreme horizontal branch (EHB; \citealt{Heber+2009+araa,Luo+etal+2021+hotsubdwarf}), whereas BLAPs exhibit a wider range of distribution in the CMD \citep{McWhirter+Lam+2022+blap_candidates} since ``BLAP'' has become a general concept synonymous with ``high-temperature, short-period, and large-amplitude pulsating stars'' \citep{lin+etal+2022+NatAs}. 
However, our recent study suggested that some classical BLAPs (i.e., low-gravity BLAPs) may be evolved hot subdwarfs that have ended their core-helium burning and started shell-helium burning \citep{lin+etal+2022+NatAs}. 
From this point of view, classical BLAPs are expected to be located above the EHB in the CMD.

Pulsating white dwarfs are also a kind of commonly observed pulsating star, which exhibit a wide range of distribution in the CMD \citep{Corsico+etal+2019+book+pulsatingWD}. 
About 90\% of them belong to the hydrogen-enriched ZZ~Ceti subclass (i.e., DAVs; \citealt{Watson+etal+2006+VSX}). 
These ZZ~Ceti variables can be easily identified, since they are generally positioned within a narrow instability strip with a temperature of around 11\,500~K \citep{VanGrootel+etal++2012+ZZ} and have very short pulsation periods ($P \lesssim 20$~min; \citealt{Corsico+etal+2019+book+pulsatingWD}).  
However, some brighter pulsating WDs (e.g., GW~Vir variables and ELMVs) are above the instability strip of ZZ~Ceti, which could overlap with CVs.
Further distinction between these two classes relies on spectroscopic analysis and oscillation identifications \citep{Hermes+etal+2012+ELMV,Uzundag+etal+2021+GWVir}.

Owing to the radiation contributed from accretion discs, CVs (except for AM~CVn variables) are usually redder and brighter than ZZ~Ceti variables, and thus they are distributed on a wide region between the main-sequence and WD branches in the CMD. 
As a category of CVs at the short-period end, the AM~CVn variables lie between the sdB stars and WDs in the CMD (see \citealt{Burdge+etal+2020+systematic}), which can be understood as they both seem to have higher effective temperatures as a result of harboring more-compact disks. 
With special light-curve profiles (e.g., outbursts, eclipses, or superhumps) and strong emission lines in the spectra (outside the outbursts), CVs can be distinguished from other short-period variable stars \citep{Drake+etal+2014+Catalina_CV,Kato+etal+2013+DN,Kato+etal+2016+SUUMa,Sun+etal+2021+LAMOST_CV}.

\begin{table*}
\centering
\caption{ Descriptions for the headers of catalogues. \label{tab:format}}
\begin{tabular}{c|l}
\hline
Heading   &  Description  \\
\hline
$N$ & Order number.\\
Source I.D. & TMTS catalogue identifier.\\
R.A. & Right Ascension in decimal degrees (J2000).\\
Dec. & Declination in decimal degrees (J2000).\\
$P_{\rm pho}$ & Photometric period corresponding to the maximum power in Lomb–Scargle periodogram of TMTS. \\
$\sigma_{\rm P,pho}$& Error of photometric period. Because the time coverage of each light curve is within one night, the period uncertainty given \\& 
by TMTS LSP is much larger than that given by long-term observations as $\sigma_f \propto 1/T $. Hence, we appended the estimates for the\\& 
errors of periods using Eq.~52 of \cite{VanderPlas+2018+LSP}.\\
$P_{\rm pho}|P_{\rm orb}$  & Orbital period of eclipsing/ellipsoidal binaries or dominant photometric period of other periodic variable stars. The orbital period\\& 
$P_{\rm orb}$ is $2/1\times P_{\rm pho}$ for eclipsing/ellipsoidal binaries. \\
$\sigma_{\rm P,pho}|\sigma_{\rm P,orb}$ & Error of $P_{\rm pho}/P_{\rm orb}$. \\
Pwr$_{\rm max}$ & Maximum power in TMTS LSP. A higher Pwr$_{\rm max}$ means that the target is more likely to be a periodic variable star. \\
SNR & Signal-to-noise ratio of variability in the light curve (see \citealt{VanderPlas+2018+LSP}). \\
amp. & Peak-to-peak amplitude obtained from best-fitting model of fourth-order Fourier series plus second-order polynomial (see \tmtsI). \\
$L_0$ & Median magnitude obtained from the best-fitting model.\\
$R_{21}$ & $R_{21}=a_2/a_1$, the amplitude ratio derived from the best-fitting  Fourier series, where $a_1$ and $a_2$ are the amplitudes of 1st and 2nd\\&  components, respectively. \\
$\phi_{21}$ & $\phi_{21}=\phi_2-2\phi_1$, the phase difference derived from the best-fitting Fourier series, where $\phi_1$ and $\phi_2$ are the phases of 1st and 2nd \\&  components, respectively \\
$T_{\rm max}$ & Modified Julian Day (MJD) corresponding to maximum light of pulsating star.\\
$T_{\rm max}|T_{\rm min}$ & MJD corresponding to the maximum light of pulsating star or the minimum light of Eclipsing/ellipsoidal binary.\\
$G_{\rm abs}$ & Absolute magnitude derived from {\it Gaia} DR2 database.\\
$(B_P-R_P)_0$ & Dereddened colour derived from {\it Gaia} DR2 database.  We did not list $G_{\rm abs}$ and $(B_P-R_P)_0$ for those targets without reliable  \\& parallax measurements.\\
$T_{\rm eff}$ & Mean effective temperature provided by LAMOST DR7.\\
$\log \, g$ &  Mean surface gravity provided by LAMOST DR7.\\
Spec. type & Spectral class provided by LAMOST DR7.\\
VSX type & VSX variability type. Detail explanations for VSX variability types can be found in\\& \url{https://www.aavso.org/vsx/index.php?view=about.top} . \\
Spec. characteristics & Description of spectroscopic characteristics. The parentheses in this column denote that the components are not significant.\\
Classification & Classification given by this work.\\
Notes & ``M'' represents that the TMTS photometry of target might be polluted by adjacent star(s), and ``C'' indicates that the  \\&
identification is absent of a reliable distance, LAMOST spectral parameters, and VSX classifications.\\
\hline
\end{tabular}
\end{table*}

Note that the distribution of short-period variable stars shows a ``desert'' region in the CMD, which coincides with that of the G-type and K-type main-sequence stars (see grey dots in the background of Fig.~\ref{fig:tmts_hrd}). The existence of this less-populated region can be caused by two reasons: one is that the short-period binaries (e.g., $P_{\rm orb} < 4$~hr) are not allowed to harbor main-sequence stars earlier than M dwarfs; the other is that the amplitudes of solar-like oscillations are as small as 100 parts per million (ppm; \citealt{Mathur+etal+2022+solar-like_oscillations}), making them very difficult to be revealed by ground-based survey telescopes. 
Although there are several short-period G-/K-type main-sequence variables that have been discovered occasionally, their classifications are still ambiguous \citep{Pojmanski+2002+ASAS_6h,Barclay+etal+2011+RTS_amcvn,Ramsay+etal+2014+Kepler_deep}.
In recent years, however, several ultrafast rotating main-sequence stars were discovered \citep{Chahal+etal+2022+BYDra_statistic,Ramsay+etal+2022+rotation}. 
These main-sequence stars can produce periodic variations faster than 2~hr owing to starspots and fast rotation.
In this paper, we also attempt to identify several candidates of short-period variable stars, which overlap with the G-/K-type main sequence in the CMD. 
However, we are not able to discriminate whether the periodic variations detected from these sources are caused by non-astrophysical effects or stellar activity because their photometric periodicities (except for TMTS~J00213079+3541576) lack additional observational evidence from other surveys.

Besides the CMD, we also identified a small fraction of TMTS short-period variable stars by combining their light curves and other parameters derived from the LAMOST spectra (i.e., spectral class; \citealt{Cui+etal+2012+LAMOST,Zhao+etal+2012+LAMOST}), or referring to the previous identifications in the VSX \citep{Watson+etal+2006+VSX}. 
In order to achieve a better understanding of the physical properties of certain variable stars, we also investigated the data from other survey missions, including ZTF \citep{ZTF+2019+first,ZTF+2019+products,Chen+etal+2020+ZTFvariable}, Transiting Exoplanet Survey Satellite ({\it TESS}; \citealt{Ricker+etal+2014+TESS,TESS+2015}). and Asteroid Terrestrial-impact Last Alert System (ATLAS; \citealt{Tonry+etal+2018+ATLAS,Heinze+etal+2018+ATLAS_variables}).
Spectroscopic follow-up observations with the Keck-I 10~m telescope, the Lick 3~m Shane telescope, and the Xinglong 2.16~m telescope, as well as X-ray imaging from {\it Swift}/XRT, were carried out for some targets of interest.

\section{Catalogues} \label{sec:catalog}

In this section, we released a catalogue containing pulsating stars, eclipsing/ellipsoidal binaries, cataclysmic variables, and BY~Dra candidates. 
We tabulated the short-period $\delta$~Scuti stars in a stand-alone catalogue owing to its large number (i.e., $N > 1\,000$). 
Since our samples contain both binaries and pulsating stars, the periods given here represent orbital period for eclipsing/ellipsoidal binaries and dominant photometric period for other periodic variable stars. 
Details of the parameters listed in the above catalogues are described in Table~\ref{tab:format}.

\subsection{$\delta$ Scuti stars} 

In Table~\ref{tab:dsct} we present the $\delta$~Scuti stars identified in the TMTS survey. Only the first 30 lines are displayed for the purpose of illustration. The complete and machine-readable version of the catalogue is available online.

\begin{landscape}
\begin{table}
\centering
\scriptsize
\caption{Example catalogue for $\delta$ Scuti stars identified from TMTS observations.
\label{tab:dsct}
}
\begin{tabular}{llccccccccccccccccccc}
\hline\hline
Source I.D.&R.A.&Dec.&$P_{\rm pho}$&$\sigma_{\rm P,pho}$&Pwr$_{\rm max}$&SNR&Amp.&$L_0$&$R_{21}$&$\phi_{21}$&$T_{\rm max}$&$G_{\rm abs}$&$(B_P-R_P)_0$&$T_{\rm eff}$&$\log \, g$&Spec. type&VSX type&Notes\\
 &degree&degree&minute&minute&&&mag&mag& & &MJD&mag&mag&K&cm$\rm \,s^{-2}$&& &\\
 \hline
TMTS J00000295+5755393&   0.01230&  57.92758&  87.60&  0.30&  68.22&  6.50& 0.048& 12.602& 0.075& 4.521& 59155.54595&  0.870& 0.408&     -&    -&          -&        -&    -\\
TMTS J00030484+5839399&   0.77017&  58.66109& 102.62&  1.20&  58.85&  2.24& 0.040& 13.885& 0.085& 8.136& 59155.55021&  1.798& 0.280&     -&    -&          -&        -&    -\\
TMTS J00032031+7119365&   0.83461&  71.32682& 107.47&  0.75&  60.45&  1.86& 0.140& 15.318& 0.357& 4.541& 59508.70499&  1.012& 0.811&     -&    -&          -&       EA&    -\\
TMTS J00034823+6003276&   0.95097&  60.05768& 115.85&  1.54&  59.33&  2.15& 0.040& 13.865& 0.129& 7.584& 59155.53209&  2.242& 0.489&     -&    -&          -&        -&    -\\
TMTS J00062727+5742415&   1.61363&  57.71152& 119.72&  0.70&  61.08&  5.21& 0.047& 12.969& 0.109& 9.401& 59155.52518&  2.389& 0.500&     -&    -&          -&        -&    -\\
TMTS J00063451+6112288&   1.64378&  61.20800& 115.85&  2.23&  40.17&  1.48& 0.200& 16.311& 0.206& 4.183& 59155.59557&  1.824& 0.336&     -&    -&          -&     DSCT&    -\\
TMTS J00075253+6111243&   1.96888&  61.19009& 112.23&  1.55&  64.91&  2.00& 0.067& 14.464& 0.122& 6.172& 59155.58169&  1.271& 0.518&     -&    -&          -&     DSCT&    -\\
TMTS J00082966+6419466&   2.12360&  64.32962&  72.23&  0.21&  79.51&  3.16& 0.028& 12.050& 0.100& 9.420& 59507.66060&  1.764& 0.384&     -&    -&          -&     DSCT&    -\\
TMTS J00083793+6213007&   2.15803&  62.21685&  81.09&  0.31&  75.84&  2.40& 0.031& 13.159& 0.073& 5.161& 59507.68222&  0.976& 0.208&     -&    -&          -&        -&    -\\
TMTS J00094541+6401239&   2.43920&  64.02331&  74.90&  0.40&  78.76&  1.81& 0.205& 15.862& 0.315& 4.190& 59507.68452&  2.612& 0.278&     -&    -&          -&        -&    M\\
TMTS J00125240+5700374&   3.21834&  57.01040&  43.84&  0.15&  31.86&  3.80& 0.015& 11.955& 0.163& 5.224& 59181.55410&  1.688& 0.195&     -&    -&          -&        -&    -\\
TMTS J00130667+6316466&   3.27780&  63.27962& 116.68&  0.94&  50.26&  1.87& 0.023& 13.129& 0.280& 6.140& 59507.64742&  2.032& 0.394&     -&    -&          -&        -&    -\\
TMTS J00131092+6317429&   3.29549&  63.29526&  49.33&  0.19&  33.85&  1.63& 0.024& 13.724& 0.217& 5.360& 59507.66081&  1.701& 0.310&     -&    -&          -&        -&    -\\
TMTS J00135451+6453172&   3.47712&  64.88810&  32.45&  0.08&  52.22&  1.69& 0.015& 12.793& 0.067& 6.843& 59507.66849&  2.076& 0.277&     -&    -&          -&        -&    -\\
TMTS J00144050+5630201&   3.66877&  56.50558& 109.61&  1.53&  72.10&  2.29& 0.387& 16.082& 0.268& 3.619& 59181.57603&  1.634& 0.289&     -&    -&          -&        -&    -\\
TMTS J00145990+6330500&   3.74957&  63.51416& 114.48&  1.23&  64.13&  1.37& 0.200& 16.213& 0.345& 4.771& 59507.70009&      -&     -&     -&    -&          -&     HADS&    -\\
TMTS J00154814+5837022&   3.95057&  58.61728&  67.48&  0.69&  42.18&  1.94& 0.013& 12.531& 0.170& 7.772& 59181.53563&  1.477& 0.193&  7699& 3.94&        A7V&        -&    -\\
TMTS J00165550+5639530&   4.23127&  56.66473&  68.51&  0.88&  36.46&  1.57& 0.029& 13.935& 0.204& 8.145& 59181.53632&  1.310& 0.312&     -&    -&          -&        -&    -\\
TMTS J00172250+6246527&   4.34373&  62.78131&  66.73&  0.39&  42.96&  1.30& 0.021& 13.868& 0.223& 7.644& 59507.68978&  1.149& 0.226&     -&    -&          -&        -&    -\\
TMTS J00173172+5752128&   4.38218&  57.87022&  96.72&  1.69&  40.97&  1.62& 0.028& 13.673& 0.057& 5.524& 59181.53739&  1.611& 0.374&     -&    -&          -&        -&    -\\
TMTS J00181589+3507585&   4.56623&  35.13293&  96.60&  1.62&  50.90&  2.52& 0.176& 15.232& 0.121& 9.383& 59148.55763&      -&     -&     -&    -&        A5V&        -&    -\\
TMTS J00201655+5803492&   5.06897&  58.06367&  86.53&  0.49&  65.63&  4.46& 0.035& 12.589& 0.079& 8.709& 59181.55057&  1.126& 0.369&     -&    -&          -&        -&    -\\
TMTS J00221703+5711194&   5.57097&  57.18872& 106.07&  2.11&  45.94&  1.71& 0.442& 17.295& 0.241& 3.420& 59181.55341&      -&     -&     -&    -&        A3V&     HADS&    -\\
TMTS J00241719+6751318&   6.07162&  67.85882& 104.06&  0.62&  37.35&  2.22& 0.020& 12.405& 0.216& 6.300& 59508.70605&  0.440& 0.124&     -&    -&          -&        -&    -\\
TMTS J00242525+5818415&   6.10523&  58.31152& 106.07&  1.55&  64.93&  2.12& 0.090& 14.606& 0.213& 4.818& 59181.52403&  0.641& 0.463&     -&    -&          -&     HADS&    -\\
TMTS J00250982+3540236&   6.29090&  35.67323&  60.37&  0.43&  45.82&  3.76& 0.040& 13.307& 0.111& 5.909& 59148.52054&  1.732& 0.124&     -&    -&       A2IV&        -&    -\\
TMTS J00252753+6922548&   6.36470&  69.38190&  70.90&  0.38&  35.85&  1.67& 0.027& 13.903& 0.258& 8.618& 59508.68314&  1.066& 0.140&     -&    -&          -&        -&    -\\
TMTS J00275878+3428136&   6.99490&  34.47045&  73.64&  0.48&  70.20&  4.87& 0.475& 15.740& 0.270& 3.610& 59148.51399&      -&     -&     -&    -&       A3IV&        -&    -\\
TMTS J00304659+5647553&   7.69413&  56.79869&  73.47&  0.99&  25.49&  1.71& 0.026& 13.375& 0.096& 8.002& 59181.56854&  1.613& 0.379&  7425& 4.01&         F0&        -&    -\\
TMTS J00304869+5717054&   7.70289&  57.28482&  44.68&  0.22&  41.72&  2.84& 0.022& 12.568& 0.251& 8.496& 59181.53668&  2.023& 0.202&     -&    -&          -&        -&    -\\

\hline
\end{tabular}
\end{table}
\end{landscape}

\begin{figure}
\centering
    \includegraphics[width=0.5\textwidth]{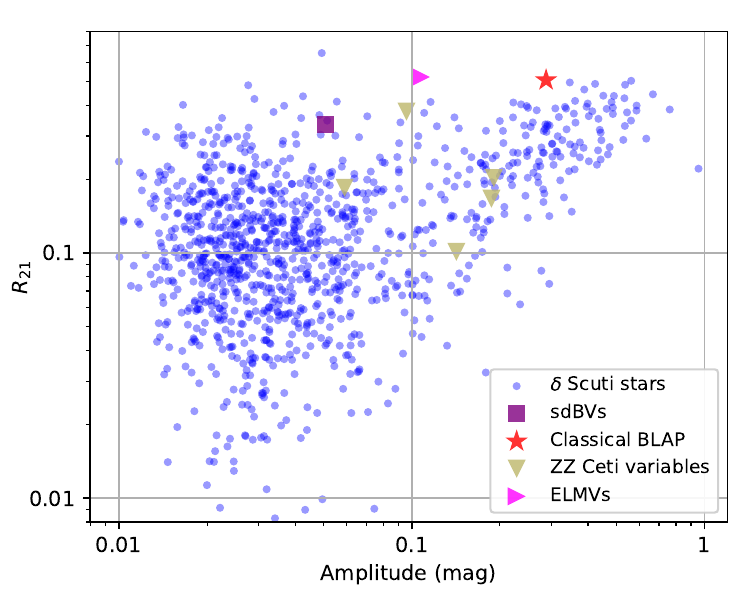}
    \caption{ 
    Relation between peak-to-peak amplitude and $R_{21}$ for pulsating stars.
    A higher $R_{21}$ means that the light-curve shapes of pulsations are more sawtooth or asymmetric.
     } 
    \label{fig:tmts_r21_amplitude}
\end{figure}

In the complete catalogue of 1076 sources, 77 targets are found to have at least two {\it Gaia} counterparts within $3\arcsec$. For these objects, the TMTS photometric results are generally consistent with those of brighter {\it Gaia} objects. Flux contamination by any fainter sources adjacent to the identified short-period variables is negligible in most cases. 
Among these $\delta$~Scuti samples, only five sources show two {\it Gaia} counterparts within $3\arcsec$ which exhibit similar $G$-band magnitudes ($\Delta\,G < 1$~mag). These five sources with non-negligible contamination were labeled with ``M'' (see the ``notes'' column in Table~\ref{tab:dsct}).
We note that 151 short-period $\delta$~Scuti stars do not have reliable parallax measurements, among which 47 stars exhibit A or F spectral types as verified by LAMOST. Besides, there are 58 sources that have been identified as $\delta$~Scuti stars in the VSX. The remaining 46 $\delta$~Scuti candidates are labeled with the letter ``C'' in the ``notes'' column of Table~\ref{tab:dsct}, implying that these stars require further confirmation.

The high-cadence observation mode of TMTS allows us to detect numerous short-period $\delta$~Scuti stars and hence provides a better understanding of the physical parameters of pulsators toward shorter periods.
Because of the amplitude variability of various oscillation modes in $\delta$~Scuti stars \citep{Bowman+etal+2016+variability}, the dominant oscillation mode (i.e., the mode corresponding to the maximum power in LSP) might switch with time \citep{Ziaali+etal+2019+PL_relation}, implying that the dominant period does not always represent the pulsation period caused by the fundamental-mode oscillation. 
The relationship between absolute magnitude and period measured for the TMTS short-period variable stars is shown in the upper panel of Fig.~\ref{fig:tmts_period_relation}, where we can see two ridges in the distribution of $\delta$~Scuti stars. 
The lower ridge corresponds to the well-known period-luminosity relation caused by the fundamental-mode oscillation in $\delta$~Scuti stars \citep{McNamara+2011+PL_relation,Ziaali+etal+2019+PL_relation}, while the upper ridge reflects the shorter-period overtones that may be excited by 2:1 resonances with the fundamental modes \citep{Ziaali+etal+2019+PL_relation}.
The overtone periods also seem to follow a period-luminosity relation.

The lower panel of Fig.~\ref{fig:tmts_period_relation} shows the relation between amplitudes and periods of different types of variables, from which we notice that the average amplitude of $\delta$~Scuti stars tends to decrease significantly toward shorter pulsation periods. Such a trend has also been seen in the samples of short-period $\delta$~Scuti stars discovered by the OmegaWhite survey (Fig.~14 of \citealt{Toma+etal+2016+omegaII}).
The relatively large amplitudes of light variations seen in other classes of pulsating stars such as ZZ~Ceti variables, sdBVs, and some ultracompact binaries \citep{Burdge+etal+2020+systematic} can help distinguish them from short-period $\delta$~Scuti stars in case their absolute magnitude and colour information are unavailable.

As shown in Fig.~\ref{fig:tmts_r21_amplitude}, a small fraction of $\delta$~Scuti stars are found to have both high amplitudes and amplitude ratios $R_{21}$, implying that they may belong to a class of high-amplitude $\delta$~Scuti (HADS) stars. 
Since HADS and low-amplitude $\delta$~Scuti (LADS) stars are only marginally separated by their amplitudes \citep{Chen+etal+2020+ZTFvariable}, we do not subdivide them in our catalogue. 
Further statistical studies of the $\delta$~Scuti stars (including those longer-period $\delta$~Scuti stars) captured by the TMTS will be presented in a forthcoming work (Chen L. et al., in prep.).

\subsection{Short-period variable stars (excluding $\delta$~Scuti stars)} 

\begin{table*}
\centering
\scriptsize
\caption{Photometric information for the TMTS short-period variable stars (excluding $\delta$ Scuti stars).
\label{tab:catalog_part1}
}
\begin{tabular}{llccccccccccccccc}
\hline\hline
N& Source I.D.& R.A.& Dec.& $P_{\rm pho}|P_{\rm orb}$& $\sigma_{\rm P,pho}|\sigma_{\rm P,orb}$& Pwr$_{\rm max}$& SNR& Amp.& $L_0$& $R_{21}$& $\phi_{21}$& $T_{\rm max}$|$T_{\rm min}$\\
 &  & degree& degree& minute& minute& & & mag& mag&  &  & MJD\\
\hline            
\multicolumn{13}{c}{Pulsating stars (excluding $\delta$ Scuti stars)}\\            
\hline            
1& TMTS J03514363+5845042&   57.93180&   58.75118&   18.89&   0.06&   48.25&   1.48&  0.288&  16.962&  0.507&  3.725&  59208.53890\\
2& TMTS J19441507+2201351&  296.06279&   22.02643&    5.97&   0.01&   16.46&   1.41&  0.051&  15.244&  0.333&  4.572&  59399.68002\\
3& TMTS J04185665+2717472&   64.73606&   27.29645&    8.26&   0.01&   25.08&   2.82&  0.059&  15.056&  0.184&  3.628&  59218.55077\\
4& TMTS J17184064+2524314&  259.66932&   25.40873&   11.96&   0.03&   17.22&   1.66&  0.187&  16.137&  0.166&  3.580&  58978.74196\\
5& TMTS J03453796+5707553&   56.40817&   57.13203&   13.58&   0.04&   14.38&   1.16&  0.142&  16.966&  0.101&  4.086&  59208.53738\\
6& TMTS J10423368+4057149&  160.64034&   40.95413&   15.26&   0.02&   20.10&   1.81&  0.096&  16.132&  0.377&  3.592&  58867.76697\\
7& TMTS J23450729+5813146&  356.28038&   58.22072&   16.16&   0.04&   37.79&   1.36&  0.191&  16.813&  0.202&  4.091&  59155.56067\\
8& TMTS J01305822+5321376&   22.74259&   53.36046&  100.69&   1.44&   56.97&   1.87&  0.107&  14.880&  0.522&  9.174&  59506.77324\\
\hline            
\multicolumn{13}{c}{Eclipsing/ellipsoidal binaries}\\            
\hline            
9& TMTS J05261043+5934451&   81.54345&   59.57919&   20.57&   0.06&   15.13&   1.23&  0.120&  17.617&  2.886&  8.283&  59201.65843\\
10& TMTS J05375319$-$0049266&   84.47161&   -0.82407&  106.62&   1.45&   57.32&   1.84&  0.180&  15.175&  0.030&  3.469&  59244.58246\\
11& TMTS J05574282+2746513&   89.42841&   27.78091&  180.35&   9.32&   44.42&   1.82&  0.239&  15.368&  3.818&  5.436&  58886.54097\\
12& TMTS J06003104+2908545&   90.12932&   29.14847&  222.45&  18.61&   32.08&   1.39&  0.024&  13.177&  2.440&  6.211&  58886.53190\\
13& TMTS J04081376+1652166&   62.05734&   16.87126&  188.70&   8.39&   58.00&   1.77&  0.429&  17.083&  1.836&  5.513&  59216.52619\\
14& TMTS J05243748+3700332&   81.15615&   37.00922&  208.56&   4.49&  105.04&   2.08&  0.396&  16.915&  3.386&  7.880&  59577.66151\\
15& TMTS J12483074+5408028&  192.12810&   54.13411&  220.70&   9.03&   79.31&   1.66&  0.239&  17.065&  6.114&  4.047&  58923.76628\\
16& TMTS J06111082$-$0622067&   92.79510&   -6.36852&  231.55&  14.70&   39.26&   2.04&  0.567&  16.884&  2.693&  5.100&  59247.52391\\
17& TMTS J15530469+4457458&  238.26953&   44.96274&  238.42&   4.23&  112.34&   5.46&  0.133&  13.159&  6.505&  3.318&  58973.64015\\
\hline            
\multicolumn{13}{c}{Cataclysmic variables}\\            
\hline            
18& TMTS J13075377+5351303&  196.97402&   53.85843&   80.25&   0.22&  125.56&   4.51&  0.669&  16.664&  0.209&  5.221& -\\
19& TMTS J07200739+4516113&  110.03081&   45.26980&   92.19&   2.01&   23.52&   3.45&      -&       -&      -&      -& -\\
20& TMTS J07112595+4404048&  107.85811&   44.06800&  118.56&   0.13&  119.32&  22.89&  1.743&  16.546&  0.095&  4.711& -\\
21& TMTS J06183036+5105550&   94.62649&   51.09888&  108.18&   0.38&   48.13&   2.60&  0.143&  16.054&  0.394&  3.800& -\\
22& TMTS J07485955+3125121&  117.24814&   31.42001&   84.56&   0.87&   30.16&   2.92&  0.449&  15.372&  0.593&  8.901& -\\
23& TMTS J09201115+3356421&  140.04647&   33.94503&  108.68&   0.42&   95.90&   6.38&  0.125&  14.416&  0.343&  4.508& -\\
24& TMTS J02461608+6217029&   41.56699&   62.28413&  110.60&   0.44&  170.38&   2.69&  0.156&  15.093&  0.256&  4.053& -\\

\hline
\end{tabular}
\end{table*}

\begin{table*}
\centering
\scriptsize
\caption{ Additional information from other catalogues and follow-up observations for the short-period variable stars (excluding $\delta$~Scuti stars).
\label{tab:catalog_part2}
}
\begin{tabular}{lcccccccccccccccc}
\hline\hline
N& Source I.D.& $G_{\rm abs}$& $(B_P-R_P)_0$& $T_{\rm eff}$& $\log \, g$& Spec. type& VSX type& Spec. characteristics& Classification\\
 &  & mag& mag& K& cm$\rm \,s^{-2}$& &  & & \\
\hline         
\multicolumn{10}{c}{Pulsating stars (excluding $\delta$ Scuti stars)}\\         
\hline         
1& TMTS J03514363+5845042&   2.966& -0.286&      -&     -&           -&         -& H/He I/He II absorption& classical BLAP\\
2& TMTS J19441507+2201351&   3.742& -0.574&      -&     -&           -&         -& sdB, H absorption& sdBV\\
3& TMTS J04185665+2717472&  11.645&  0.025&      -&     -&          WD&       ZZA& DA& ZZ Ceti\\
4& TMTS J17184064+2524314&  11.870&  0.012&      -&     -&           -&         -& DA& ZZ Ceti\\
5& TMTS J03453796+5707553&  11.477&  0.105&      -&     -&           -&         -& DA& ZZ Ceti\\
6& TMTS J10423368+4057149&  11.840&  0.032&      -&     -&          WD&       ZZA& DA& ZZ Ceti\\
7& TMTS J23450729+5813146&  11.948&  0.046&      -&     -&           -&         -& DA& ZZ Ceti\\
8& TMTS J01305822+5321376&   8.962&  0.223&      -&     -&           -&         -& wide H absorption& ELMV\\
\hline         
\multicolumn{10}{c}{Eclipsing/ellipsoidal binaries}\\         
\hline         
9& TMTS J05261043+5934451&   6.857& -0.408&      -&     -&           -&         -& H absorption& Ultracompact ellipsoidal binary\\
10& TMTS J05375319$-$0049266&   4.169& -0.517&      -&     -&           -&     EA/HW& sdB, H absorption& HW Vir\\
11& TMTS J05574282+2746513&   2.939& -0.365&      -&     -&       A0III&       VAR& sdB, H/He I/He II absorption& hot subdwarf binary\\
12& TMTS J06003104+2908545&   2.854& -0.531&   7959&  2.00&          B6&         -& sdB, H/He I/He II absorption& hot subdwarf binary\\
13& TMTS J04081376+1652166&   6.236&  0.424&      -&     -&           -&       VAR& H absorption& proto-ELM WD\\
14& TMTS J05243748+3700332&   7.175&  0.678&   5683&  4.39&          F9&       VAR& H absorption& proto-ELM WD\\
15& TMTS J12483074+5408028&  10.178&  2.444&   3400&  5.19&         dM5&       VAR& H emission& M-dwarf binary\\
16& TMTS J06111082$-$0622067&  10.095&  2.628&      -&     -&           -&         -& weak H emission& M-dwarf binary\\
17& TMTS J15530469+4457458&  10.309&  2.772&      -&     -&           -&      DSCT& H emission, double-lined& M-dwarf binary\\
\hline         
\multicolumn{10}{c}{Cataclysmic variables}\\         
\hline         
18& TMTS J13075377+5351303&   9.476&  0.176&      -&     -&          CV&        AM& H/He I/He II emission& Polar\\
19& TMTS J07200739+4516113&       -&      -&      -&     -&          CV&        UG& H/He I emission& Polar\\
20& TMTS J07112595+4404048&  10.698&  0.683&      -&     -&          CV&      AM+E& H/He I/He II emission& Polar\\
21& TMTS J06183036+5105550&   5.808&  0.040&      -&     -&           -&         -& H/He I(/He II) emission& intermediate polar candidate\\
22& TMTS J07485955+3125121&   9.570&  0.113&      -&     -&       A0III&    UGSU+E& Double-peaked H emission& dwarf Nova\\
23& TMTS J09201115+3356421&   5.973&  0.020&      -&     -&           -& NA:+UGER+NL& -& dwarf Nova\\
24& TMTS J02461608+6217029&   5.046& -0.165&      -&     -&           -&     UGER:& -& dwarf Nova\\

\hline
\end{tabular}
\end{table*}

Apart from $\delta$~Scuti stars, we also discovered some interesting short-period variable stars during the first-two-year survey, including BLAPs, sdBVs, pulsating white dwarfs, ultracompact/short-period eclipsing/ellipsoidal binaries, polars, and dwarf novae, which are summarised in Table~\ref{tab:catalog_part1} and Table~\ref{tab:catalog_part2}.
\begin{itemize}
    \item \textbf{Pulsating stars (with $\delta$~Scuti stars excluded)}. Among the list of pulsating stars, two ZZ~Ceti variables, TMTS~J04185665+2717472 (= V0411~Tau; \citealt{Landolt+1968+new_blue_ZZ,Fitch+1973+ZZ_period,Bognar+Sodor+2016+ZZ_catalog}) and TMTS~J10423368+4057149 (= WD~1039+412; \citealt{Bognar+Sodor+2016+ZZ_catalog}), have been discovered by previous studies and reported to the VSX. TMTS~J01305822+5321376 (= GD~278) has recently been identified as an extremely low-mass WD variable (ELMV) with {\it TESS} observations \citep{Lopez+etal+2021+GD278}.
    \item \textbf{Eclipsing/ellipsoidal binaries}. Most of the eclipsing/ellipsoidal binaries observed by the TMTS survey have already been identified as periodic variable stars by observations of ATLAS \citep{Heinze+etal+2018+ATLAS_variables}. However, only one of them (TMTS~J05375319$-$0049266 = ATO~J084.4719$-$00.8239) has been classified as a binary (HW~Vir) in the VSX \citep{Schaffenroth+etal+2019+Short_binary}. The ``proto-ELM WDs'' belong to a new class of binary systems which can be characterised by the presence of an evolved secondary star. Such a binary system is expected to fall below the main sequence in the CMD and shares similar temperatures with the AFGK-type main-sequence stars \citep{El-Badry+etal+2021+proto-ELM}.
    \item \textbf{Cataclysmic variables}. Many CVs listed in the catalogue were already discovered in the past, owing to their dramatic luminosity variations during outbursts of dwarf novae and strong emission lines generated from ionised material in the accretion disk \citep{Sun+etal+2021+LAMOST_CV}. The short-period variations seen in these CVs can be caused by eclipses, superhumps \citep{Kato+etal+2013+DN,Kato+etal+2016+SUUMa}, or inhomogeneous accretion \citep{Liebert+Stockman+1985+inho_accretion}.
    We notice that the CV candidate TMTS~J07200739+4516113 (= ZTF18aabrbjr; \citealt{Forster+etal+2021+ALeRCE}) cannot be well fitted by the Fourier series, owing to its low apparent brightness that is close to the detection limit of TMTS.
\end{itemize}

\begin{table*}
\tiny
\caption{Catalogue for short-period BY~Dra candidates identified by the TMTS survey.
\label{tab:by_dra}
}
\begin{tabular}{lcccccccccc}
\hline\hline
Source I.D.&R.A.&Dec.&$P_{\rm pho}$&$\sigma_{\rm P,pho}$&Pwr$_{\rm max}$&SNR&Amp.&$L_0$&$G_{\rm abs}$&$(B_P-R_P)_0$\\
 &degree&degree&minute&minute&&&mag&mag&mag&mag\\
 \hline
TMTS J08242980+0309554& 126.12416&   3.16539&  52.21&  0.45&  29.03&  1.32& 0.072& 16.149&  5.618& 0.853\\
TMTS J07262275+4619091& 111.59480&  46.31918&  62.68&  0.59&  31.93&  1.42& 0.194& 17.079&  6.713& 1.230\\
TMTS J06350757+3337132&  98.78155&  33.62034&  63.02&  0.39&  27.47&  1.31& 0.133& 16.636&  5.724& 0.987\\
TMTS J06164885+5008501&  94.20354&  50.14726&  99.33&  0.60&  55.86&  1.35& 0.033& 15.445&  5.861& 0.976\\
TMTS J06053482+5012020&  91.39510&  50.20082& 100.73&  0.58&  56.04&  1.43& 0.138& 16.872&  6.393& 1.017\\
TMTS J08140431+0227281& 123.51794&   2.45781& 119.18&  2.05&  54.16&  1.54& 0.244& 17.060&  5.878& 0.954\\
\hline
\end{tabular}
\end{table*}

As the periodic variations of the BY~Dra stars from the TMTS were not confirmed, we list these candidates in a separate catalogue (Table~\ref{tab:by_dra}) to avoid ambiguity. 
Considering the irregular light-curve shapes of BY~Dra stars (see \citealt{Chen+etal+2020+ZTFvariable}), we did not estimate the parameters related to the shapes of their light curves in the table.

\section{Individual Variable Stars} \label{sec:stars}

\subsection{Eclipsing binaries with a $\delta$~Scuti component}
\label{sec:dsct_binary}
\begin{figure*}
\centering
    \includegraphics[width=0.9\textwidth]{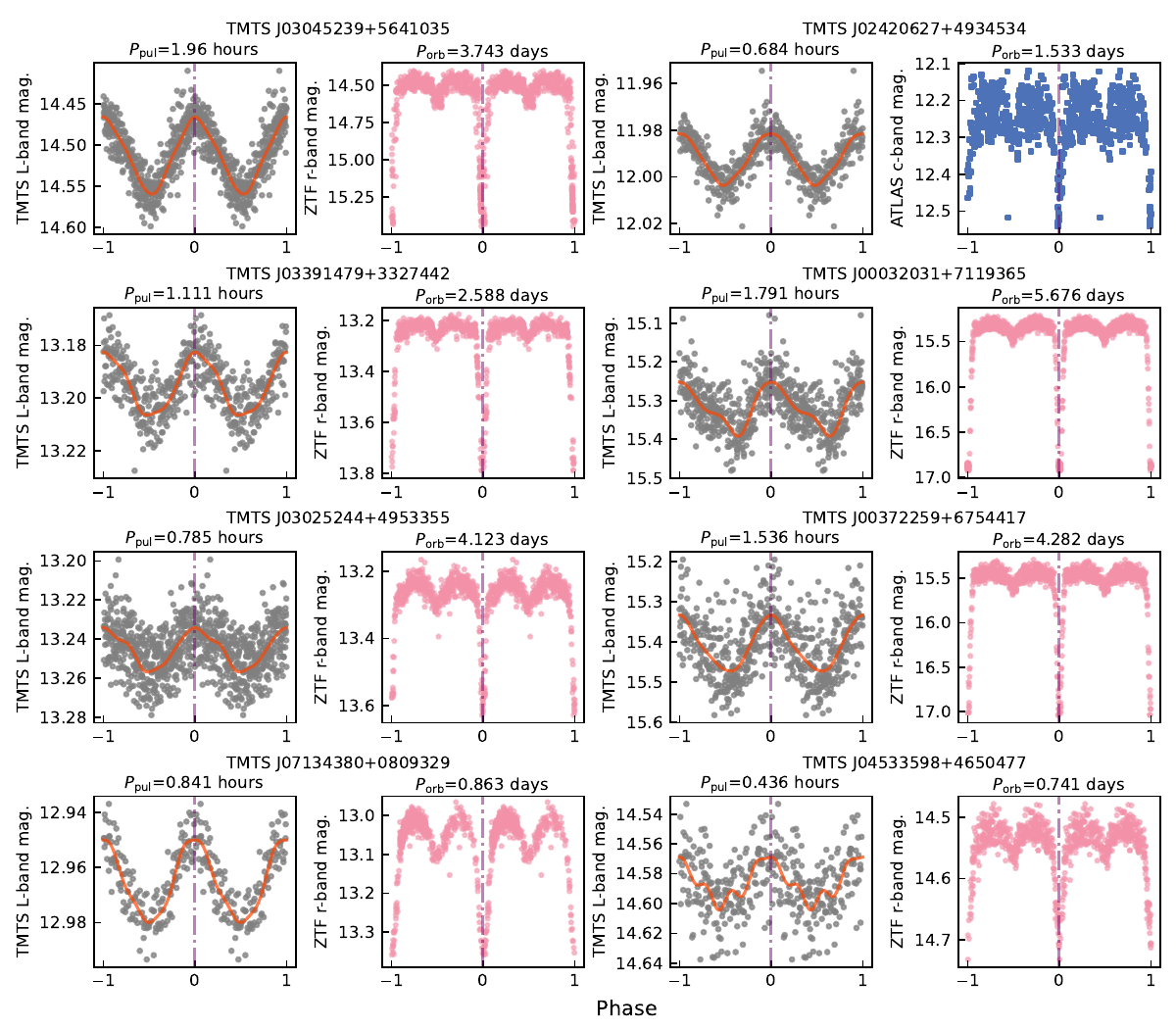}
    \caption{Phase-folded light curves of eclipsing binaries with a $\delta$~Scuti component. The TMTS light curves (grey points) were folded with the pulsation period $P_{\rm pul}$, while the long-cadence survey data from the ZTF $r$ band (pink points) or ATLAS $c$ band (blue points) were folded with the orbital period $P_{\rm orb}$. The red solid lines represent the best-fitting models of Fourier series truncated at the fourth harmonic.
     } 
    \label{fig:dsct_binary}
\end{figure*}

Binary systems are important astrophysical tools for deriving stellar mass, radius, and luminosity. 
Hence, the binaries harboring a component of $\delta$~Scuti type provide a window to test stellar evolution and oscillation theories. 
Although tens of thousands of $\delta$~Scuti stars have been discovered in past decades \citep{Chen+etal+2020+ZTFvariable,Pietrukowicz+etal+2020+deltascuti,Soszynski+etal+2021+24000deltascuti}, only about 200 $\delta$~Scuti stars in binary systems have been reported \citep{Rodriguez+etal+2000+DSCT_EDSCT,Zhou+2010+oscillating_binary,Liakos+Niarchos+2017+EDSCT_cat}.

In our analysis, eclipsing binaries containing a $\delta$~Scuti component will be first identified as $\delta$~Scuti stars rather than eclipsing binaries according to the threshold of the period ($P<2$~hr) adopted in this work, even though the light variation caused by eclipses is usually much stronger than that from stellar oscillations.
By cross-matching the list of TMTS $\delta$~Scuti stars with the VSX, we found that some $\delta$~Scuti stars were previously classified as eclipsing binaries, implying that these $\delta$~Scuti stars have a companion. 
By utilising the long-term light curves from ZTF and ATLAS, we found that 8 TMTS $\delta$~Scuti stars reside in eclipsing binary systems. 
Fig.~\ref{fig:dsct_binary} presents their phase-folded light curves. 
More detailed information about these stars can be found in Table~\ref{tab:dsct}.

\subsection{Blue large-amplitude pulsators}

\begin{figure*}
\centering
    \includegraphics[width=0.95\textwidth]{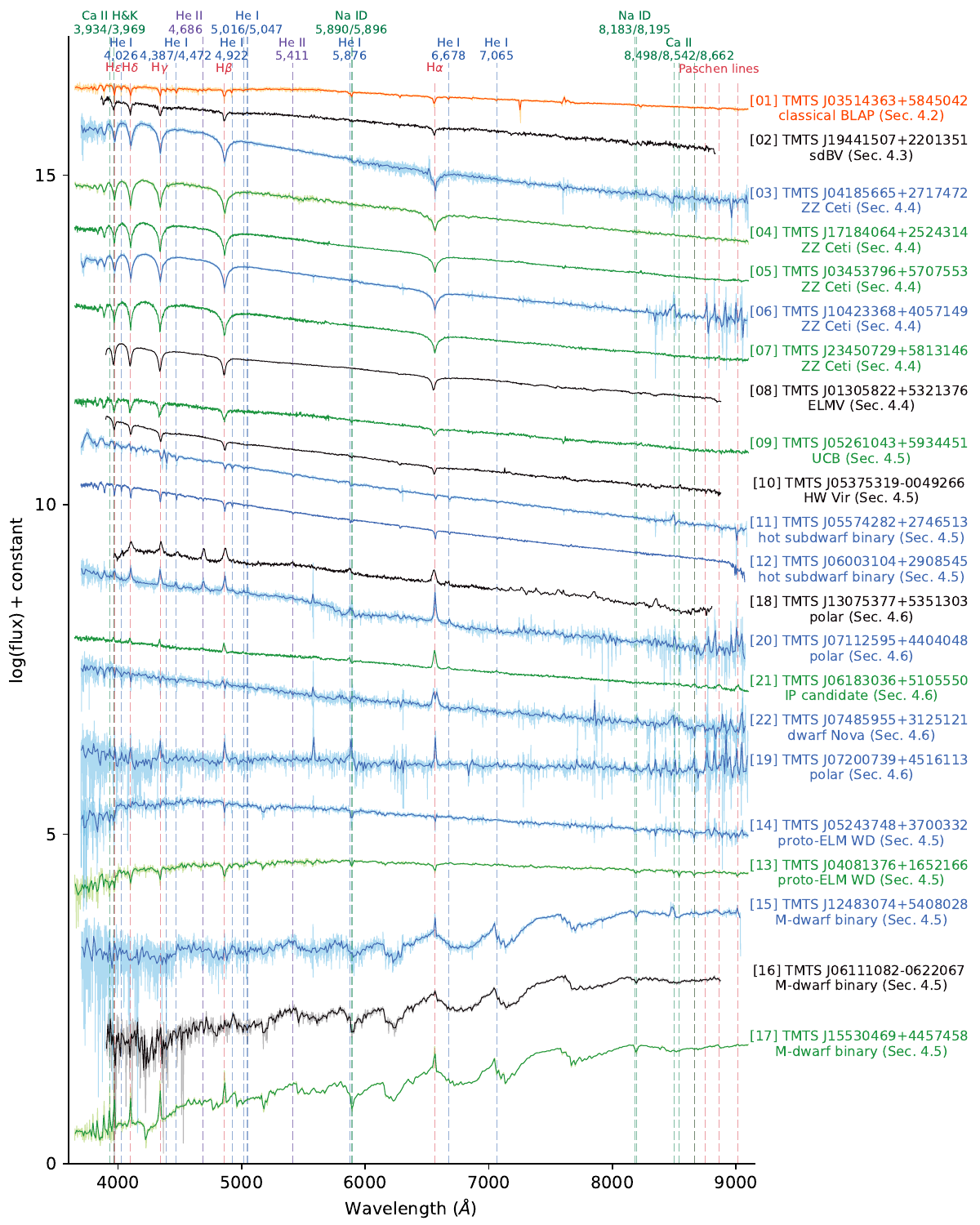}
    \caption{
    Spectra of some short-period variable stars identified by the TMTS. The spectra were obtained by the Xinglong 2.16~m telescope (black), Lick 3~m Shane telescope (green), Keck-I 10~m telescope (orange), and LAMOST (blue). Some typical absorption/emission lines are indicated by vertical dashed lines. The order numbers, source identifiers, classifications and involved Sections are marked next to the spectra.
     } 
    \label{fig:spectra_list}
\end{figure*}

\begin{figure}
\centering
    \includegraphics[width=0.45\textwidth]{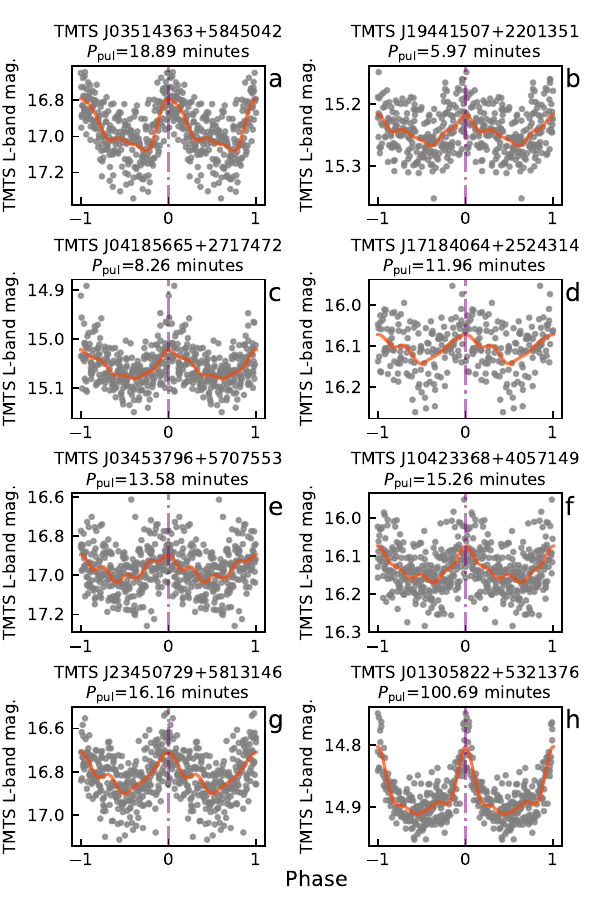}
    \caption{ Phase-folded light curves of pulsating stars (excluding $\delta$ Scuti stars) identified from TMTS observations.
     } 
    \label{fig:pulsating_stars}
\end{figure}

\begin{figure}
\centering
    \includegraphics[width=0.45\textwidth]{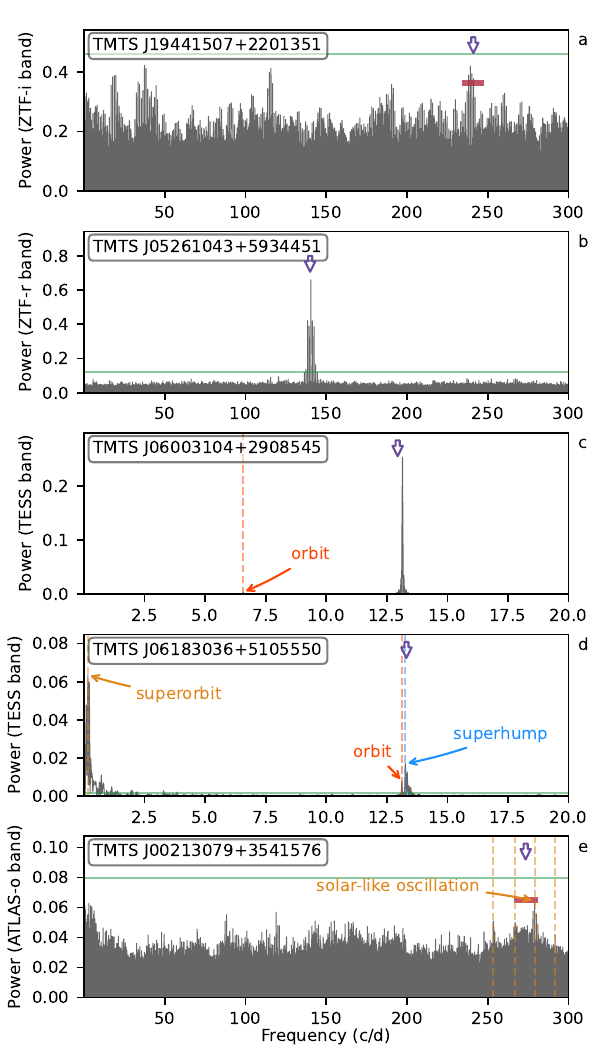}
    \caption{
    Lomb–Scargle periodograms for several short-period variable stars. 
    The Lomb–Scargle periodograms were computed from light curves provided by ATLAS, ZTF, and {\it TESS}.
    The purple arrows indicate the frequency of the maximum LSP peak ($f_{\rm max}$) revealed by the TMTS observations.
    The horizontal green solid lines represent the 1\% significance level corresponding to whole frequency ranges of the periodograms, while the horizontal red bars denote the 1\% significance level derived from the periodicity detection for $\pm 2\%$ frequency ranges around TMTS frequency $f_{\rm max}$.
     } 
    \label{fig:lsp}
\end{figure}

 BLAPs represent a new and mysterious class of hot pulsating stars with unusually large amplitudes and short periods \citep{Pietrukowicz+etal+2017+BLAPs}. 
 Only 24 confirmed BLAPs in our Galaxy have been discovered thus far \citep{lin+etal+2022+NatAs}, making them an extremely rare class of pulsating stars. 
 These BLAPs can be subdivided into two groups: the classical BLAPs with lower surface gravity ($\log\,g \approx 4.5$) and longer pulsation period ($P \gtrsim 20$~min), and the high-gravity BLAPs with higher surface gravity ($\log\,g > 5.0$) and shorter pulsation period ($P \lesssim 10$~min).

TMTS~J03514363+5845042 (= TMTS-BLAP-1) is an 18.9~min BLAP captured by TMTS on 24 and 25 December 2020 \citep{tmtsI+2022}. 
A moderately high helium abundance and a surface gravity $\log\,{g}$ from 4.4 to 4.9 have been inferred from the phase-resolved spectra obtained by the Low-Resolution Imaging Spectrometer (LRIS; \citealt{Oke+etal+1995+Keck+LRIS,McCarthy+etal+1998_LRIS}; see also Fig~\ref{fig:spectra_list}) mounted on the Keck-I telescope. 
Such properties suggest that TMTS-BLAP-1 is a member of classical BLAPs at the shortest-period end. 
The unusually large rate of period change, $\dot{P}/P=2.2 \times 10^{-6}\,{\rm yr^{-1}}$, derived from long-term monitoring data from the ZTF and ATLAS projects, supports the conclusion that TMTS-BLAP-1 is an evolved hot subdwarf star at the short-lived phase of unstable shell-helium burning \citep{lin+etal+2022+NatAs}.

\subsection{Pulsating hot subdwarfs}

Pulsating subdwarf B stars (sdBs) include three subclasses of multiperiodic, low-amplitude pulsators \citep{Heber+2009+araa}: \textbf{(a)} the V361~Hya stars with short-period ($\lesssim 600$~s) light variations caused by pressure-mode (p-mode) oscillations \citep{Kilkenny+etal+1997+361_discovery}; \textbf{(b)} the V1093~Her stars characterised by gravity-mode (g-mode) oscillations with a longer period (2000 to 9000~s; \citealt{Green+etal+2003+sdB}); and \textbf{(c)} 
the DW~Lyn stars with hybrid pulsations from both p-mode and g-mode oscillations \citep{Schuh+etal+2006+DWLyn}.
Furthermore, some pulsating hot subdwarfs present only a single, short-period, and large-amplitude oscillation, such as CS~1246 \citep{Barlow+etal+2010+CS1246}, which indeed blur the boundary between sdBVs and (high-gravity) BLAPs (see also \citealt{Kupfer+etal+2021+ZTFhighcadence}).

TMTS~J19441507+2201351 is a hot subluminous star identified by the {\it Gaia} mission \citep{Geier+etal+2019+hotsd_Gaia}. 
TMTS detected a 6.0~min periodic variation for this source, as can be seen from Fig.~\ref{fig:pulsating_stars}b. 
As this periodic signal is not very significant, we also examined its LSP based on the ZTF $i$-band data. 
As shown in Fig.~\ref{fig:lsp}a, all powers computed across the entire frequency range (0 to 300 cycles/day) of the periodogram are below the 1\% significance level. 
However, the threshold power depends on the attempt frequency range proposed to search for periodic signals \citep{VanderPlas+2018+LSP}. 
Since an approximate frequency was already obtained from the TMTS observation, we set the attempt frequency range to be narrower (i.e., $\pm2\%$ around the TMTS frequency). 
This approach led to the detection of a prominent periodic signal that is consistent with the frequency identified using the TMTS light curves, suggesting that the periodicity is not occasional. 
Furthermore, the spectroscopic observation reveals a series of hydrogen Balmer lines, and no significant helium features are present (see Fig.~\ref{fig:spectra_list}). 
Therefore, TMTS~J19441507+2201351 may be a helium-poor p-mode sdBV.

\subsection{Pulsating white dwarfs}

The broad category of pulsating WDs includes pulsating DA, DB, DQ, ELM, accreting WDs, and pre-WDs \citep{Corsico+etal+2019+book+pulsatingWD}. 
Unlike sdBVs, all subclasses of WD pulsators can be well distinguished by surface chemical composition, effective temperature, and surface gravity derived from spectroscopy.

In the first-two-year survey, TMTS has revealed at least six WD pulsators (see Table~\ref{tab:catalog_part2}), three of which have previously been reported \citep{Landolt+1968+new_blue_ZZ,Fitch+1973+ZZ_period,Bognar+Sodor+2016+ZZ_catalog,Lopez+etal+2021+GD278}. 
Among them, TMTS~J01305822+5321376 (= GD~278) is a 4.61~hr single-lined spectroscopic binary harboring a 0.19~$\rm M_\odot$ ELM WD \citep{Brown+etal+2020+ELMVIII}, and it has been further identified as an ELMV by recent {\it TESS} observations \citep{Lopez+etal+2021+GD278}.
The spectra of all these 6 WD pulsators are shown in Fig.~\ref{fig:spectra_list}, which can be characterised by broad hydrogen absorption lines and absence of obvious helium features, implying that they are hydrogen-rich (ELM) WDs. 
Except for TMTS~J01305822+5321376, these hydrogen-rich WDs are located around the ZZ~Ceti instability strip (see Fig.~\ref{fig:tmts_hrd}) with pulsation periods of about 10~min (see Table~\ref{tab:catalog_part1} and Fig.~\ref{fig:pulsating_stars}), supporting that they are ZZ~Ceti variables (also DAVs). 

We have performed detailed studies for one of these DAVs, TMTS~J23450729+5813146 (Guo et al. 2023). 
Fitting the spectrum to a theoretical WD model \citep{Guo+etal+2015+WD,Guo+etal+2022+WD} suggests an effective temperature of $T_{\rm eff}=11,800$~K and a mass of $M=0.84~{\rm M_\odot}$. 
Additional photometric observations would allow an asteroseismic analysis of this object, which enables a more accurate estimation of its stellar parameters and evolution history.

\subsection{Eclipsing/ellipsoidal binaries}

\begin{figure*}
\centering
    \includegraphics[width=0.9\textwidth]{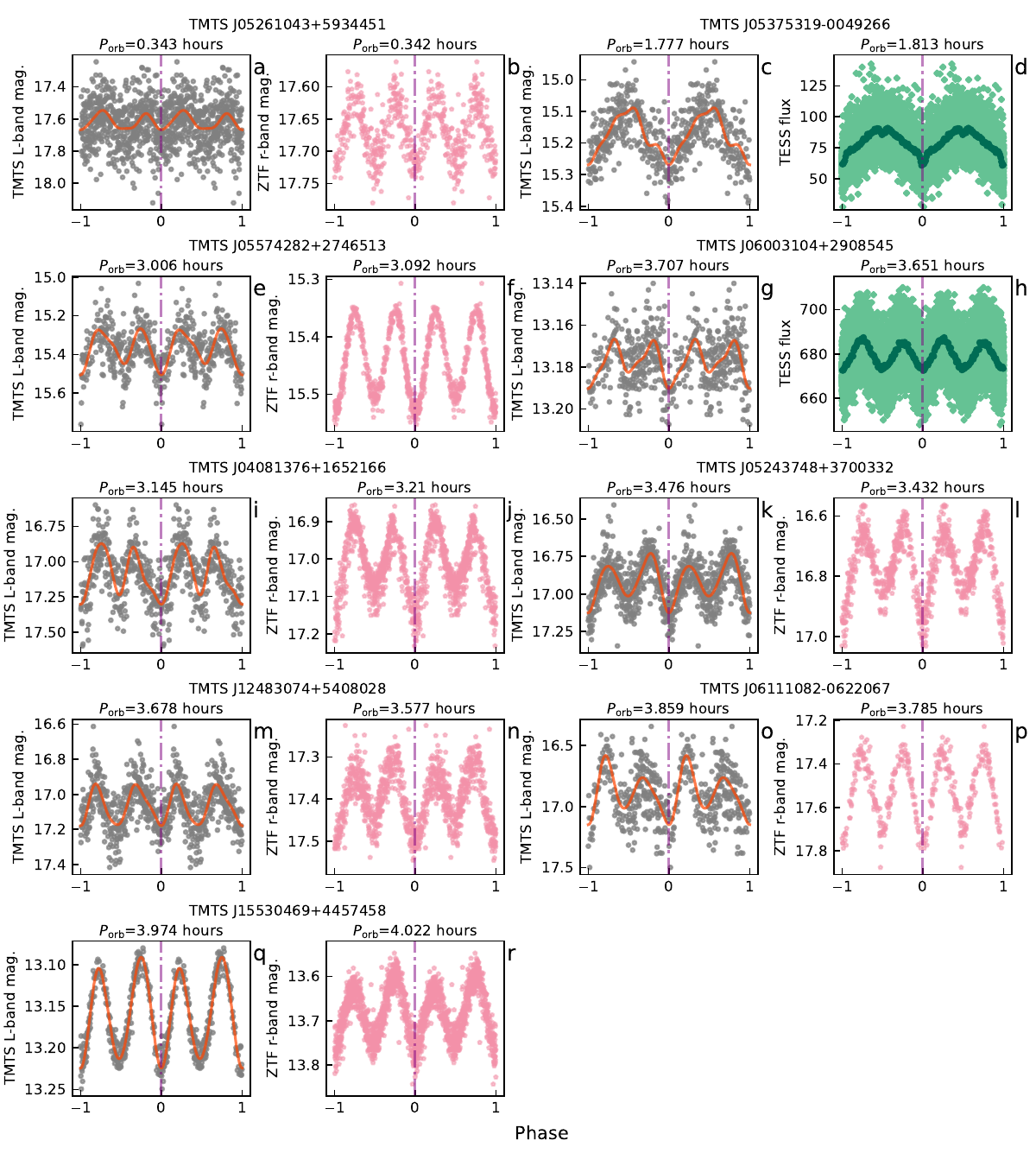}
    \caption{Phase-folded light curves of short-period eclipsing/ellipsoidal binaries. 
    The TMTS (grey points), ZTF $r$ band (pink points), and {\it TESS} (light-green points) light curves were folded with the orbital period $P_{\rm orb}$ derived from the LSPs. The dark-green points overlapped on the {\it TESS} light curves denote the mean of data points within each 0.01 phase bin.
     The red solid lines represent the best-fitting models of Fourier series truncated at the fourth harmonic.
     } 
    \label{fig:eclipsing_binaries}
\end{figure*}

Owing to the constraints from Roche-lobe sizes, components of short-period binaries with orbital period shorter than 4~hr should be ``small-size'' stars, such as late-type main-sequence stars, subdwarfs, and degenerate stars. 
Fig.~\ref{fig:eclipsing_binaries} shows the phase-folded light curves for nine ultracompact/short-period binaries discovered by TMTS. 
As the light-curve shape plays a key role in classifying binaries, we also investigated the phase-folded light curves from ZTF and {\it TESS} observations. 

TMTS J05261043+5934451 is a WD candidate identified by {\it Gaia} observations . 
However, its probability of being a WD is very low ($p_{\rm WD}=0.00460$; \citealt{GentileFusillo+etal+2019+gaia_wd}) because this candidate is more luminous and bluer than most WDs. 
A 10.3~min periodicity of this source has been detected by TMTS on 18 December 2020, and its ZTF ($r$ band) LSP is dominated by a single periodicity, as seen in Fig.~\ref{fig:lsp}b.
However, the RVs obtained from time-resolved spectroscopic observations with the Keck and GTC telescopes support the idea that TMTS J05261043+5934451 is an ultracompact ellipsoidal binary with an orbital period of 20.5~min. 
A more detailed analysis of the nature of this new member of GW verification binaries will be presented in a forthcoming paper (Lin et al. in prep.).

TMTS~J05375319$-$0049266 (= ATO~J084.4719$-$00.8239) was previously discovered by cross-matching the catalogue of hot subdwarfs \citep{Geier+etal+2019+hotsd_Gaia} with the periodic variable stars observed by ATLAS \citep{Heinze+etal+2018+ATLAS_variables}.
This object was identified as an HW~Vir-type star \citep{Schaffenroth+etal+2019+Short_binary}, a binary system consisting of a hot subdwarf with a red or brown dwarf companion.
Such a system is characterised by a light-curve profile that has been affected significantly by strong reflection, and it often presents additional eclipses overlapped on the minima and maxima of the light curve as shown in Fig.~\ref{fig:eclipsing_binaries}d.

TMTS~J05574282+2746513 (= ATO~J089.4285+27.7808) has been identified as a hot subdwarf by {\it Gaia} \citep{Geier+etal+2019+hotsd_Gaia} and as a periodic variable star by ATLAS \citep{Heinze+etal+2018+ATLAS_variables}.
The spectrum presented in Fig.~\ref{fig:spectra_list} clearly reveals the presence of H, He~I, and He~II absorption lines. 
The appearance of noticeable ionised helium lines implies a very high effective temperature (e.g., $T_{\rm eff}>30,000$~K) consistent with that of hot subdwarf stars. 
Since its light curve shows little resemblance to those of HW~Vir stars, the secondary component of TMTS~J05574282+2746513 may not be a cool star irradiated by the primary hot subdwarf. 
We hereby classified TMTS~J05574282+2746513 as a hot subdwarf binary. 
The properties of its secondary component should be determined by modeling its light curve. 

TMTS~J06003104+2908545 is a faint ultraviolet-bright star (i.e., Lanning~11; \citealt{Lanning+1973+UV_sources}) in the Galactic plane, and it has been identified as a hot subdwarf based on its LAMOST spectra and {\it Gaia} photometric observations \citep{Lei+etal+2018+Lamost_subdwarf,Geier+etal+2019+hotsd_Gaia}.
As a hot subdwarf, TMTS~J06003104+2908545 clearly displays H, He~I, and He~II absorption lines in the low-resolution spectrum of LAMOST (see Fig.~\ref{fig:spectra_list}). 
Besides a low-resolution spectrum, LAMOST also collected a time-series of 30 medium-resolution spectra (MRS) of this object, which allows modelling of the light curves with the aid of RVs. 
The RVs inferred from the MRS can be folded with a double photometric period (i.e., $P_{\rm orb}=2\times P_{\rm pho}=3.65$~hr; see panel~\emph{c} of Fig.~\ref{fig:lsp}), supporting that TMTS~J06003104+2908545 is a binary system. 
The semi-amplitude of RV variation is about 250~km~s$^{-1}$, and its mass function is estimated to be about 0.18~$\rm M_\odot$.
We remark that the amplitude of light variation is only 0.024~mag from TMTS observations and about 2.3\% from {\it TESS} data, an order of magnitude smaller than all other short-period binaries discussed in this paper. 
Such a low amplitude of variation may indicate a rather small orbital inclination. Furthermore, the sinusoidal light curves could be caused by tidal deformation of the primary component, and the small but significantly unequal maxima are a possible result of the beaming effect \citep{Loeb+Gaudi+2003+beaming,Zucker+etal+2007+beaming}.
A detailed analysis of this object will be presented in a forthcoming paper.

TMTS~J04081376+1652166 (= ATO~J062.0573+16.8712) and TMTS~J05243748+3700332 (= ATO~J081.1562+37.0091) have recently been identified as progenitors of ELM WDs \citep{El-Badry+etal+2021+proto-ELM}, by cross-matching of the samples selected from {\it Gaia} EDR3 database with the ZTF light curves. 
The periodic variations (see also Fig.~\ref{fig:eclipsing_binaries}) were believed to be caused by corotating ellipsoidal donors, which are envelope-stripped dwarf stars that have ended their first Roche-lobe overflow (RLOF). 
The unseen WD companions are speculated based on the high mass functions computed from semi-amplitudes of RVs and orbital periods \citep{El-Badry+etal+2021+proto-ELM}. 
The spectra of the above two sources present strong Ca~II near-infrared triplets without Paschen lines (see Fig.~\ref{fig:spectra_list}), suggesting that their donors are likely  F/G-type subdwarfs. 
The absent emission lines favour that they are associated with detached CVs.

TMTS~J12483074+5408028 (= ATO~J192.1282+54.1343), TMTS~J06111082$-$0622067 (= ATO~J092.7953$-$06.3689), and TMTS~J15530469+4457458 (= RX~J1553.0+4457) are periodic variable stars identified by ATLAS \citep{Heinze+etal+2018+ATLAS_variables}. 
Their optical spectra presented in Fig.~\ref{fig:spectra_list} show obvious TiO bands and H$\alpha$ emission, suggesting that they are M-type dwarfs with significant chromospheric activity. 
All these M-dwarf binaries have a similar orbital period from 220~min to 238~min, significantly shorter than the $\sim 0.22$~day period cutoff \citep{Rucinski+1992+cutoff,Zhang+Qian+2020+cutoff}. 
The short-period orbits imply that these binaries might have experienced a larger angular momentum loss (AML) due to potential third-body interaction or underestimated magnetic braking in M-dwarf binaries \citep{Nefs+etal+2012+Mdwarfs}. 

As one of the nearby M-dwarfs \citep{Daemgen+etal+2007+Mdwarf}, TMTS~J15530469+4457458 was also detected by the {\it ROSAT} X-ray satellite mission \citep{Zickgraf+etal+2003+Rosat} owing to its magnetic activity \citep{Wright+etal+2011+stellar_activity}. 
The unequal minima and maxima seen in its sinusoidal-like light curve should be caused by tidal deformation and hot spots on the visible component. 
However, the double H$\alpha$ emission lines revealed by LAMOST MRS spectra indicate that this object is a double-line spectroscopic binary (see \citealt{Li+etal+2021+double_spectroscopic_binaries}). 
As a binary system consisting of two late-type stars with prominent chromospheric activity, TMTS~J15530469+4457458 is similar to RS~CVn binaries, but its light variations are dominated by gravity darkening rather than hot spots, and its orbital period is much shorter.

\subsection{Cataclysmic variables}

\begin{figure*}
\centering
    \includegraphics[width=0.9\textwidth]{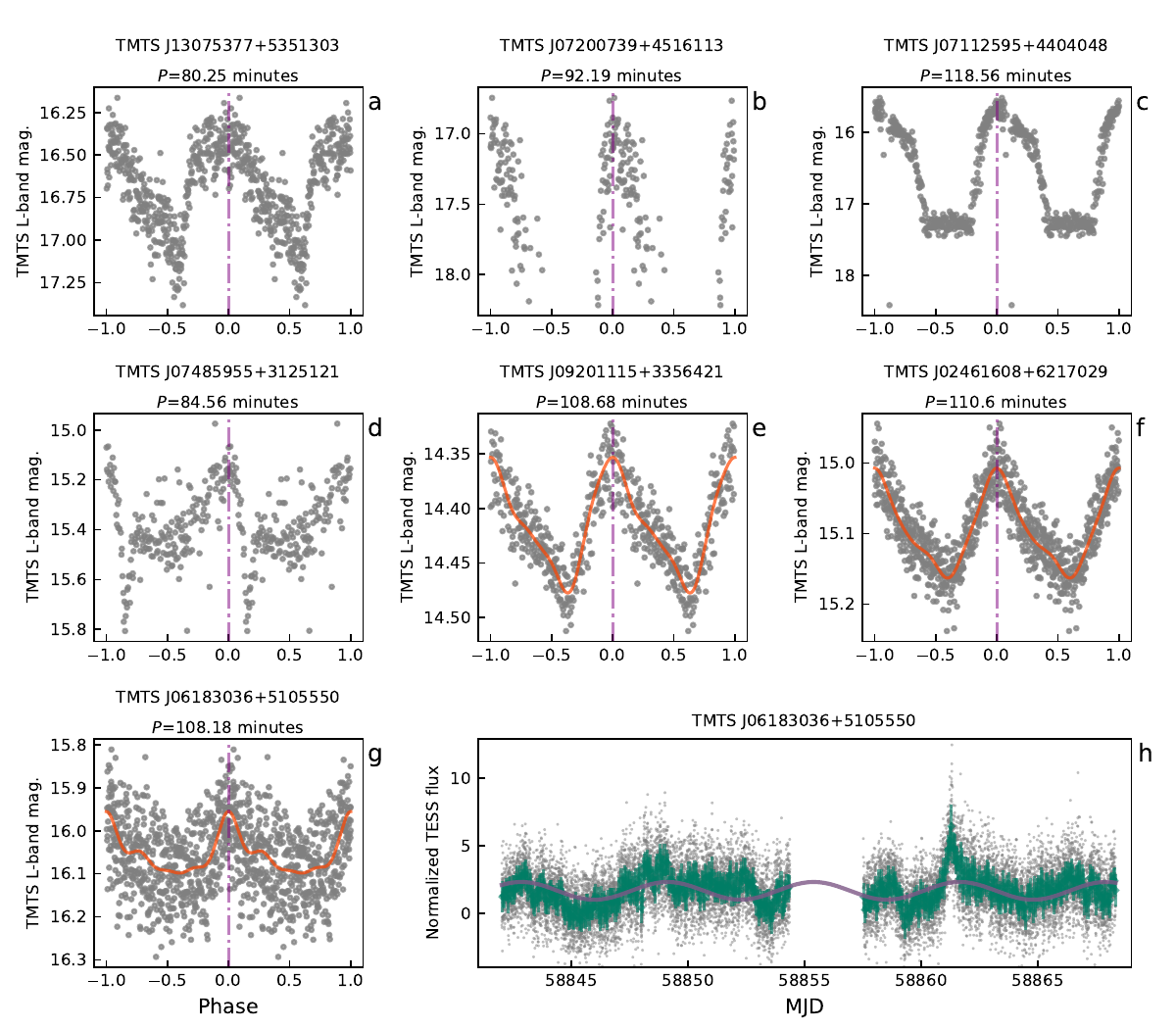}
    \caption{ Phase-folded light curves of CVs identified from TMTS observations 
    and unfolded {\it TESS} light curve for TMTS~J06183036+5105550.
    The {\it TESS} fluxes were normalised relative to the pre-(out)burst fluxes around MJD~58860. The green points overlapped on the {\it TESS} fluxes (grey points) represent the average {\it TESS} fluxes over each 0.05~day bin. The purple line represents the best-fitting sinusoidal model with the superorbit period for the {\it TESS} fluxes.
     } 
    \label{fig:CVs}
\end{figure*}

The phase-folded light curves of seven CVs are shown in Fig.~\ref{fig:CVs}. Among them, six CVs have been previously discovered and studied for at least a decade \citep{Skillman+Patterson+1993+0917,Osborne+etal+1994+RE1307,Downes+etal+1993+CVcatalog,Downes+etal+2001+ATLAS_CV,Gansicke+etal+2009+CV_minimum,Zellem+etal+2009+BKLyn,Drake+etal+2009+catalina+realtime,Drake+etal+2014+Catalina_CV,Thorne+etal+2010+CSS081231:071126+440405,Wils+etal+2010+DN_mining,Carter+etal+2013+AMCVn_SDSS,Coppejans+etal+2016+DN,Kato+etal+2016+SUUMa,Forster+etal+2021+ALeRCE}. 

From ZTF light curves, the light-variation amplitude of TMTS~J07200739+4516113 (= MASTER~OT~J072007.30+451611.6; \citealt{Pogrosheva+etal+2018+MASTER_CV}) can reach up to 3~mag! 
Such large periodic light variations near the detection limit of ZTF observations led to misidentification of this source as an U~Gem star (i.e., a dwarf nova) in the view of long-cadence observations. 
\cite{Burdge+etal+2022+Nature+widow} suggest that such a large amplitude of periodic variations in short-period binaries can be caused by an extreme reflection effect from a tidally-locked cool companion irradiated by a neutron star with strong X-ray radiation (i.e., black widow). 
However, we do not identify similar colour-dependent modulations from ZTF $r$-band and $g$-band observations. 
Therefore, TMTS~J07200739+451611 is more likely a polar \citep{Denisenko+2018+0720_polar}, since an intense accretion column at the magnetic poles of WDs or inhomogeneous accretion can be responsible for periodic light variations larger than 2~mag (e.g., \citealt{Littlefield+etal+2015+highamplitudepolar}).

Superhumps represent the light variations in CVs at a period slightly displaced from the orbital period, which were thought to arise when the outer rim of an accretion disk reaches the 3:1 resonance radius of the binary orbit, and thus became a hallmark of superoutbursts in short-period dwarf novae \citep{Osaki+1996+DN_review,Kato+etal+2016+SUUMa}. 
However, there is increasing evidence that superhumps have also been detected in nova-like stars \citep{Bruch+etal+2022+novalike} and intermediate polars (IPs, \citealt{Miguel+etal+2017+superhump_ip}).
The TMTS observations present the first case of capturing the superhump variations from TMTS J02461608+6217029 (= V495 Cas).
Because TMTS~J02461608+6217029 is located in a crowded field of the Galactic disk, we must confirm that the periodic light variation is intrinsic to this CV, rather than from a close potential variable star (e.g., a $\delta$ Scuti star). 
We checked all 12 nearby stars within $30\arcsec$ using {\it Gaia} DR3 database. 
These stars are all fainter than 17.9~mag in the {\it Gaia} $G$ band (similar to TMTS $L$ band) and are thus negligible because TMTS~J02461608+6217029 is about 15.1~mag during the outbursts (see Fig.~\ref{fig:CVs}).
The light-curve shape of TMTS~J02461608+6217029 shows a striking resemblance to that of TMTS~J09201115+3356421 (= BK~Lyn; see Fig.~\ref{fig:CVs}) and other SU~UMa stars \citep{Kato+etal+2013+DN,Kato+etal+2016+SUUMa}, supporting the notion that the periodic variations in this source are caused by superhumps.

TMTS~J06183036+5105550 is a new short-period CV discovered by TMTS. 
The 12~hr uninterrupted observing sequence taken on 19 December 2020 reveals the presence of periodic variations of 108~min and 11~min. 
The amplitude and profile of the long-period variation agree with the features of superhumps, while the short-period variation may be caused by a rapidly rotating WD or flickering. 
As shown in Fig.~\ref{fig:lsp}d, the Lomb-Scargle periodogram inferred from the {\it TESS} data presents three very significant frequencies, corresponding to superorbit, orbit, and superhump. 
The 108~min periodicity obtained from TMTS observations is exactly identical to the frequency of superhumps, favouring the existence of a precessing accretion disk in the CV system. 
However, the {\it TESS} observation did not reveal any prominent signal with a frequency between 20~cycles/day and 360~cycles/day (the Nyquist frequency).
The noticeable H/He~I emission lines in its optical spectrum (see Fig.~\ref{fig:spectra_list}) indicate that the accretion disk was ionised by its ultraviolet radiation arising from the inner regions of disk, in agreement with the properties of CVs.
A faint He~II emission line around 4686~\AA\ implies that the system may harbor a rotating WD with an EUV/X-ray searchlight beam, which irradiates the disk and powers the lines with high ionisation potential (e.g., $>50$~eV; \citealt{Patterson+etal+1978+DQ_model,Patterson+1994+DQ}). 
Furthermore, we proposed an X-ray observation with {\it Swift}/XRT for this target on 29 October 2022.
The XRT detected an X-ray source (SNR $\approx 3.6$) at the position of TMTS~J06183036+5105550. 
With the distance given by {\it Gaia} DR3, we roughly estimate the 0.5--10~keV band luminosity of this object to be $\sim 10^{31}$~erg~s$^{-1}$, fainter than most IPs but stronger than DQ~Her \citep{Patterson+1994+DQ}.

\begin{figure}
\centering
    \includegraphics[width=0.45\textwidth]{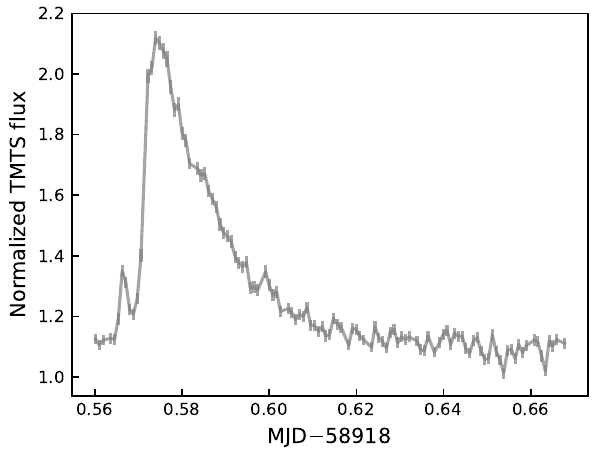}
    \caption{ A flare with precursor from an M dwarf TMTS~J09024535+2207408. The TMTS fluxes are normalized with the post-flare fluxes.
     } 
    \label{fig:precursor}
\end{figure}

Furthermore, with the long-term monitoring data from ATLAS, TMTS~J06183036+5105550 was found to be brighter by about 0.3~mag in the $o$ band before 2017, and it then transitioned to lower luminosity.
For this system, a short (out)burst was recorded around 13 January 2020 by {\it TESS} (see Fig.~\ref{fig:CVs}h). 
The $\sim 1$~day duration of the (out)burst does not follow the same manner of typical dwarf-nova outbursts \citep{Hameury+Lasota+2017+DN_outburst_IP}. 
However, it is very similar to the so-called ``localised thermonuclear bursts'' from accreting magnetic WDs \citep{Scaringi+etal+2022+nature+localthermoburst_CV}.
At this point, we remain skeptical about whether these bursts were caused by runaway thermonuclear burning \citep{Galloway+Keek+2021+thermonuclear_bursts} on the WD surface. 
The short bursts prior to a main burst (i.e., precursor) are commonly seen in intermediate-duration bursts of helium-rich accreting neutron stars \citep{Galloway+etal+2020+MINBAR,Lin+Yu+2020+1712} and are believed to be caused by a photospheric superexpansion when the burst luminosities reach the Eddington limit \citep{Zand+weinberg+2010+superexpansion}.
The precursor in the burst of ASASSN-19bh on MJD~59349 (see the bottom panel of Fig.~2 of \citealt{Scaringi+etal+2022+nature+localthermoburst_CV}) was obviously not caused by the same mechanism, since its peak luminosity was about 4 orders of magnitude lower than the Eddington limit. 
Thus, it is not reasonable to take the precursor as evidence of localised thermonuclear bursts. 
Moreover, TMTS observations captured a similar precursor in an optical flare of the M-dwarf star TMTS~J09024535+2207408, as shown in Fig.~\ref{fig:precursor}, implying that similar precursors can be produced by stellar magnetic activity. 
To understand the exact nature of TMTS~J06183036+5105550, we have performed polarization observations of this object and further analysis is undergoing.

\subsection{Short-period light variations from G- and K-type main sequence}
\begin{figure}
\centering
    \includegraphics[width=0.45\textwidth]{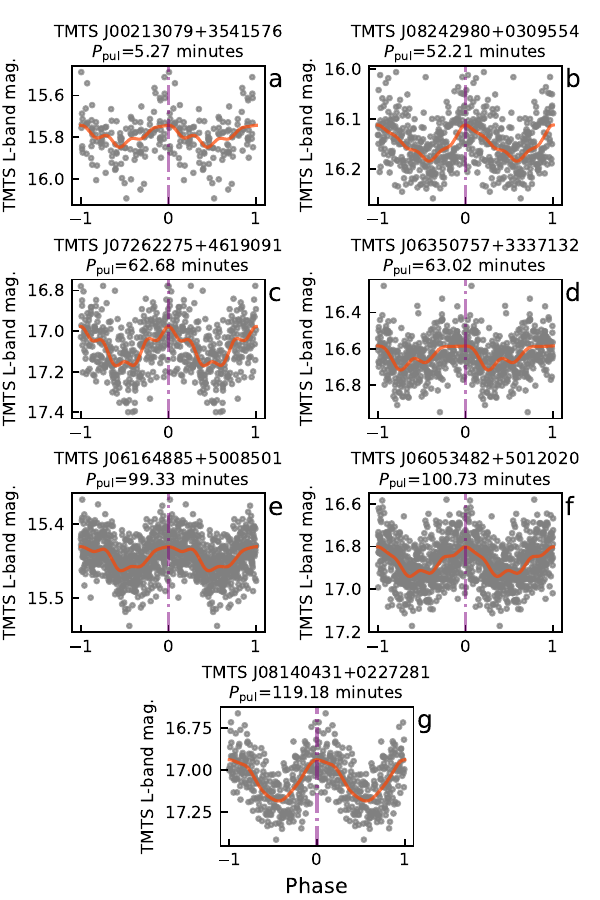}
    \caption{ Phase-folded light curves of main-sequence G-/K-type periodic variable stars identified from TMTS observations.
     } 
    \label{fig:unspecific}
\end{figure}

As discussed in Section~\ref{sec:data}, G/K-type main-sequence stars were rarely discovered as short-period variable stars in the VSX. 
But this does not mean that most G/K-type main-sequence stars do not exhibit any short-period variations. 
In fact, minute-long oscillations have been detected in a large number of solar-like stars by some space-based survey missions such as {\it Kepler} \citep{Mathur+etal+2022+solar-like_oscillations} and {\it TESS} \citep{Hatt+etal+2022+TESS_oscillations}. 
However, these stars were not included in the VSX, owing to their very low amplitudes of oscillations of tens to hundreds of ppm. 
We detect weak 5.3-min variability in a solar-like star, TMTS~J00213079+3541576. 
Because the variability is not significant in TMTS observations (see Fig.\ref{fig:unspecific}a), we also investigated the LSPs from other survey missions. 
As shown in Fig.~\ref{fig:lsp}e, the frequency of maximum power $f_{\rm max}=279.04$~cycle/day (or $\nu_{\rm max}=3229.6\,\mu$Hz) and the mean large frequency separation $\Delta f=12.74$~cycle/day (or $\Delta\nu=147.5~\mu$Hz) were revealed by the LSP derived from ATLAS $o$-band observations. 
These frequency features follow the $\nu_{\rm max}$--$\Delta\nu$ relation in solar-like oscillations from {\it Kepler} and {\it TESS} observations, and are located at the extremely high-frequency end of the relation \citep{Chaplin+etal+2014+solar,Mathur+etal+2022+solar-like_oscillations,Hatt+etal+2022+TESS_oscillations}. 
Since solar-like oscillations are very rarely detected by current ground-based survey missions, the detections of solar-like oscillations from TMTS~J00213079+3541576 may be driven by its unusually large amplitudes. 
We expect further discussion and study of TMTS~J00213079+3541576.

Besides the solar-like oscillations, the G/K-type main-sequence stars with inhomogeneous starspots induced by chromospheric activity (e.g., BY~Dra and RS~CVn stars) can also contribute to short-period light variations. 
From \cite{Chen+etal+2020+ZTFvariable}, about 0.15\% of the 165,789 chromospheric variable stars have periods shorter than 2~hr, although all of these short-period variable stars were reclassified as unspecified variables in the VSX.
In the ZTF catalogue of periodic variable stars, the number of chromospheric variable stars is second only to the EW-type eclipsing binaries \citep{Chen+etal+2020+ZTFvariable}. 
In this work, we have avoided most chromospheric variable stars by adopting a 2~hr threshold. 
However, a very tiny fraction of chromospheric variable stars can present ultrafast rotations and thus short-period light variations \citep{Chahal+etal+2022+BYDra_statistic,Doyle+etal+2022+rotation,Ramsay+etal+2022+rotation}. 
Here we report the discovery of six G/K-type main-sequence stars with unusually short period variations, as shown in Fig.~\ref{fig:unspecific}. 
We checked the neighbouring targets of these 6 stars within 10$\arcsec$ with the {\it Gaia} DR3 database and found that they are all fainter by more than 1.5~mag than the six main-sequence stars, implying that the adjacent sources are not likely to be responsible for the short-period variations.
Recent works suggested that BY~Dra stars are not only limited to K/M dwarfs but also cover F-/G-type main-sequence stars \citep{Chahal+etal+2022+BYDra_statistic}. 
For this reason, we listed these G/K-type main-sequence stars with extreme short-period variations as possible candidates of BY~Dra stars (see Table~\ref{tab:by_dra}). 
If further observations and studies confirm the BY~Dra properties for these candidates, their ultrafast rotation would suggest a nature of very young star without experiencing rotational braking \citep{Skumanich+1972+rotation}.

\section{Summary} \label{sec:summary}

In the first-two-year survey, TMTS has monitored the LAMOST fields covering a valid sky area of 3\,857~deg$^2$. 
About 1\,100 short-period variable stars with periods shorter than 2~hr were selected from a total of 10,856,233 uninterrupted light curves generated from the two-year observations of TMTS, using new and period-dependent thresholds for the maximum powers in Lomb–Scargle periodograms (LSPs). 
With the {\it Gaia} parallaxes, $G$-band absolute magnitudes, and $B-R$ colours, we tried to classify these variables with the colour-magnitude diagram (CMD). 
Most of these variables are distributed near the main sequence and its left side, and different classes of variables tend to be located in different areas in the CMD.
In addition, the LAMOST spectral parameters, VSX classifications, and archived observations from other survey missions helped us classify a small fraction of TMTS sources. 
Moreover, we proposed some optical and X-ray observations for further classifications of some interesting variable stars.

Among the samples with short-period light variations, $\delta$~Scuti stars are dominant and most of them were not recorded in the VSX. 
Based on our sample of the light curves of $\delta$~Scuti stars, we found that the overtones of $\delta$~Scuti stars also seem to follow a period-luminosity relation. 
The upper limit of pulsation amplitude tends to decline toward shorter period, which can be used as a convenient tool to distinguish other short-period variable stars. 
By cross-matching the TMTS $\delta$~Scuti stars with the eclipsing binaries in the VSX, we identified eight $\delta$~Scuti stars in eclipsing binary systems.

Apart from the $\delta$~Scuti stars at the short-period end, TMTS also discovered some other categories of short-period variable stars, including classical/high-gravity BLAPs, sdBVs, pulsating WDs, CVs below the period gap, and short-period eclipsing/ellipsoidal binaries with an orbital period shorter than the period cutoff. 
We also presented optical spectra, Lomb-Scargle periodograms, light-curve parameters, and classifications for them. 

In forthcoming papers, we will investigate several interesting short-period variables in more detail, and discuss the upper limit of amplitudes for $\delta$~Scuti stars with the help of asteroseismology. 
A catalogue for eclipsing binaries captured by TMTS and their detailed stellar parameters will be published in separate papers. 
Utilising the short-period variables identified in this study, we will construct training and test datasets and adopt machine-learning techniques to generate a more complete catalogue of periodic variables based on the current and future TMTS observations.

\section*{Acknowledgements}
We are grateful to Tianrui Sun (PMO), Xiaodian Chen (NAOC), Tanda Li (BNU), and Lorenzo Rimoldini (UNIGE) for very useful suggestions and discussions.
We acknowledge the support of the staffs from Xinglong Observatory of NAOC during the installation, commissioning, and operation of the TMTS system. We are grateful to the staffs of various telescopes and observatories with which the data were obtained (Xinglong 2.16~m Telescope, Lick 3~m Shane telescope, Keck-I 10~m telescope, Neil Gehrels {\it Swift} Observatory). This work is supported by the National Science Foundation of China (NSFC grants 12288102, 12033003 and 12203006), the Ma Huateng Foundation, the Scholar Program of Beijing Academy of Science and Technology (DZ:BS202002), and the Tencent Xplorer Prize. 
This paper is sponsored by the Experiments for Space Exploration Program and the Qian Xuesen Laboratory, China Academy of Space Technology (Grant No. TKTSPY-2020-5-3).
A.V.F.'s group at U.C. Berkeley is grateful for financial support from the Christopher R. Redlich Fund and many individual donors. The research of Y.Y. is supported through a Bengier-Winslow-Robertson Fellowship. P.N. acknowledges support from the Grant Agency of the Czech Republic (GA\v{C}R 22-34467S). The Astronomical Institute in Ond\v{r}ejov is supported by the project RVO:67985815.

Some of the data presented herein were obtained at the W. M. Keck Observatory, which is operated as a scientific partnership among the California Institute of Technology, the University of California, and NASA; the observatory was made possible by the generous financial support of the W. M. Keck Foundation.
A major upgrade of the Kast spectrograph on the Shane 3~m telescope at Lick Observatory, led by Brad Holden, was made possible through gifts from the Heising-Simons Foundation, William and Marina Kast, and the University of California Observatories. Research at Lick Observatory is partially supported by a generous gift from Google.

Guoshoujing Telescope (the Large Sky Area Multi-Object Fiber Spectroscopic Telescope LAMOST) is a National Major Scientific Project built by the Chinese Academy of Sciences. Funding for the project has been provided by the National Development and Reform Commission. LAMOST is operated and managed by the National Astronomical Observatories, Chinese Academy of Sciences.

This research has used the services of \mbox{\url{www.Astroserver.org}}. 
This work has made use of data from the European Space Agency (ESA) mission
{\it Gaia} (\url{https://www.cosmos.esa.int/gaia}), processed by the {\it Gaia}
Data Processing and Analysis Consortium (DPAC,
\url{https://www.cosmos.esa.int/web/gaia/dpac/consortium}). Funding for the DPAC
has been provided by national institutions, in particular the institutions
participating in the {\it Gaia} Multilateral Agreement.

Based in part on observations obtained with the Samuel Oschin 48-inch telescope and the 60-inch telescope at Palomar Observatory as part of the Zwicky Transient Facility project. ZTF is supported by the National Science Foundation under grants AST-1440341 and AST-2034437, and a collaboration including current partners Caltech, IPAC, the Weizmann Institute for Science, the Oskar Klein Center at Stockholm University, the University of Maryland, Deutsches Elektronen-Synchrotron and Humboldt University, the TANGO Consortium of Taiwan, the University of Wisconsin at Milwaukee, Trinity College Dublin, Lawrence Livermore National Laboratories, IN2P3, University of Warwick, Ruhr University Bochum, and Northwestern University, and former partners the University of Washington, Los Alamos National Laboratories, and Lawrence Berkeley National Laboratories. Operations are conducted by COO, IPAC, and UW.

This paper includes data collected with the {\it TESS} mission, obtained from the MAST data archive at the Space Telescope Science Institute (STScI). Funding for the {\it TESS} mission is provided by the NASA Explorer Program. STScI is operated by the Association of Universities for Research in Astronomy, Inc., under NASA contract NAS 5–26555.

This work has made use of data from the Asteroid Terrestrial-impact Last Alert System (ATLAS) project. The ATLAS project is primarily funded to search for near-Earth objects (NEOs) through NASA grants NN12AR55G, 80NSSC18K0284, and 80NSSC18K1575; byproducts of the NEO search include images and catalogues from the survey area. This work was partially funded by Kepler/K2 grant J1944/80NSSC19K0112 and {\it HST} GO-15889, and STFC grants ST/T000198/1 and ST/S006109/1. The ATLAS science products have been made possible through the contributions of the University of Hawaii Institute for Astronomy, the Queen’s University Belfast, the Space Telescope Science Institute, the South African Astronomical Observatory, and The Millennium Institute of Astrophysics (MAS), Chile.

This research has made use of the International Variable Star Index (VSX, \citealt{Watson+etal+2006+VSX}) database, operated at AAVSO, Cambridge, Massachusetts, USA.
Some of the results in this paper have been derived using the HEALPix \citep{Gorski+etal+2005} package.

\section*{Data Availability}
The catalogues are all available from this paper and CDS.




\begin{thebibliography}{}
\makeatletter
\relax
\def\mn@urlcharsother{\let\do\@makeother \do\$\do\&\do\#\do\^\do\_\do\%\do\~}
\def\mn@doi{\begingroup\mn@urlcharsother \@ifnextchar [ {\mn@doi@}
  {\mn@doi@[]}}
\def\mn@doi@[#1]#2{\def\@tempa{#1}\ifx\@tempa\@empty \href
  {http://dx.doi.org/#2} {doi:#2}\else \href {http://dx.doi.org/#2} {#1}\fi
  \endgroup}
\def\mn@eprint#1#2{\mn@eprint@#1:#2::\@nil}
\def\mn@eprint@arXiv#1{\href {http://arxiv.org/abs/#1} {{\tt arXiv:#1}}}
\def\mn@eprint@dblp#1{\href {http://dblp.uni-trier.de/rec/bibtex/#1.xml}
  {dblp:#1}}
\def\mn@eprint@#1:#2:#3:#4\@nil{\def\@tempa {#1}\def\@tempb {#2}\def\@tempc
  {#3}\ifx \@tempc \@empty \let \@tempc \@tempb \let \@tempb \@tempa \fi \ifx
  \@tempb \@empty \def\@tempb {arXiv}\fi \@ifundefined
  {mn@eprint@\@tempb}{\@tempb:\@tempc}{\expandafter \expandafter \csname
  mn@eprint@\@tempb\endcsname \expandafter{\@tempc}}}

\bibitem[\protect\citeauthoryear{{Arentoft}, {Sterken}, {Knudsen},
  {Freyhammer}, {Duerbeck}, {Pompei}, {Delahodde}  \& {Clasen}}{{Arentoft}
  et~al.}{2001}]{Arentoft+etal+2001+beta_ceph}
{Arentoft} T.,  {Sterken} C.,  {Knudsen} M.~R.,  {Freyhammer} L.~M.,
  {Duerbeck} H.~W.,  {Pompei} E.,  {Delahodde} C.~E.,   {Clasen} J.~W.,  2001,
  \mn@doi [\aap] {10.1051/0004-6361:20011472}, \href
  {https://ui.adsabs.harvard.edu/abs/2001A&A...380..599A} {380, 599}

\bibitem[\protect\citeauthoryear{{Bahramian} et~al.,}{{Bahramian}
  et~al.}{2017}]{Bahramian+etal+2017+BH_UCXB}
{Bahramian} A.,  et~al., 2017, \mn@doi [\mnras] {10.1093/mnras/stx166}, \href
  {https://ui.adsabs.harvard.edu/abs/2017MNRAS.467.2199B} {467, 2199}

\bibitem[\protect\citeauthoryear{{Barclay}, {Ramsay}, {Hakala}, {Napiwotzki},
  {Nelemans}, {Potter}  \& {Todd}}{{Barclay}
  et~al.}{2011}]{Barclay+etal+2011+RTS_amcvn}
{Barclay} T.,  {Ramsay} G.,  {Hakala} P.,  {Napiwotzki} R.,  {Nelemans} G.,
  {Potter} S.,   {Todd} I.,  2011, \mn@doi [\mnras]
  {10.1111/j.1365-2966.2011.18345.x}, \href
  {https://ui.adsabs.harvard.edu/abs/2011MNRAS.413.2696B} {413, 2696}

\bibitem[\protect\citeauthoryear{{Barlow} et~al.,}{{Barlow}
  et~al.}{2010}]{Barlow+etal+2010+CS1246}
{Barlow} B.~N.,  et~al., 2010, \mn@doi [\mnras]
  {10.1111/j.1365-2966.2009.16119.x}, \href
  {https://ui.adsabs.harvard.edu/abs/2010MNRAS.403..324B} {403, 324}

\bibitem[\protect\citeauthoryear{{Bellm} et~al.,}{{Bellm}
  et~al.}{2019}]{ZTF+2019+first}
{Bellm} E.~C.,  et~al., 2019, \mn@doi [\pasp] {10.1088/1538-3873/aaecbe}, \href
  {https://ui.adsabs.harvard.edu/abs/2019PASP..131a8002B} {131, 018002}

\bibitem[\protect\citeauthoryear{{Bognar} \& {Sodor}}{{Bognar} \&
  {Sodor}}{2016}]{Bognar+Sodor+2016+ZZ_catalog}
{Bognar} Z.,  {Sodor} A.,  2016, Information Bulletin on Variable Stars, \href
  {https://ui.adsabs.harvard.edu/abs/2016IBVS.6184....1B} {6184, 1}

\bibitem[\protect\citeauthoryear{{Bowman} \& {Kurtz}}{{Bowman} \&
  {Kurtz}}{2018}]{Bowman+Kurtz+2018+dsct_strip}
{Bowman} D.~M.,  {Kurtz} D.~W.,  2018, \mn@doi [\mnras] {10.1093/mnras/sty449},
  \href {https://ui.adsabs.harvard.edu/abs/2018MNRAS.476.3169B} {476, 3169}

\bibitem[\protect\citeauthoryear{{Bowman}, {Kurtz}, {Breger}, {Murphy}  \&
  {Holdsworth}}{{Bowman} et~al.}{2016}]{Bowman+etal+2016+variability}
{Bowman} D.~M.,  {Kurtz} D.~W.,  {Breger} M.,  {Murphy} S.~J.,   {Holdsworth}
  D.~L.,  2016, \mn@doi [\mnras] {10.1093/mnras/stw1153}, \href
  {https://ui.adsabs.harvard.edu/abs/2016MNRAS.460.1970B} {460, 1970}

\bibitem[\protect\citeauthoryear{{Brown}, {Kilic}, {Hermes}, {Allende Prieto},
  {Kenyon}  \& {Winget}}{{Brown} et~al.}{2011}]{Brown+etal+2011+12min_EWD}
{Brown} W.~R.,  {Kilic} M.,  {Hermes} J.~J.,  {Allende Prieto} C.,  {Kenyon}
  S.~J.,   {Winget} D.~E.,  2011, \mn@doi [\apjl]
  {10.1088/2041-8205/737/1/L23}, \href
  {https://ui.adsabs.harvard.edu/abs/2011ApJ...737L..23B} {737, L23}

\bibitem[\protect\citeauthoryear{{Brown}, {Kilic}, {Kosakowski}  \&
  {Gianninas}}{{Brown} et~al.}{2017}]{Brown+etal+2017+40minWD}
{Brown} W.~R.,  {Kilic} M.,  {Kosakowski} A.,   {Gianninas} A.,  2017, \mn@doi
  [\apj] {10.3847/1538-4357/aa8724}, \href
  {https://ui.adsabs.harvard.edu/abs/2017ApJ...847...10B} {847, 10}

\bibitem[\protect\citeauthoryear{{Brown} et~al.,}{{Brown}
  et~al.}{2020}]{Brown+etal+2020+ELMVIII}
{Brown} W.~R.,  et~al., 2020, \mn@doi [\apj] {10.3847/1538-4357/ab63cd}, \href
  {https://ui.adsabs.harvard.edu/abs/2020ApJ...889...49B} {889, 49}

\bibitem[\protect\citeauthoryear{{Brown}, {Kilic}, {Kosakowski}  \&
  {Gianninas}}{{Brown} et~al.}{2022}]{Brown+etal+2022+ELMIX}
{Brown} W.~R.,  {Kilic} M.,  {Kosakowski} A.,   {Gianninas} A.,  2022, \mn@doi
  [\apj] {10.3847/1538-4357/ac72ac}, \href
  {https://ui.adsabs.harvard.edu/abs/2022ApJ...933...94B} {933, 94}

\bibitem[\protect\citeauthoryear{{Bruch}}{{Bruch}}{2022}]{Bruch+etal+2022+novalike}
{Bruch} A.,  2022, arXiv e-prints, \href
  {https://ui.adsabs.harvard.edu/abs/2022arXiv221204424B} {p. arXiv:2212.04424}

\bibitem[\protect\citeauthoryear{{Burdge} et~al.,}{{Burdge}
  et~al.}{2019}]{Burdge+etal+2019+Nature+7minWD}
{Burdge} K.~B.,  et~al., 2019, \mn@doi [\nat] {10.1038/s41586-019-1403-0},
  \href {https://ui.adsabs.harvard.edu/abs/2019Natur.571..528B} {571, 528}

\bibitem[\protect\citeauthoryear{{Burdge} et~al.,}{{Burdge}
  et~al.}{2020a}]{Burdge+etal+2020+systematic}
{Burdge} K.~B.,  et~al., 2020a, \mn@doi [\apj] {10.3847/1538-4357/abc261},
  \href {https://ui.adsabs.harvard.edu/abs/2020ApJ...905...32B} {905, 32}

\bibitem[\protect\citeauthoryear{{Burdge} et~al.,}{{Burdge}
  et~al.}{2020b}]{Burdge+etal+2020+9minute}
{Burdge} K.~B.,  et~al., 2020b, \mn@doi [\apjl] {10.3847/2041-8213/abca91},
  \href {https://ui.adsabs.harvard.edu/abs/2020ApJ...905L...7B} {905, L7}

\bibitem[\protect\citeauthoryear{{Burdge} et~al.,}{{Burdge}
  et~al.}{2022a}]{Burdge+etal+2022+Nature+widow}
{Burdge} K.~B.,  et~al., 2022a, \mn@doi [\nat] {10.1038/s41586-022-04551-1},
  \href {https://ui.adsabs.harvard.edu/abs/2022Natur.605...41B} {605, 41}

\bibitem[\protect\citeauthoryear{{Burdge} et~al.,}{{Burdge}
  et~al.}{2022b}]{Burdge+etal+2022+TCV+nature}
{Burdge} K.~B.,  et~al., 2022b, \mn@doi [\nat] {10.1038/s41586-022-05195-x},
  \href {https://ui.adsabs.harvard.edu/abs/2022Natur.610..467B} {610, 467}

\bibitem[\protect\citeauthoryear{{Burggraaff} et~al.,}{{Burggraaff}
  et~al.}{2018}]{Burggraaff+etal+2018+bright_variables}
{Burggraaff} O.,  et~al., 2018, \mn@doi [\aap] {10.1051/0004-6361/201833142},
  \href {https://ui.adsabs.harvard.edu/abs/2018A&A...617A..32B} {617, A32}

\bibitem[\protect\citeauthoryear{{Byrne} \& {Jeffery}}{{Byrne} \&
  {Jeffery}}{2018}]{Byrne+etal+2018+BLAPs}
{Byrne} C.~M.,  {Jeffery} C.~S.,  2018, \mn@doi [\mnras]
  {10.1093/mnras/sty2545}, \href
  {https://ui.adsabs.harvard.edu/abs/2018MNRAS.481.3810B} {481, 3810}

\bibitem[\protect\citeauthoryear{{Byrne} \& {Jeffery}}{{Byrne} \&
  {Jeffery}}{2020}]{Byrne+etal+2020+faint_BLAP}
{Byrne} C.~M.,  {Jeffery} C.~S.,  2020, \mn@doi [\mnras]
  {10.1093/mnras/stz3486}, \href
  {https://ui.adsabs.harvard.edu/abs/2020MNRAS.492..232B} {492, 232}

\bibitem[\protect\citeauthoryear{{Byrne}, {Stanway}  \& {Eldridge}}{{Byrne}
  et~al.}{2021}]{Byrne+etal+2021+population_BLAP}
{Byrne} C.~M.,  {Stanway} E.~R.,   {Eldridge} J.~J.,  2021, \mn@doi [\mnras]
  {10.1093/mnras/stab2115}, \href
  {https://ui.adsabs.harvard.edu/abs/2021MNRAS.507..621B} {507, 621}

\bibitem[\protect\citeauthoryear{{Caiazzo} et~al.,}{{Caiazzo}
  et~al.}{2021}]{Caiazzo+etal+2021+moon}
{Caiazzo} I.,  et~al., 2021, \mn@doi [\nat] {10.1038/s41586-021-03615-y}, \href
  {https://ui.adsabs.harvard.edu/abs/2021Natur.595...39C} {595, 39}

\bibitem[\protect\citeauthoryear{{Carter} et~al.,}{{Carter}
  et~al.}{2013}]{Carter+etal+2013+AMCVn_SDSS}
{Carter} P.~J.,  et~al., 2013, \mn@doi [\mnras] {10.1093/mnras/sts485}, \href
  {https://ui.adsabs.harvard.edu/abs/2013MNRAS.429.2143C} {429, 2143}

\bibitem[\protect\citeauthoryear{{Castanheira} et~al.,}{{Castanheira}
  et~al.}{2006}]{Castanheira+etal+2006+newZZ}
{Castanheira} B.~G.,  et~al., 2006, \mn@doi [\aap]
  {10.1051/0004-6361:20053500}, \href
  {https://ui.adsabs.harvard.edu/abs/2006A&A...450..227C} {450, 227}

\bibitem[\protect\citeauthoryear{{Castanheira}, {Kepler}, {Kleinman}, {Nitta}
  \& {Fraga}}{{Castanheira} et~al.}{2010}]{Castanheira+etal+2010+newZZ}
{Castanheira} B.~G.,  {Kepler} S.~O.,  {Kleinman} S.~J.,  {Nitta} A.,   {Fraga}
  L.,  2010, \mn@doi [\mnras] {10.1111/j.1365-2966.2010.16633.x}, \href
  {https://ui.adsabs.harvard.edu/abs/2010MNRAS.405.2561C} {405, 2561}

\bibitem[\protect\citeauthoryear{{Chahal}, {de Grijs}, {Kamath}  \&
  {Chen}}{{Chahal} et~al.}{2022}]{Chahal+etal+2022+BYDra_statistic}
{Chahal} D.,  {de Grijs} R.,  {Kamath} D.,   {Chen} X.,  2022, \mn@doi [\mnras]
  {10.1093/mnras/stac1660}, \href
  {https://ui.adsabs.harvard.edu/abs/2022MNRAS.514.4932C} {514, 4932}

\bibitem[\protect\citeauthoryear{{Chang}, {Byun}  \& {Hartman}}{{Chang}
  et~al.}{2015}]{Chang+etal+2015+beta_ceph}
{Chang} S.~W.,  {Byun} Y.~I.,   {Hartman} J.~D.,  2015, \mn@doi [\aj]
  {10.1088/0004-6256/150/1/27}, \href
  {https://ui.adsabs.harvard.edu/abs/2015AJ....150...27C} {150, 27}

\bibitem[\protect\citeauthoryear{{Chanmugam} \& {Ray}}{{Chanmugam} \&
  {Ray}}{1984}]{Chanmugam+Ray+1984+AMDQ}
{Chanmugam} G.,  {Ray} A.,  1984, \mn@doi [\apj] {10.1086/162499}, \href
  {https://ui.adsabs.harvard.edu/abs/1984ApJ...285..252C} {285, 252}

\bibitem[\protect\citeauthoryear{{Chaplin} et~al.,}{{Chaplin}
  et~al.}{2014}]{Chaplin+etal+2014+solar}
{Chaplin} W.~J.,  et~al., 2014, \mn@doi [\apjs] {10.1088/0067-0049/210/1/1},
  \href {https://ui.adsabs.harvard.edu/abs/2014ApJS..210....1C} {210, 1}

\bibitem[\protect\citeauthoryear{{Chen}, {Wang}, {Deng}, {de Grijs}, {Yang}  \&
  {Tian}}{{Chen} et~al.}{2020a}]{Chen+etal+2020+ZTFvariable}
{Chen} X.,  {Wang} S.,  {Deng} L.,  {de Grijs} R.,  {Yang} M.,   {Tian} H.,
  2020a, \mn@doi [\apjs] {10.3847/1538-4365/ab9cae}, \href
  {https://ui.adsabs.harvard.edu/abs/2020ApJS..249...18C} {249, 18}

\bibitem[\protect\citeauthoryear{{Chen}, {Liu}  \& {Wang}}{{Chen}
  et~al.}{2020b}]{Chen+etal+2020+UCXB_LISA}
{Chen} W.-C.,  {Liu} D.-D.,   {Wang} B.,  2020b, \mn@doi [\apjl]
  {10.3847/2041-8213/abae66}, \href
  {https://ui.adsabs.harvard.edu/abs/2020ApJ...900L...8C} {900, L8}

\bibitem[\protect\citeauthoryear{{Coppejans}, {K{\"o}rding}, {Knigge},
  {Pretorius}, {Woudt}, {Groot}, {Van Eck}  \& {Drake}}{{Coppejans}
  et~al.}{2016}]{Coppejans+etal+2016+DN}
{Coppejans} D.~L.,  {K{\"o}rding} E.~G.,  {Knigge} C.,  {Pretorius} M.~L.,
  {Woudt} P.~A.,  {Groot} P.~J.,  {Van Eck} C.~L.,   {Drake} A.~J.,  2016,
  \mn@doi [\mnras] {10.1093/mnras/stv2921}, \href
  {https://ui.adsabs.harvard.edu/abs/2016MNRAS.456.4441C} {456, 4441}

\bibitem[\protect\citeauthoryear{{C{\'o}rsico}, {Althaus}, {Miller Bertolami}
  \& {Kepler}}{{C{\'o}rsico} et~al.}{2019}]{Corsico+etal+2019+book+pulsatingWD}
{C{\'o}rsico} A.~H.,  {Althaus} L.~G.,  {Miller Bertolami} M.~M.,   {Kepler}
  S.~O.,  2019, \mn@doi [\aapr] {10.1007/s00159-019-0118-4}, \href
  {https://ui.adsabs.harvard.edu/abs/2019A&ARv..27....7C} {27, 7}

\bibitem[\protect\citeauthoryear{{Cui} et~al.,}{{Cui}
  et~al.}{2012}]{Cui+etal+2012+LAMOST}
{Cui} X.-Q.,  et~al., 2012, \mn@doi [Research in Astronomy and Astrophysics]
  {10.1088/1674-4527/12/9/003}, \href
  {https://ui.adsabs.harvard.edu/abs/2012RAA....12.1197C} {12, 1197}

\bibitem[\protect\citeauthoryear{{Daemgen}, {Siegler}, {Reid}  \&
  {Close}}{{Daemgen} et~al.}{2007}]{Daemgen+etal+2007+Mdwarf}
{Daemgen} S.,  {Siegler} N.,  {Reid} I.~N.,   {Close} L.~M.,  2007, \mn@doi
  [\apj] {10.1086/509109}, \href
  {https://ui.adsabs.harvard.edu/abs/2007ApJ...654..558D} {654, 558}

\bibitem[\protect\citeauthoryear{{Dan}, {Rosswog}, {Br{\"u}ggen}  \&
  {Podsiadlowski}}{{Dan} et~al.}{2014}]{Dan+etal+2014+WD_remnants}
{Dan} M.,  {Rosswog} S.,  {Br{\"u}ggen} M.,   {Podsiadlowski} P.,  2014,
  \mn@doi [\mnras] {10.1093/mnras/stt1766}, \href
  {https://ui.adsabs.harvard.edu/abs/2014MNRAS.438...14D} {438, 14}

\bibitem[\protect\citeauthoryear{{Darragh} \& {Murphy}}{{Darragh} \&
  {Murphy}}{2012}]{Darragh+Murphy+2012+SXPHE}
{Darragh} A.~N.,  {Murphy} B.~W.,  2012, Journal of the Southeastern
  Association for Research in Astronomy, \href
  {https://ui.adsabs.harvard.edu/abs/2012JSARA...6...72D} {6, 72}

\bibitem[\protect\citeauthoryear{{Denisenko}}{{Denisenko}}{2018}]{Denisenko+2018+0720_polar}
{Denisenko} D.,  2018, The Astronomer's Telegram, \href
  {https://ui.adsabs.harvard.edu/abs/2018ATel11626....1D} {11626, 1}

\bibitem[\protect\citeauthoryear{{Downes} \& {Shara}}{{Downes} \&
  {Shara}}{1993}]{Downes+etal+1993+CVcatalog}
{Downes} R.~A.,  {Shara} M.~M.,  1993, \mn@doi [\pasp] {10.1086/133139}, \href
  {https://ui.adsabs.harvard.edu/abs/1993PASP..105..127D} {105, 127}

\bibitem[\protect\citeauthoryear{{Downes}, {Webbink}, {Shara}, {Ritter}, {Kolb}
   \& {Duerbeck}}{{Downes} et~al.}{2001}]{Downes+etal+2001+ATLAS_CV}
{Downes} R.~A.,  {Webbink} R.~F.,  {Shara} M.~M.,  {Ritter} H.,  {Kolb} U.,
  {Duerbeck} H.~W.,  2001, \mn@doi [\pasp] {10.1086/320802}, \href
  {https://ui.adsabs.harvard.edu/abs/2001PASP..113..764D} {113, 764}

\bibitem[\protect\citeauthoryear{{Doyle}, {Bagnulo}, {Ramsay}, {Doyle}  \&
  {Hakala}}{{Doyle} et~al.}{2022}]{Doyle+etal+2022+rotation}
{Doyle} L.,  {Bagnulo} S.,  {Ramsay} G.,  {Doyle} J.~G.,   {Hakala} P.,  2022,
  \mn@doi [\mnras] {10.1093/mnras/stac464}, \href
  {https://ui.adsabs.harvard.edu/abs/2022MNRAS.512..979D} {512, 979}

\bibitem[\protect\citeauthoryear{{Drake} et~al.,}{{Drake}
  et~al.}{2009}]{Drake+etal+2009+catalina+realtime}
{Drake} A.~J.,  et~al., 2009, \mn@doi [\apj] {10.1088/0004-637X/696/1/870},
  \href {https://ui.adsabs.harvard.edu/abs/2009ApJ...696..870D} {696, 870}

\bibitem[\protect\citeauthoryear{{Drake} et~al.,}{{Drake}
  et~al.}{2014}]{Drake+etal+2014+Catalina_CV}
{Drake} A.~J.,  et~al., 2014, \mn@doi [\mnras] {10.1093/mnras/stu639}, \href
  {https://ui.adsabs.harvard.edu/abs/2014MNRAS.441.1186D} {441, 1186}

\bibitem[\protect\citeauthoryear{{El-Badry}, {Rix}, {Quataert}, {Kupfer}  \&
  {Shen}}{{El-Badry} et~al.}{2021}]{El-Badry+etal+2021+proto-ELM}
{El-Badry} K.,  {Rix} H.-W.,  {Quataert} E.,  {Kupfer} T.,   {Shen} K.~J.,
  2021, \mn@doi [\mnras] {10.1093/mnras/stab2583}, \href
  {https://ui.adsabs.harvard.edu/abs/2021MNRAS.508.4106E} {508, 4106}

\bibitem[\protect\citeauthoryear{{Ferrario}, {Vennes}, {Wickramasinghe},
  {Bailey}  \& {Christian}}{{Ferrario} et~al.}{1997}]{Ferrario+etal+1997+MWD}
{Ferrario} L.,  {Vennes} S.,  {Wickramasinghe} D.~T.,  {Bailey} J.~A.,
  {Christian} D.~J.,  1997, \mn@doi [\mnras] {10.1093/mnras/292.2.205}, \href
  {https://ui.adsabs.harvard.edu/abs/1997MNRAS.292..205F} {292, 205}

\bibitem[\protect\citeauthoryear{{Fitch}}{{Fitch}}{1973}]{Fitch+1973+ZZ_period}
{Fitch} W.~S.,  1973, \mn@doi [\apjl] {10.1086/181193}, \href
  {https://ui.adsabs.harvard.edu/abs/1973ApJ...181L..95F} {181, L95}

\bibitem[\protect\citeauthoryear{{F{\"o}rster} et~al.,}{{F{\"o}rster}
  et~al.}{2021}]{Forster+etal+2021+ALeRCE}
{F{\"o}rster} F.,  et~al., 2021, \mn@doi [\aj] {10.3847/1538-3881/abe9bc},
  \href {https://ui.adsabs.harvard.edu/abs/2021AJ....161..242F} {161, 242}

\bibitem[\protect\citeauthoryear{{Gaia Collaboration} et~al.,}{{Gaia
  Collaboration} et~al.}{2016}]{Gaia_Collaboration+2016+performance}
{Gaia Collaboration} et~al., 2016, \mn@doi [\aap]
  {10.1051/0004-6361/201629272}, \href
  {https://ui.adsabs.harvard.edu/abs/2016A&A...595A...1G} {595, A1}

\bibitem[\protect\citeauthoryear{{Gaia Collaboration} et~al.,}{{Gaia
  Collaboration} et~al.}{2018}]{Gaia_collaboration+2018+data}
{Gaia Collaboration} et~al., 2018, \mn@doi [\aap]
  {10.1051/0004-6361/201833051}, \href
  {https://ui.adsabs.harvard.edu/abs/2018A&A...616A...1G} {616, A1}

\bibitem[\protect\citeauthoryear{{Galloway} \& {Keek}}{{Galloway} \&
  {Keek}}{2021}]{Galloway+Keek+2021+thermonuclear_bursts}
{Galloway} D.~K.,  {Keek} L.,  2021, in {Belloni} T.~M.,  {M{\'e}ndez} M.,
  {Zhang} C.,  eds,  Astrophysics and Space Science Library Vol. 461,
  Astrophysics and Space Science Library. pp 209--262 (\mn@eprint {arXiv}
  {1712.06227}), \mn@doi{10.1007/978-3-662-62110-3_5}

\bibitem[\protect\citeauthoryear{{Galloway} et~al.,}{{Galloway}
  et~al.}{2020}]{Galloway+etal+2020+MINBAR}
{Galloway} D.~K.,  et~al., 2020, \mn@doi [\apjs] {10.3847/1538-4365/ab9f2e},
  \href {https://ui.adsabs.harvard.edu/abs/2020ApJS..249...32G} {249, 32}

\bibitem[\protect\citeauthoryear{{G{\"a}nsicke} et~al.,}{{G{\"a}nsicke}
  et~al.}{2009}]{Gansicke+etal+2009+CV_minimum}
{G{\"a}nsicke} B.~T.,  et~al., 2009, \mn@doi [\mnras]
  {10.1111/j.1365-2966.2009.15126.x}, \href
  {https://ui.adsabs.harvard.edu/abs/2009MNRAS.397.2170G} {397, 2170}

\bibitem[\protect\citeauthoryear{{Geier}, {Raddi}, {Gentile Fusillo}  \&
  {Marsh}}{{Geier} et~al.}{2019}]{Geier+etal+2019+hotsd_Gaia}
{Geier} S.,  {Raddi} R.,  {Gentile Fusillo} N.~P.,   {Marsh} T.~R.,  2019,
  \mn@doi [\aap] {10.1051/0004-6361/201834236}, \href
  {https://ui.adsabs.harvard.edu/abs/2019A&A...621A..38G} {621, A38}

\bibitem[\protect\citeauthoryear{{Gentile Fusillo} et~al.,}{{Gentile Fusillo}
  et~al.}{2019}]{GentileFusillo+etal+2019+gaia_wd}
{Gentile Fusillo} N.~P.,  et~al., 2019, \mn@doi [\mnras]
  {10.1093/mnras/sty3016}, \href
  {https://ui.adsabs.harvard.edu/abs/2019MNRAS.482.4570G} {482, 4570}

\bibitem[\protect\citeauthoryear{{G{\'o}rski}, {Hivon}, {Banday}, {Wandelt},
  {Hansen}, {Reinecke}  \& {Bartelmann}}{{G{\'o}rski}
  et~al.}{2005}]{Gorski+etal+2005}
{G{\'o}rski} K.~M.,  {Hivon} E.,  {Banday} A.~J.,  {Wandelt} B.~D.,  {Hansen}
  F.~K.,  {Reinecke} M.,   {Bartelmann} M.,  2005, \mn@doi [\apj]
  {10.1086/427976}, \href
  {https://ui.adsabs.harvard.edu/abs/2005ApJ...622..759G} {622, 759}

\bibitem[\protect\citeauthoryear{{Green}}{{Green}}{2018}]{Green+2018+python}
{Green} G.~M.,  2018, \mn@doi [The Journal of Open Source Software]
  {10.21105/joss.00695}, \href
  {https://ui.adsabs.harvard.edu/abs/2018JOSS....3..695G} {3, 695}

\bibitem[\protect\citeauthoryear{{Green} et~al.,}{{Green}
  et~al.}{2003}]{Green+etal+2003+sdB}
{Green} E.~M.,  et~al., 2003, \mn@doi [\apjl] {10.1086/367929}, \href
  {https://ui.adsabs.harvard.edu/abs/2003ApJ...583L..31G} {583, L31}

\bibitem[\protect\citeauthoryear{{Green}, {Schlafly}, {Zucker}, {Speagle}  \&
  {Finkbeiner}}{{Green} et~al.}{2019}]{Green+etal+2019+3dmap}
{Green} G.~M.,  {Schlafly} E.,  {Zucker} C.,  {Speagle} J.~S.,   {Finkbeiner}
  D.,  2019, \mn@doi [\apj] {10.3847/1538-4357/ab5362}, \href
  {https://ui.adsabs.harvard.edu/abs/2019ApJ...887...93G} {887, 93}

\bibitem[\protect\citeauthoryear{{Green} et~al.,}{{Green}
  et~al.}{2020}]{Green+etal+2020+UCABs}
{Green} M.~J.,  et~al., 2020, \mn@doi [\mnras] {10.1093/mnras/staa1509}, \href
  {https://ui.adsabs.harvard.edu/abs/2020MNRAS.496.1243G} {496, 1243}

\bibitem[\protect\citeauthoryear{{Guo}, {Zhao}, {Tziamtzis}, {Liu}, {Li},
  {Zhang}, {Hou}  \& {Wang}}{{Guo} et~al.}{2015}]{Guo+etal+2015+WD}
{Guo} J.,  {Zhao} J.,  {Tziamtzis} A.,  {Liu} J.,  {Li} L.,  {Zhang} Y.,  {Hou}
  Y.,   {Wang} Y.,  2015, \mn@doi [\mnras] {10.1093/mnras/stv2104}, \href
  {https://ui.adsabs.harvard.edu/abs/2015MNRAS.454.2787G} {454, 2787}

\bibitem[\protect\citeauthoryear{{Guo}, {Zhao}, {Zhang}, {Zhang}, {Bai},
  {Walters}, {Yang}  \& {Liu}}{{Guo} et~al.}{2022}]{Guo+etal+2022+WD}
{Guo} J.,  {Zhao} J.,  {Zhang} H.,  {Zhang} J.,  {Bai} Y.,  {Walters} N.,
  {Yang} Y.,   {Liu} J.,  2022, \mn@doi [\mnras] {10.1093/mnras/stab3151},
  \href {https://ui.adsabs.harvard.edu/abs/2022MNRAS.509.2674G} {509, 2674}

\bibitem[\protect\citeauthoryear{{Hameury} \& {Lasota}}{{Hameury} \&
  {Lasota}}{2017}]{Hameury+Lasota+2017+DN_outburst_IP}
{Hameury} J.~M.,  {Lasota} J.~P.,  2017, \mn@doi [\aap]
  {10.1051/0004-6361/201730760}, \href
  {https://ui.adsabs.harvard.edu/abs/2017A&A...602A.102H} {602, A102}

\bibitem[\protect\citeauthoryear{{Handler} \& {Paunzen}}{{Handler} \&
  {Paunzen}}{1999}]{Handler+Paunzen+1999+roAp}
{Handler} G.,  {Paunzen} E.,  1999, \mn@doi [\aaps] {10.1051/aas:1999159},
  \href {https://ui.adsabs.harvard.edu/abs/1999A&AS..135...57H} {135, 57}

\bibitem[\protect\citeauthoryear{{Hatt} et~al.,}{{Hatt}
  et~al.}{2022}]{Hatt+etal+2022+TESS_oscillations}
{Hatt} E.,  et~al., 2022, arXiv e-prints, \href
  {https://ui.adsabs.harvard.edu/abs/2022arXiv221009109H} {p. arXiv:2210.09109}

\bibitem[\protect\citeauthoryear{{Heber}}{{Heber}}{2009}]{Heber+2009+araa}
{Heber} U.,  2009, \mn@doi [\araa] {10.1146/annurev-astro-082708-101836}, \href
  {https://ui.adsabs.harvard.edu/abs/2009ARA&A..47..211H} {47, 211}

\bibitem[\protect\citeauthoryear{{Heinke}, {Edmonds}  \& {Grindlay}}{{Heinke}
  et~al.}{2001}]{Heinke+etal+2001+XB1832}
{Heinke} C.~O.,  {Edmonds} P.~D.,   {Grindlay} J.~E.,  2001, \mn@doi [\apj]
  {10.1086/323493}, \href
  {https://ui.adsabs.harvard.edu/abs/2001ApJ...562..363H} {562, 363}

\bibitem[\protect\citeauthoryear{{Heinze} et~al.,}{{Heinze}
  et~al.}{2018}]{Heinze+etal+2018+ATLAS_variables}
{Heinze} A.~N.,  et~al., 2018, \mn@doi [\aj] {10.3847/1538-3881/aae47f}, \href
  {https://ui.adsabs.harvard.edu/abs/2018AJ....156..241H} {156, 241}

\bibitem[\protect\citeauthoryear{{Hermes}, {Kilic}, {Brown}, {Montgomery}  \&
  {Winget}}{{Hermes} et~al.}{2012a}]{Hermes+etal+2012+ell_WD}
{Hermes} J.~J.,  {Kilic} M.,  {Brown} W.~R.,  {Montgomery} M.~H.,   {Winget}
  D.~E.,  2012a, \mn@doi [\apj] {10.1088/0004-637X/749/1/42}, \href
  {https://ui.adsabs.harvard.edu/abs/2012ApJ...749...42H} {749, 42}

\bibitem[\protect\citeauthoryear{{Hermes}, {Montgomery}, {Winget}, {Brown},
  {Kilic}  \& {Kenyon}}{{Hermes} et~al.}{2012b}]{Hermes+etal+2012+ELMV}
{Hermes} J.~J.,  {Montgomery} M.~H.,  {Winget} D.~E.,  {Brown} W.~R.,  {Kilic}
  M.,   {Kenyon} S.~J.,  2012b, \mn@doi [\apjl] {10.1088/2041-8205/750/2/L28},
  \href {https://ui.adsabs.harvard.edu/abs/2012ApJ...750L..28H} {750, L28}

\bibitem[\protect\citeauthoryear{{Holdsworth} et~al.,}{{Holdsworth}
  et~al.}{2014}]{Holdsworth+etal+2014+roapam}
{Holdsworth} D.~L.,  et~al., 2014, \mn@doi [\mnras] {10.1093/mnras/stu094},
  \href {https://ui.adsabs.harvard.edu/abs/2014MNRAS.439.2078H} {439, 2078}

\bibitem[\protect\citeauthoryear{{Jakate}}{{Jakate}}{1979}]{Jakate+1979+bceps}
{Jakate} S.~M.,  1979, \mn@doi [\aj] {10.1086/112510}, \href
  {https://ui.adsabs.harvard.edu/abs/1979AJ.....84.1042J} {84, 1042}

\bibitem[\protect\citeauthoryear{{Jayasinghe} et~al.,}{{Jayasinghe}
  et~al.}{2020}]{Jayasinghe+etal+2020+deltascuti}
{Jayasinghe} T.,  et~al., 2020, \mn@doi [\mnras] {10.1093/mnras/staa499}, \href
  {https://ui.adsabs.harvard.edu/abs/2020MNRAS.493.4186J} {493, 4186}

\bibitem[\protect\citeauthoryear{{Kaluzny} \& {Krzeminski}}{{Kaluzny} \&
  {Krzeminski}}{1993}]{Kaluzny+Krzeminski+1993+SXPHE}
{Kaluzny} J.,  {Krzeminski} W.,  1993, \mn@doi [\mnras]
  {10.1093/mnras/264.4.785}, \href
  {https://ui.adsabs.harvard.edu/abs/1993MNRAS.264..785K} {264, 785}

\bibitem[\protect\citeauthoryear{{Kato} et~al.,}{{Kato}
  et~al.}{2013}]{Kato+etal+2013+DN}
{Kato} T.,  et~al., 2013, \mn@doi [\pasj] {10.1093/pasj/65.1.23}, \href
  {https://ui.adsabs.harvard.edu/abs/2013PASJ...65...23K} {65, 23}

\bibitem[\protect\citeauthoryear{{Kato} et~al.,}{{Kato}
  et~al.}{2016}]{Kato+etal+2016+SUUMa}
{Kato} T.,  et~al., 2016, \mn@doi [\pasj] {10.1093/pasj/psw064}, \href
  {https://ui.adsabs.harvard.edu/abs/2016PASJ...68...65K} {68, 65}

\bibitem[\protect\citeauthoryear{{Keller}, {Breedt}, {Hodgkin}, {Belokurov},
  {Wild}, {Garc{\'\i}a-Soriano}  \& {Wise}}{{Keller}
  et~al.}{2022}]{Keller+etal+2022+eclipsingWD}
{Keller} P.~M.,  {Breedt} E.,  {Hodgkin} S.,  {Belokurov} V.,  {Wild} J.,
  {Garc{\'\i}a-Soriano} I.,   {Wise} J.~L.,  2022, \mn@doi [\mnras]
  {10.1093/mnras/stab3293}, \href
  {https://ui.adsabs.harvard.edu/abs/2022MNRAS.509.4171K} {509, 4171}

\bibitem[\protect\citeauthoryear{{Kilic}, {Brown}, {Gianninas}, {Hermes},
  {Allende Prieto}  \& {Kenyon}}{{Kilic} et~al.}{2014}]{Kilic+etal+2014+EWD}
{Kilic} M.,  {Brown} W.~R.,  {Gianninas} A.,  {Hermes} J.~J.,  {Allende Prieto}
  C.,   {Kenyon} S.~J.,  2014, \mn@doi [\mnras] {10.1093/mnrasl/slu093}, \href
  {https://ui.adsabs.harvard.edu/abs/2014MNRAS.444L...1K} {444, L1}

\bibitem[\protect\citeauthoryear{{Kilkenny}, {Koen}, {O'Donoghue}  \&
  {Stobie}}{{Kilkenny} et~al.}{1997}]{Kilkenny+etal+1997+361_discovery}
{Kilkenny} D.,  {Koen} C.,  {O'Donoghue} D.,   {Stobie} R.~S.,  1997, \mn@doi
  [\mnras] {10.1093/mnras/285.3.640}, \href
  {https://ui.adsabs.harvard.edu/abs/1997MNRAS.285..640K} {285, 640}

\bibitem[\protect\citeauthoryear{{Knevitt}, {Wynn}, {Vaughan}  \&
  {Watson}}{{Knevitt} et~al.}{2014}]{Knevitt+etal+2014+inefficient_accretion}
{Knevitt} G.,  {Wynn} G.~A.,  {Vaughan} S.,   {Watson} M.~G.,  2014, \mn@doi
  [\mnras] {10.1093/mnras/stt2008}, \href
  {https://ui.adsabs.harvard.edu/abs/2014MNRAS.437.3087K} {437, 3087}

\bibitem[\protect\citeauthoryear{{Krzesinski}, {{\c{S}}ener}, {Zola}  \&
  {Siwak}}{{Krzesinski} et~al.}{2022}]{Krzesinski+etal+2022+hst_binary}
{Krzesinski} J.,  {{\c{S}}ener} H.~T.,  {Zola} S.,   {Siwak} M.,  2022, \mn@doi
  [\mnras] {10.1093/mnras/stac2088}, \href
  {https://ui.adsabs.harvard.edu/abs/2022MNRAS.516.1509K} {516, 1509}

\bibitem[\protect\citeauthoryear{{Kupfer} et~al.,}{{Kupfer}
  et~al.}{2019}]{Kupfer+2019+high-g_BLAPs}
{Kupfer} T.,  et~al., 2019, \mn@doi [\apjl] {10.3847/2041-8213/ab263c}, \href
  {https://ui.adsabs.harvard.edu/abs/2019ApJ...878L..35K} {878, L35}

\bibitem[\protect\citeauthoryear{{Kupfer} et~al.,}{{Kupfer}
  et~al.}{2021}]{Kupfer+etal+2021+ZTFhighcadence}
{Kupfer} T.,  et~al., 2021, \mn@doi [\mnras] {10.1093/mnras/stab1344}, \href
  {https://ui.adsabs.harvard.edu/abs/2021MNRAS.505.1254K} {505, 1254}

\bibitem[\protect\citeauthoryear{{Landolt}}{{Landolt}}{1968}]{Landolt+1968+new_blue_ZZ}
{Landolt} A.~U.,  1968, \mn@doi [\apj] {10.1086/149645}, \href
  {https://ui.adsabs.harvard.edu/abs/1968ApJ...153..151L} {153, 151}

\bibitem[\protect\citeauthoryear{{Lanning}}{{Lanning}}{1973}]{Lanning+1973+UV_sources}
{Lanning} H.~H.,  1973, \mn@doi [\pasp] {10.1086/129406}, \href
  {https://ui.adsabs.harvard.edu/abs/1973PASP...85...70L} {85, 70}

\bibitem[\protect\citeauthoryear{{Lei}, {Zhao}, {N{\'e}meth}  \& {Zhao}}{{Lei}
  et~al.}{2018}]{Lei+etal+2018+Lamost_subdwarf}
{Lei} Z.,  {Zhao} J.,  {N{\'e}meth} P.,   {Zhao} G.,  2018, \mn@doi [\apj]
  {10.3847/1538-4357/aae82b}, \href
  {https://ui.adsabs.harvard.edu/abs/2018ApJ...868...70L} {868, 70}

\bibitem[\protect\citeauthoryear{{Li}, {Kim}, {Xia}, {Michel}, {Hu}, {Gao},
  {Guo}  \& {Chen}}{{Li} et~al.}{2020}]{Li+etal+2020+cutoff}
{Li} K.,  {Kim} C.-H.,  {Xia} Q.-Q.,  {Michel} R.,  {Hu} S.-M.,  {Gao} X.,
  {Guo} D.-F.,   {Chen} X.,  2020, \mn@doi [\aj] {10.3847/1538-3881/ab7cda},
  \href {https://ui.adsabs.harvard.edu/abs/2020AJ....159..189L} {159, 189}

\bibitem[\protect\citeauthoryear{{Li}, {Shi}, {Yan}, {Fu}, {Li}  \& {Hou}}{{Li}
  et~al.}{2021}]{Li+etal+2021+double_spectroscopic_binaries}
{Li} C.-q.,  {Shi} J.-r.,  {Yan} H.-l.,  {Fu} J.-N.,  {Li} J.-d.,   {Hou}
  Y.-H.,  2021, \mn@doi [\apjs] {10.3847/1538-4365/ac22a8}, \href
  {https://ui.adsabs.harvard.edu/abs/2021ApJS..256...31L} {256, 31}

\bibitem[\protect\citeauthoryear{{Liakos} \& {Niarchos}}{{Liakos} \&
  {Niarchos}}{2017}]{Liakos+Niarchos+2017+EDSCT_cat}
{Liakos} A.,  {Niarchos} P.,  2017, \mn@doi [\mnras] {10.1093/mnras/stw2756},
  \href {https://ui.adsabs.harvard.edu/abs/2017MNRAS.465.1181L} {465, 1181}

\bibitem[\protect\citeauthoryear{{Liebert} \& {Stockman}}{{Liebert} \&
  {Stockman}}{1985}]{Liebert+Stockman+1985+inho_accretion}
{Liebert} J.,  {Stockman} H.~S.,  1985, in {Lamb} D.~Q.,  {Patterson} J.,  eds,
  Cataclysmic Variables and Low-Mass X-ray Binaries. p.~151,
  \mn@doi{10.1007/978-94-009-5319-2_20}

\bibitem[\protect\citeauthoryear{{Lin} \& {Yu}}{{Lin} \&
  {Yu}}{2018}]{Lin+Yu+2018+UCXB}
{Lin} J.,  {Yu} W.,  2018, \mn@doi [\mnras] {10.1093/mnras/stx2818}, \href
  {https://ui.adsabs.harvard.edu/abs/2018MNRAS.474.1922L} {474, 1922}

\bibitem[\protect\citeauthoryear{{Lin} \& {Yu}}{{Lin} \&
  {Yu}}{2020}]{Lin+Yu+2020+1712}
{Lin} J.,  {Yu} W.,  2020, \mn@doi [\apj] {10.3847/1538-4357/abb76f}, \href
  {https://ui.adsabs.harvard.edu/abs/2020ApJ...903...37L} {903, 37}

\bibitem[\protect\citeauthoryear{{Lin}, {Yan}, {Han}  \& {Yu}}{{Lin}
  et~al.}{2019}]{Lin+etal+2019+outbursts}
{Lin} J.,  {Yan} Z.,  {Han} Z.,   {Yu} W.,  2019, \mn@doi [\apj]
  {10.3847/1538-4357/aaf39b}, \href
  {https://ui.adsabs.harvard.edu/abs/2019ApJ...870..126L} {870, 126}

\bibitem[\protect\citeauthoryear{{Lin} et~al.,}{{Lin}
  et~al.}{2022}]{tmtsI+2022}
{Lin} J.,  et~al., 2022, \mn@doi [\mnras] {10.1093/mnras/stab2812}, \href
  {https://ui.adsabs.harvard.edu/abs/2022MNRAS.509.2362L} {509, 2362}

\bibitem[\protect\citeauthoryear{{Lin} et~al.,}{{Lin}
  et~al.}{2023}]{lin+etal+2022+NatAs}
{Lin} J.,  et~al., 2023, \mn@doi [Nature Astronomy]
  {10.1038/s41550-022-01783-z}, \href
  {https://ui.adsabs.harvard.edu/abs/2023NatAs...7..223L} {7, 223}

\bibitem[\protect\citeauthoryear{{Littlefield}, {Garnavich}, {Magno},
  {Murison}, {Deal}, {McClelland}  \& {Rose}}{{Littlefield}
  et~al.}{2015}]{Littlefield+etal+2015+highamplitudepolar}
{Littlefield} C.,  {Garnavich} P.,  {Magno} K.,  {Murison} M.,  {Deal} S.,
  {McClelland} C.,   {Rose} B.,  2015, Information Bulletin on Variable Stars,
  \href {https://ui.adsabs.harvard.edu/abs/2015IBVS.6129....1L} {6129, 1}

\bibitem[\protect\citeauthoryear{{Liu} et~al.,}{{Liu}
  et~al.}{2019}]{Liu+etal+2019+Nature+LB}
{Liu} J.,  et~al., 2019, \mn@doi [\nat] {10.1038/s41586-019-1766-2}, \href
  {https://ui.adsabs.harvard.edu/abs/2019Natur.575..618L} {575, 618}

\bibitem[\protect\citeauthoryear{{Loeb} \& {Gaudi}}{{Loeb} \&
  {Gaudi}}{2003}]{Loeb+Gaudi+2003+beaming}
{Loeb} A.,  {Gaudi} B.~S.,  2003, \mn@doi [\apjl] {10.1086/375551}, \href
  {https://ui.adsabs.harvard.edu/abs/2003ApJ...588L.117L} {588, L117}

\bibitem[\protect\citeauthoryear{{Lomb}}{{Lomb}}{1976}]{Lomb+1976}
{Lomb} N.~R.,  1976, \mn@doi [\apss] {10.1007/BF00648343}, \href
  {https://ui.adsabs.harvard.edu/abs/1976Ap&SS..39..447L} {39, 447}

\bibitem[\protect\citeauthoryear{{Lopez}, {Hermes}, {Calcaferro}, {Bell},
  {Samuels}, {Vanderbosch}, {C{\'o}rsico}  \& {Istrate}}{{Lopez}
  et~al.}{2021}]{Lopez+etal+2021+GD278}
{Lopez} I.~D.,  {Hermes} J.~J.,  {Calcaferro} L.~M.,  {Bell} K.~J.,  {Samuels}
  A.,  {Vanderbosch} Z.~P.,  {C{\'o}rsico} A.~H.,   {Istrate} A.~G.,  2021,
  \mn@doi [\apj] {10.3847/1538-4357/ac2d28}, \href
  {https://ui.adsabs.harvard.edu/abs/2021ApJ...922..220L} {922, 220}

\bibitem[\protect\citeauthoryear{{Luo}, {N{\'e}meth}, {Wang}, {Wang}  \&
  {Han}}{{Luo} et~al.}{2021}]{Luo+etal+2021+hotsubdwarf}
{Luo} Y.,  {N{\'e}meth} P.,  {Wang} K.,  {Wang} X.,   {Han} Z.,  2021, \mn@doi
  [\apjs] {10.3847/1538-4365/ac11f6}, \href
  {https://ui.adsabs.harvard.edu/abs/2021ApJS..256...28L} {256, 28}

\bibitem[\protect\citeauthoryear{{Macfarlane}, {Toma}, {Ramsay}, {Groot},
  {Woudt}, {Drew}, {Barentsen}  \& {Eisl{\"o}ffel}}{{Macfarlane}
  et~al.}{2015}]{Macfarlane+etal+2015+omega_I}
{Macfarlane} S.~A.,  {Toma} R.,  {Ramsay} G.,  {Groot} P.~J.,  {Woudt} P.~A.,
  {Drew} J.~E.,  {Barentsen} G.,   {Eisl{\"o}ffel} J.,  2015, \mn@doi [\mnras]
  {10.1093/mnras/stv1989}, \href
  {https://ui.adsabs.harvard.edu/abs/2015MNRAS.454..507M} {454, 507}

\bibitem[\protect\citeauthoryear{{Masci} et~al.,}{{Masci}
  et~al.}{2019}]{ZTF+2019+products}
{Masci} F.~J.,  et~al., 2019, \mn@doi [\pasp] {10.1088/1538-3873/aae8ac}, \href
  {https://ui.adsabs.harvard.edu/abs/2019PASP..131a8003M} {131, 018003}

\bibitem[\protect\citeauthoryear{{Mathur} et~al.,}{{Mathur}
  et~al.}{2022}]{Mathur+etal+2022+solar-like_oscillations}
{Mathur} S.,  et~al., 2022, \mn@doi [\aap] {10.1051/0004-6361/202141168}, \href
  {https://ui.adsabs.harvard.edu/abs/2022A&A...657A..31M} {657, A31}

\bibitem[\protect\citeauthoryear{{Maxted} et~al.,}{{Maxted}
  et~al.}{2013}]{Maxted+etal+2013+preELMV}
{Maxted} P. F.~L.,  et~al., 2013, \mn@doi [\nat] {10.1038/nature12192}, \href
  {https://ui.adsabs.harvard.edu/abs/2013Natur.498..463M} {498, 463}

\bibitem[\protect\citeauthoryear{{McCarthy} et~al.,}{{McCarthy}
  et~al.}{1998}]{McCarthy+etal+1998_LRIS}
{McCarthy} J.~K.,  et~al., 1998, in {D'Odorico} S.,  ed.,  Society of
  Photo-Optical Instrumentation Engineers (SPIE) Conference Series Vol. 3355,
  Optical Astronomical Instrumentation. pp 81--92, \mn@doi{10.1117/12.316831}

\bibitem[\protect\citeauthoryear{{McNamara}}{{McNamara}}{2011}]{McNamara+2011+PL_relation}
{McNamara} D.~H.,  2011, \mn@doi [\aj] {10.1088/0004-6256/142/4/110}, \href
  {https://ui.adsabs.harvard.edu/abs/2011AJ....142..110M} {142, 110}

\bibitem[\protect\citeauthoryear{{McWhirter} \& {Lam}}{{McWhirter} \&
  {Lam}}{2022}]{McWhirter+Lam+2022+blap_candidates}
{McWhirter} P.~R.,  {Lam} M.~C.,  2022, \mn@doi [\mnras]
  {10.1093/mnras/stac291}, \href
  {https://ui.adsabs.harvard.edu/abs/2022MNRAS.511.4971M} {511, 4971}

\bibitem[\protect\citeauthoryear{{Menou}, {Narayan}  \& {Lasota}}{{Menou}
  et~al.}{1999}]{Menou+etal+1999}
{Menou} K.,  {Narayan} R.,   {Lasota} J.-P.,  1999, \mn@doi [\apj]
  {10.1086/306878}, \href
  {https://ui.adsabs.harvard.edu/abs/1999ApJ...513..811M} {513, 811}

\bibitem[\protect\citeauthoryear{{Motch} et~al.,}{{Motch}
  et~al.}{2010}]{Motch+etal+2010+XMM_sources}
{Motch} C.,  et~al., 2010, \mn@doi [\aap] {10.1051/0004-6361/200913570}, \href
  {https://ui.adsabs.harvard.edu/abs/2010A&A...523A..92M} {523, A92}

\bibitem[\protect\citeauthoryear{{Muno} et~al.,}{{Muno}
  et~al.}{2003}]{Muno+etal+2003+Chandra_sources}
{Muno} M.~P.,  et~al., 2003, \mn@doi [\apj] {10.1086/374639}, \href
  {https://ui.adsabs.harvard.edu/abs/2003ApJ...589..225M} {589, 225}

\bibitem[\protect\citeauthoryear{{Murphy}, {Hey}, {Van Reeth}  \&
  {Bedding}}{{Murphy} et~al.}{2019}]{Murphy+etal+2019+ds}
{Murphy} S.~J.,  {Hey} D.,  {Van Reeth} T.,   {Bedding} T.~R.,  2019, \mn@doi
  [\mnras] {10.1093/mnras/stz590}, \href
  {https://ui.adsabs.harvard.edu/abs/2019MNRAS.485.2380M} {485, 2380}

\bibitem[\protect\citeauthoryear{{Napiwotzki}}{{Napiwotzki}}{2009}]{Napiwotzki+2009+pop_WD}
{Napiwotzki} R.,  2009, in Journal of Physics Conference Series. p. 012004
  (\mn@eprint {arXiv} {0903.2159}), \mn@doi{10.1088/1742-6596/172/1/012004}

\bibitem[\protect\citeauthoryear{{Nebot G{\'o}mez-Mor{\'a}n} et~al.,}{{Nebot
  G{\'o}mez-Mor{\'a}n} et~al.}{2011}]{Gomez-Moran+2011+post_common_binary}
{Nebot G{\'o}mez-Mor{\'a}n} A.,  et~al., 2011, \mn@doi [\aap]
  {10.1051/0004-6361/201117514}, \href
  {https://ui.adsabs.harvard.edu/abs/2011A&A...536A..43N} {536, A43}

\bibitem[\protect\citeauthoryear{{Nefs} et~al.,}{{Nefs}
  et~al.}{2012}]{Nefs+etal+2012+Mdwarfs}
{Nefs} S.~V.,  et~al., 2012, \mn@doi [\mnras]
  {10.1111/j.1365-2966.2012.21338.x}, \href
  {https://ui.adsabs.harvard.edu/abs/2012MNRAS.425..950N} {425, 950}

\bibitem[\protect\citeauthoryear{{Nelemans}, {Yungelson}, {Portegies Zwart}  \&
  {Verbunt}}{{Nelemans} et~al.}{2001}]{Nelemans+etal+2001+pop_WD}
{Nelemans} G.,  {Yungelson} L.~R.,  {Portegies Zwart} S.~F.,   {Verbunt} F.,
  2001, \mn@doi [\aap] {10.1051/0004-6361:20000147}, \href
  {https://ui.adsabs.harvard.edu/abs/2001A&A...365..491N} {365, 491}

\bibitem[\protect\citeauthoryear{{Norton}, {Butters}, {Parker}  \&
  {Wynn}}{{Norton} et~al.}{2008}]{Norton+etal+2008+EXHYA}
{Norton} A.~J.,  {Butters} O.~W.,  {Parker} T.~L.,   {Wynn} G.~A.,  2008,
  \mn@doi [\apj] {10.1086/523932}, \href
  {https://ui.adsabs.harvard.edu/abs/2008ApJ...672..524N} {672, 524}

\bibitem[\protect\citeauthoryear{{Oke} et~al.,}{{Oke}
  et~al.}{1995}]{Oke+etal+1995+Keck+LRIS}
{Oke} J.~B.,  et~al., 1995, \mn@doi [\pasp] {10.1086/133562}, \href
  {https://ui.adsabs.harvard.edu/abs/1995PASP..107..375O} {107, 375}

\bibitem[\protect\citeauthoryear{{Osaki}}{{Osaki}}{1996}]{Osaki+1996+DN_review}
{Osaki} Y.,  1996, \mn@doi [\pasp] {10.1086/133689}, \href
  {https://ui.adsabs.harvard.edu/abs/1996PASP..108...39O} {108, 39}

\bibitem[\protect\citeauthoryear{{Osborne}, {Beardmore}, {Wheatley}, {Hakala},
  {Watson}, {Mason}, {Hassall}  \& {King}}{{Osborne}
  et~al.}{1994}]{Osborne+etal+1994+RE1307}
{Osborne} J.~P.,  {Beardmore} A.~P.,  {Wheatley} P.~J.,  {Hakala} P.,  {Watson}
  M.~G.,  {Mason} K.~O.,  {Hassall} B.~J.~M.,   {King} A.~R.,  1994, \mn@doi
  [\mnras] {10.1093/mnras/270.3.650}, \href
  {https://ui.adsabs.harvard.edu/abs/1994MNRAS.270..650O} {270, 650}

\bibitem[\protect\citeauthoryear{{{\O}stensen} et~al.,}{{{\O}stensen}
  et~al.}{2010}]{Ostensen+etal+2010+pmodesdBVs}
{{\O}stensen} R.~H.,  et~al., 2010, \mn@doi [\aap]
  {10.1051/0004-6361/200913480}, \href
  {https://ui.adsabs.harvard.edu/abs/2010A&A...513A...6O} {513, A6}

\bibitem[\protect\citeauthoryear{{Parsons} et~al.,}{{Parsons}
  et~al.}{2013}]{Parsons+etal+2013+R_ell_binary}
{Parsons} S.~G.,  et~al., 2013, \mn@doi [\mnras] {10.1093/mnras/sts332}, \href
  {https://ui.adsabs.harvard.edu/abs/2013MNRAS.429..256P} {429, 256}

\bibitem[\protect\citeauthoryear{{Patterson}}{{Patterson}}{1994}]{Patterson+1994+DQ}
{Patterson} J.,  1994, \mn@doi [\pasp] {10.1086/133375}, \href
  {https://ui.adsabs.harvard.edu/abs/1994PASP..106..209P} {106, 209}

\bibitem[\protect\citeauthoryear{{Patterson}, {Robinson}  \&
  {Nather}}{{Patterson} et~al.}{1978}]{Patterson+etal+1978+DQ_model}
{Patterson} J.,  {Robinson} E.~L.,   {Nather} R.~E.,  1978, \mn@doi [\apj]
  {10.1086/156405}, \href
  {https://ui.adsabs.harvard.edu/abs/1978ApJ...224..570P} {224, 570}

\bibitem[\protect\citeauthoryear{{Pelisoli} et~al.,}{{Pelisoli}
  et~al.}{2021}]{Pelisoli+etal+2021_NatAst_ell}
{Pelisoli} I.,  et~al., 2021, \mn@doi [Nature Astronomy]
  {10.1038/s41550-021-01413-0}, \href
  {https://ui.adsabs.harvard.edu/abs/2021NatAs...5.1052P} {5, 1052}

\bibitem[\protect\citeauthoryear{{Pietrukowicz} et~al.,}{{Pietrukowicz}
  et~al.}{2017}]{Pietrukowicz+etal+2017+BLAPs}
{Pietrukowicz} P.,  et~al., 2017, \mn@doi [Nature Astronomy]
  {10.1038/s41550-017-0166}, \href
  {https://ui.adsabs.harvard.edu/abs/2017NatAs...1E.166P} {1, 0166}

\bibitem[\protect\citeauthoryear{{Pietrukowicz} et~al.,}{{Pietrukowicz}
  et~al.}{2020}]{Pietrukowicz+etal+2020+deltascuti}
{Pietrukowicz} P.,  et~al., 2020, \mn@doi [\actaa] {10.32023/0001-5237/70.4.1},
  \href {https://ui.adsabs.harvard.edu/abs/2020AcA....70..241P} {70, 241}

\bibitem[\protect\citeauthoryear{{Pigulski} \& {Pojma{\'n}ski}}{{Pigulski} \&
  {Pojma{\'n}ski}}{2008}]{Pigulski+Pojmanski+2008+beta_cep}
{Pigulski} A.,  {Pojma{\'n}ski} G.,  2008, \mn@doi [\aap]
  {10.1051/0004-6361:20078581}, \href
  {https://ui.adsabs.harvard.edu/abs/2008A&A...477..917P} {477, 917}

\bibitem[\protect\citeauthoryear{{Pigulski}, {Kotysz}  \&
  {Ko{\l}aczek-Szyma{\'n}ski}}{{Pigulski}
  et~al.}{2022}]{Pigulski+Kolaczek-Szymanski+2022+TESS_BLAP}
{Pigulski} A.,  {Kotysz} K.,   {Ko{\l}aczek-Szyma{\'n}ski} P.~A.,  2022,
  \mn@doi [\aap] {10.1051/0004-6361/202243293}, \href
  {https://ui.adsabs.harvard.edu/abs/2022A&A...663A..62P} {663, A62}

\bibitem[\protect\citeauthoryear{{Pogrosheva} et~al.,}{{Pogrosheva}
  et~al.}{2018}]{Pogrosheva+etal+2018+MASTER_CV}
{Pogrosheva} T.,  et~al., 2018, The Astronomer's Telegram, \href
  {https://ui.adsabs.harvard.edu/abs/2018ATel11620....1P} {11620, 1}

\bibitem[\protect\citeauthoryear{{Pojmanski}}{{Pojmanski}}{2002}]{Pojmanski+2002+ASAS_6h}
{Pojmanski} G.,  2002, \actaa, \href
  {https://ui.adsabs.harvard.edu/abs/2002AcA....52..397P} {52, 397}

\bibitem[\protect\citeauthoryear{{Pretorius} \& {Mukai}}{{Pretorius} \&
  {Mukai}}{2014}]{Pretorius+etal+2014+IP_density}
{Pretorius} M.~L.,  {Mukai} K.,  2014, \mn@doi [\mnras] {10.1093/mnras/stu990},
  \href {https://ui.adsabs.harvard.edu/abs/2014MNRAS.442.2580P} {442, 2580}

\bibitem[\protect\citeauthoryear{{Pretorius}, {Knigge}  \&
  {Schwope}}{{Pretorius} et~al.}{2013}]{Pretorius+etal+2013+mCV_density}
{Pretorius} M.~L.,  {Knigge} C.,   {Schwope} A.~D.,  2013, \mn@doi [\mnras]
  {10.1093/mnras/stt499}, \href
  {https://ui.adsabs.harvard.edu/abs/2013MNRAS.432..570P} {432, 570}

\bibitem[\protect\citeauthoryear{{Pshirkov} et~al.,}{{Pshirkov}
  et~al.}{2020}]{Pshirkov+etal+2020+WD1832}
{Pshirkov} M.~S.,  et~al., 2020, \mn@doi [\mnras] {10.1093/mnrasl/slaa149},
  \href {https://ui.adsabs.harvard.edu/abs/2020MNRAS.499L..21P} {499, L21}

\bibitem[\protect\citeauthoryear{{Ramsay}, {Napiwotzki}, {Barclay}, {Hakala},
  {Potter}  \& {Cropper}}{{Ramsay} et~al.}{2011}]{Ramsay+etal+2011+SXPHE}
{Ramsay} G.,  {Napiwotzki} R.,  {Barclay} T.,  {Hakala} P.,  {Potter} S.,
  {Cropper} M.,  2011, \mn@doi [\mnras] {10.1111/j.1365-2966.2011.19275.x},
  \href {https://ui.adsabs.harvard.edu/abs/2011MNRAS.417..400R} {417, 400}

\bibitem[\protect\citeauthoryear{{Ramsay} et~al.,}{{Ramsay}
  et~al.}{2014}]{Ramsay+etal+2014+Kepler_deep}
{Ramsay} G.,  et~al., 2014, \mn@doi [\mnras] {10.1093/mnras/stt1863}, \href
  {https://ui.adsabs.harvard.edu/abs/2014MNRAS.437..132R} {437, 132}

\bibitem[\protect\citeauthoryear{{Ramsay}, {Hakala}, {Doyle}, {Doyle}  \&
  {Bagnulo}}{{Ramsay} et~al.}{2022a}]{Ramsay+etal+2022+rotation}
{Ramsay} G.,  {Hakala} P.,  {Doyle} J.~G.,  {Doyle} L.,   {Bagnulo} S.,  2022a,
  \mn@doi [\mnras] {10.1093/mnras/stac188}, \href
  {https://ui.adsabs.harvard.edu/abs/2022MNRAS.511.2755R} {511, 2755}

\bibitem[\protect\citeauthoryear{{Ramsay} et~al.,}{{Ramsay}
  et~al.}{2022b}]{Ramsay+etal+2022+BLAP_sdbWD}
{Ramsay} G.,  et~al., 2022b, \mn@doi [\mnras] {10.1093/mnras/stac1000}, \href
  {https://ui.adsabs.harvard.edu/abs/2022MNRAS.513.2215R} {513, 2215}

\bibitem[\protect\citeauthoryear{{Reed} et~al.,}{{Reed}
  et~al.}{2010}]{Reed+etal+2010+gmodesdBVs}
{Reed} M.~D.,  et~al., 2010, \mn@doi [\mnras]
  {10.1111/j.1365-2966.2010.17423.x}, \href
  {https://ui.adsabs.harvard.edu/abs/2010MNRAS.409.1496R} {409, 1496}

\bibitem[\protect\citeauthoryear{{Ricker} et~al.,}{{Ricker}
  et~al.}{2014}]{Ricker+etal+2014+TESS}
{Ricker} G.~R.,  et~al., 2014, in {Oschmann} Jacobus~M. J.,  {Clampin} M.,
  {Fazio} G.~G.,   {MacEwen} H.~A.,  eds,  Society of Photo-Optical
  Instrumentation Engineers (SPIE) Conference Series Vol. 9143, Space
  Telescopes and Instrumentation 2014: Optical, Infrared, and Millimeter Wave.
  p. 914320 (\mn@eprint {arXiv} {1406.0151}), \mn@doi{10.1117/12.2063489}

\bibitem[\protect\citeauthoryear{{Ricker} et~al.,}{{Ricker}
  et~al.}{2015}]{TESS+2015}
{Ricker} G.~R.,  et~al., 2015, \mn@doi [Journal of Astronomical Telescopes,
  Instruments, and Systems] {10.1117/1.JATIS.1.1.014003}, \href
  {https://ui.adsabs.harvard.edu/abs/2015JATIS...1a4003R} {1, 014003}

\bibitem[\protect\citeauthoryear{{Rimoldini} et~al.,}{{Rimoldini}
  et~al.}{2022}]{Rimoldini+etal+2022+GaiaDR3}
{Rimoldini} L.,  et~al., 2022, arXiv e-prints, \href
  {https://ui.adsabs.harvard.edu/abs/2022arXiv221117238R} {p. arXiv:2211.17238}

\bibitem[\protect\citeauthoryear{{Rodr{\'\i}guez}, {L{\'o}pez-Gonz{\'a}lez}  \&
  {L{\'o}pez de Coca}}{{Rodr{\'\i}guez}
  et~al.}{2000}]{Rodriguez+etal+2000+DSCT_EDSCT}
{Rodr{\'\i}guez} E.,  {L{\'o}pez-Gonz{\'a}lez} M.~J.,   {L{\'o}pez de Coca} P.,
   2000, \mn@doi [\aaps] {10.1051/aas:2000221}, \href
  {https://ui.adsabs.harvard.edu/abs/2000A&AS..144..469R} {144, 469}

\bibitem[\protect\citeauthoryear{{Rowan}, {Stanek}, {Jayasinghe}, {Kochanek},
  {Thompson}, {Shappee}, {Holoien}  \& {Prieto}}{{Rowan}
  et~al.}{2021}]{Rowan+etal+2021+Ell_ASASSN}
{Rowan} D.~M.,  {Stanek} K.~Z.,  {Jayasinghe} T.,  {Kochanek} C.~S.,
  {Thompson} T.~A.,  {Shappee} B.~J.,  {Holoien} T.~W.~S.,   {Prieto} J.~L.,
  2021, \mn@doi [\mnras] {10.1093/mnras/stab2126}, \href
  {https://ui.adsabs.harvard.edu/abs/2021MNRAS.507..104R} {507, 104}

\bibitem[\protect\citeauthoryear{{Rucinski}}{{Rucinski}}{1992}]{Rucinski+1992+cutoff}
{Rucinski} S.~M.,  1992, \mn@doi [\aj] {10.1086/116118}, \href
  {https://ui.adsabs.harvard.edu/abs/1992AJ....103..960R} {103, 960}

\bibitem[\protect\citeauthoryear{{Scargle}}{{Scargle}}{1982}]{Scargle+1982}
{Scargle} J.~D.,  1982, \mn@doi [\apj] {10.1086/160554}, \href
  {https://ui.adsabs.harvard.edu/abs/1982ApJ...263..835S} {263, 835}

\bibitem[\protect\citeauthoryear{{Scaringi} et~al.,}{{Scaringi}
  et~al.}{2022}]{Scaringi+etal+2022+nature+localthermoburst_CV}
{Scaringi} S.,  et~al., 2022, \mn@doi [\nat] {10.1038/s41586-022-04495-6},
  \href {https://ui.adsabs.harvard.edu/abs/2022Natur.604..447S} {604, 447}

\bibitem[\protect\citeauthoryear{{Schaefer}}{{Schaefer}}{2010}]{Schaefer+2010+novae}
{Schaefer} B.~E.,  2010, \mn@doi [\apjs] {10.1088/0067-0049/187/2/275}, \href
  {https://ui.adsabs.harvard.edu/abs/2010ApJS..187..275S} {187, 275}

\bibitem[\protect\citeauthoryear{{Schaefer}}{{Schaefer}}{2022}]{Schaefer+2022+Novae}
{Schaefer} B.~E.,  2022, \mn@doi [\mnras] {10.1093/mnras/stac2089}, \href
  {https://ui.adsabs.harvard.edu/abs/2022MNRAS.tmp.2002S} {}

\bibitem[\protect\citeauthoryear{{Schaffenroth} et~al.,}{{Schaffenroth}
  et~al.}{2019}]{Schaffenroth+etal+2019+Short_binary}
{Schaffenroth} V.,  et~al., 2019, \mn@doi [\aap] {10.1051/0004-6361/201936019},
  \href {https://ui.adsabs.harvard.edu/abs/2019A&A...630A..80S} {630, A80}

\bibitem[\protect\citeauthoryear{{Schuh}, {Huber}, {Dreizler}, {Heber},
  {O'Toole}, {Green}  \& {Fontaine}}{{Schuh}
  et~al.}{2006}]{Schuh+etal+2006+DWLyn}
{Schuh} S.,  {Huber} J.,  {Dreizler} S.,  {Heber} U.,  {O'Toole} S.~J.,
  {Green} E.~M.,   {Fontaine} G.,  2006, \mn@doi [\aap]
  {10.1051/0004-6361:200500210}, \href
  {https://ui.adsabs.harvard.edu/abs/2006A&A...445L..31S} {445, L31}

\bibitem[\protect\citeauthoryear{{Shah}, {van der Sluys}  \& {Nelemans}}{{Shah}
  et~al.}{2012}]{Shah+etal+2012+GW_aid}
{Shah} S.,  {van der Sluys} M.,   {Nelemans} G.,  2012, \mn@doi [\aap]
  {10.1051/0004-6361/201219309}, \href
  {https://ui.adsabs.harvard.edu/abs/2012A&A...544A.153S} {544, A153}

\bibitem[\protect\citeauthoryear{{Shahbaz}, {Watson}, {Zurita}, {Villaver}  \&
  {Hernandez-Peralta}}{{Shahbaz} et~al.}{2008}]{Shahbaz+etal+2008+0614}
{Shahbaz} T.,  {Watson} C.~A.,  {Zurita} C.,  {Villaver} E.,
  {Hernandez-Peralta} H.,  2008, \mn@doi [\pasp] {10.1086/590505}, \href
  {https://ui.adsabs.harvard.edu/abs/2008PASP..120..848S} {120, 848}

\bibitem[\protect\citeauthoryear{{Skillman} \& {Patterson}}{{Skillman} \&
  {Patterson}}{1993}]{Skillman+Patterson+1993+0917}
{Skillman} D.~R.,  {Patterson} J.,  1993, \mn@doi [\apj] {10.1086/173312},
  \href {https://ui.adsabs.harvard.edu/abs/1993ApJ...417..298S} {417, 298}

\bibitem[\protect\citeauthoryear{{Skumanich}}{{Skumanich}}{1972}]{Skumanich+1972+rotation}
{Skumanich} A.,  1972, \mn@doi [\apj] {10.1086/151310}, \href
  {https://ui.adsabs.harvard.edu/abs/1972ApJ...171..565S} {171, 565}

\bibitem[\protect\citeauthoryear{{Smalley} et~al.,}{{Smalley}
  et~al.}{2011}]{Smalley+etal+2011+Superroam}
{Smalley} B.,  et~al., 2011, \mn@doi [\aap] {10.1051/0004-6361/201117230},
  \href {https://ui.adsabs.harvard.edu/abs/2011A&A...535A...3S} {535, A3}

\bibitem[\protect\citeauthoryear{{Soszy{\'n}ski} et~al.,}{{Soszy{\'n}ski}
  et~al.}{2015}]{Soszynski+etal+2015+USPbinary_OGLE}
{Soszy{\'n}ski} I.,  et~al., 2015, \actaa, \href
  {https://ui.adsabs.harvard.edu/abs/2015AcA....65...39S} {65, 39}

\bibitem[\protect\citeauthoryear{{Soszy{\'n}ski} et~al.,}{{Soszy{\'n}ski}
  et~al.}{2016}]{Soszynski+etal+2016+ell}
{Soszy{\'n}ski} I.,  et~al., 2016, \actaa, \href
  {https://ui.adsabs.harvard.edu/abs/2016AcA....66..405S} {66, 405}

\bibitem[\protect\citeauthoryear{{Soszy{\'n}ski} et~al.,}{{Soszy{\'n}ski}
  et~al.}{2021}]{Soszynski+etal+2021+24000deltascuti}
{Soszy{\'n}ski} I.,  et~al., 2021, \mn@doi [\actaa]
  {10.32023/0001-5237/71.3.1}, \href
  {https://ui.adsabs.harvard.edu/abs/2021AcA....71..189S} {71, 189}

\bibitem[\protect\citeauthoryear{{Southworth}, {G{\"a}nsicke}, {Marsh}, {de
  Martino}  \& {Aungwerojwit}}{{Southworth}
  et~al.}{2007}]{Southworth+etal+2007+longspin_IP}
{Southworth} J.,  {G{\"a}nsicke} B.~T.,  {Marsh} T.~R.,  {de Martino} D.,
  {Aungwerojwit} A.,  2007, \mn@doi [\mnras]
  {10.1111/j.1365-2966.2007.11796.x}, \href
  {https://ui.adsabs.harvard.edu/abs/2007MNRAS.378..635S} {378, 635}

\bibitem[\protect\citeauthoryear{{Spruit} \& {Ritter}}{{Spruit} \&
  {Ritter}}{1983}]{Spruit+Ritter+1983+period_gap}
{Spruit} H.~C.,  {Ritter} H.,  1983, \aap, \href
  {https://ui.adsabs.harvard.edu/abs/1983A&A...124..267S} {124, 267}

\bibitem[\protect\citeauthoryear{{Steele} et~al.,}{{Steele}
  et~al.}{2013}]{Steele+etal+2013+shortest_Rbinary}
{Steele} P.~R.,  et~al., 2013, \mn@doi [\mnras] {10.1093/mnras/sts620}, \href
  {https://ui.adsabs.harvard.edu/abs/2013MNRAS.429.3492S} {429, 3492}

\bibitem[\protect\citeauthoryear{{Sun} et~al.,}{{Sun}
  et~al.}{2021}]{Sun+etal+2021+LAMOST_CV}
{Sun} Y.,  et~al., 2021, \mn@doi [\apjs] {10.3847/1538-4365/ac283a}, \href
  {https://ui.adsabs.harvard.edu/abs/2021ApJS..257...65S} {257, 65}

\bibitem[\protect\citeauthoryear{{Thompson} et~al.,}{{Thompson}
  et~al.}{2019}]{Thompson+etal+2019+Science+noninteractive}
{Thompson} T.~A.,  et~al., 2019, \mn@doi [Science] {10.1126/science.aau4005},
  \href {https://ui.adsabs.harvard.edu/abs/2019Sci...366..637T} {366, 637}

\bibitem[\protect\citeauthoryear{{Thorne}, {Garnavich}  \& {Mohrig}}{{Thorne}
  et~al.}{2010}]{Thorne+etal+2010+CSS081231:071126+440405}
{Thorne} K.,  {Garnavich} P.,   {Mohrig} K.,  2010, Information Bulletin on
  Variable Stars, \href {https://ui.adsabs.harvard.edu/abs/2010IBVS.5923....1T}
  {5923, 1}

\bibitem[\protect\citeauthoryear{{Toma} et~al.,}{{Toma}
  et~al.}{2016}]{Toma+etal+2016+omegaII}
{Toma} R.,  et~al., 2016, \mn@doi [\mnras] {10.1093/mnras/stw2079}, \href
  {https://ui.adsabs.harvard.edu/abs/2016MNRAS.463.1099T} {463, 1099}

\bibitem[\protect\citeauthoryear{{Tonry} et~al.,}{{Tonry}
  et~al.}{2018}]{Tonry+etal+2018+ATLAS}
{Tonry} J.~L.,  et~al., 2018, \mn@doi [\pasp] {10.1088/1538-3873/aabadf}, \href
  {https://ui.adsabs.harvard.edu/abs/2018PASP..130f4505T} {130, 064505}

\bibitem[\protect\citeauthoryear{{Tudor} et~al.,}{{Tudor}
  et~al.}{2018}]{Tudor+etal+2018+BH_UCXB_HST}
{Tudor} V.,  et~al., 2018, \mn@doi [\mnras] {10.1093/mnras/sty284}, \href
  {https://ui.adsabs.harvard.edu/abs/2018MNRAS.476.1889T} {476, 1889}

\bibitem[\protect\citeauthoryear{{Uzundag} et~al.,}{{Uzundag}
  et~al.}{2021a}]{Uzundag+etal+2021+TESS_gmode_sdBVs}
{Uzundag} M.,  et~al., 2021a, \mn@doi [\aap] {10.1051/0004-6361/202140961},
  \href {https://ui.adsabs.harvard.edu/abs/2021A&A...651A.121U} {651, A121}

\bibitem[\protect\citeauthoryear{{Uzundag} et~al.,}{{Uzundag}
  et~al.}{2021b}]{Uzundag+etal+2021+GWVir}
{Uzundag} M.,  et~al., 2021b, \mn@doi [\aap] {10.1051/0004-6361/202141253},
  \href {https://ui.adsabs.harvard.edu/abs/2021A&A...655A..27U} {655, A27}

\bibitem[\protect\citeauthoryear{{Van Grootel}, {Dupret}, {Fontaine},
  {Brassard}, {Grigahc{\`e}ne}  \& {Quirion}}{{Van Grootel}
  et~al.}{2012}]{VanGrootel+etal++2012+ZZ}
{Van Grootel} V.,  {Dupret} M.~A.,  {Fontaine} G.,  {Brassard} P.,
  {Grigahc{\`e}ne} A.,   {Quirion} P.~O.,  2012, \mn@doi [\aap]
  {10.1051/0004-6361/201118371}, \href
  {https://ui.adsabs.harvard.edu/abs/2012A&A...539A..87V} {539, A87}

\bibitem[\protect\citeauthoryear{{Van Grootel}, {Fontaine}, {Brassard}  \&
  {Dupret}}{{Van Grootel} et~al.}{2013}]{Grootel+etal+2013+newZZ}
{Van Grootel} V.,  {Fontaine} G.,  {Brassard} P.,   {Dupret} M.~A.,  2013,
  \mn@doi [\apj] {10.1088/0004-637X/762/1/57}, \href
  {https://ui.adsabs.harvard.edu/abs/2013ApJ...762...57V} {762, 57}

\bibitem[\protect\citeauthoryear{{VanderPlas}}{{VanderPlas}}{2018}]{VanderPlas+2018+LSP}
{VanderPlas} J.~T.,  2018, \mn@doi [\apjs] {10.3847/1538-4365/aab766}, \href
  {https://ui.adsabs.harvard.edu/abs/2018ApJS..236...16V} {236, 16}

\bibitem[\protect\citeauthoryear{{Varadi}, {Eyer}, {Jordan}, {Mowlavi}  \&
  {Koester}}{{Varadi} et~al.}{2009}]{Varadi+etal+2009+Gaia_spv}
{Varadi} M.,  {Eyer} L.,  {Jordan} S.,  {Mowlavi} N.,   {Koester} D.,  2009, in
  {Guzik} J.~A.,  {Bradley} P.~A.,  eds,  American Institute of Physics
  Conference Series Vol. 1170, Stellar Pulsation: Challenges for Theory and
  Observation. pp 330--332 (\mn@eprint {arXiv} {0907.4084}),
  \mn@doi{10.1063/1.3246507}

\bibitem[\protect\citeauthoryear{{Vaughan}}{{Vaughan}}{2005}]{Vaughan+2005+rednoise}
{Vaughan} S.,  2005, \mn@doi [\aap] {10.1051/0004-6361:20041453}, \href
  {https://ui.adsabs.harvard.edu/abs/2005A&A...431..391V} {431, 391}

\bibitem[\protect\citeauthoryear{{Vaughan}}{{Vaughan}}{2010}]{Vaughan+2010+rednoise}
{Vaughan} S.,  2010, \mn@doi [\mnras] {10.1111/j.1365-2966.2009.15868.x}, \href
  {https://ui.adsabs.harvard.edu/abs/2010MNRAS.402..307V} {402, 307}

\bibitem[\protect\citeauthoryear{{Vennes}, {Kawka}, {O'Toole}, {N{\'e}meth}  \&
  {Burton}}{{Vennes} et~al.}{2012}]{Vennes+etal+2012+sd-WD}
{Vennes} S.,  {Kawka} A.,  {O'Toole} S.~J.,  {N{\'e}meth} P.,   {Burton} D.,
  2012, \mn@doi [\apjl] {10.1088/2041-8205/759/1/L25}, \href
  {https://ui.adsabs.harvard.edu/abs/2012ApJ...759L..25V} {759, L25}

\bibitem[\protect\citeauthoryear{{Walter}, {Battisti}, {Towers}, {Bond}  \&
  {Stringfellow}}{{Walter} et~al.}{2012}]{Walter+etal+2012+novae}
{Walter} F.~M.,  {Battisti} A.,  {Towers} S.~E.,  {Bond} H.~E.,
  {Stringfellow} G.~S.,  2012, \mn@doi [\pasp] {10.1086/668404}, \href
  {https://ui.adsabs.harvard.edu/abs/2012PASP..124.1057W} {124, 1057}

\bibitem[\protect\citeauthoryear{{Wang} \& {Chakrabarty}}{{Wang} \&
  {Chakrabarty}}{2004}]{Wang+Chakrabarty+2004+4u1543}
{Wang} Z.,  {Chakrabarty} D.,  2004, \mn@doi [\apjl] {10.1086/426787}, \href
  {https://ui.adsabs.harvard.edu/abs/2004ApJ...616L.139W} {616, L139}

\bibitem[\protect\citeauthoryear{{Wang}, {Chen}, {Liu}, {Chen}, {Wu}, {Tang},
  {Guo}  \& {Han}}{{Wang} et~al.}{2021}]{Wang+etal+2021+UCXB}
{Wang} B.,  {Chen} W.-C.,  {Liu} D.-D.,  {Chen} H.-L.,  {Wu} C.-Y.,  {Tang}
  W.-S.,  {Guo} Y.-L.,   {Han} Z.-W.,  2021, \mn@doi [\mnras]
  {10.1093/mnras/stab2032}, \href
  {https://ui.adsabs.harvard.edu/abs/2021MNRAS.506.4654W} {506, 4654}

\bibitem[\protect\citeauthoryear{{Warner}}{{Warner}}{1995}]{Warner+1995+book_CV}
{Warner} B.,  1995, {Cataclysmic variable stars}.
 Cambridge Astrophysics Series Vol. 28

\bibitem[\protect\citeauthoryear{{Watson}, {Henden}  \& {Price}}{{Watson}
  et~al.}{2006}]{Watson+etal+2006+VSX}
{Watson} C.~L.,  {Henden} A.~A.,   {Price} A.,  2006, Society for Astronomical
  Sciences Annual Symposium, \href
  {https://ui.adsabs.harvard.edu/abs/2006SASS...25...47W} {25, 47}

\bibitem[\protect\citeauthoryear{{Williams}, {Hermes}  \&
  {Vanderbosch}}{{Williams} et~al.}{2022}]{Williams+etal+2022+MWD}
{Williams} K.~A.,  {Hermes} J.~J.,   {Vanderbosch} Z.~P.,  2022, arXiv
  e-prints, \href {https://ui.adsabs.harvard.edu/abs/2022arXiv220713763W} {p.
  arXiv:2207.13763}

\bibitem[\protect\citeauthoryear{{Wils}, {G{\"a}nsicke}, {Drake}  \&
  {Southworth}}{{Wils} et~al.}{2010}]{Wils+etal+2010+DN_mining}
{Wils} P.,  {G{\"a}nsicke} B.~T.,  {Drake} A.~J.,   {Southworth} J.,  2010,
  \mn@doi [\mnras] {10.1111/j.1365-2966.2009.15894.x}, \href
  {https://ui.adsabs.harvard.edu/abs/2010MNRAS.402..436W} {402, 436}

\bibitem[\protect\citeauthoryear{{Woudt} \& {Warner}}{{Woudt} \&
  {Warner}}{2002}]{Woudt+Warner+2002+novae}
{Woudt} P.~A.,  {Warner} B.,  2002, \mn@doi [\mnras]
  {10.1046/j.1365-8711.2002.05613.x}, \href
  {https://ui.adsabs.harvard.edu/abs/2002MNRAS.335...44W} {335, 44}

\bibitem[\protect\citeauthoryear{{Wright}, {Drake}, {Mamajek}  \&
  {Henry}}{{Wright} et~al.}{2011}]{Wright+etal+2011+stellar_activity}
{Wright} N.~J.,  {Drake} J.~J.,  {Mamajek} E.~E.,   {Henry} G.~W.,  2011,
  \mn@doi [\apj] {10.1088/0004-637X/743/1/48}, \href
  {https://ui.adsabs.harvard.edu/abs/2011ApJ...743...48W} {743, 48}

\bibitem[\protect\citeauthoryear{{Wu} \& {Li}}{{Wu} \&
  {Li}}{2018}]{Wu+etal+2018+CHeB_BLAP}
{Wu} T.,  {Li} Y.,  2018, \mn@doi [\mnras] {10.1093/mnras/sty1347}, \href
  {https://ui.adsabs.harvard.edu/abs/2018MNRAS.478.3871W} {478, 3871}

\bibitem[\protect\citeauthoryear{{Wu}, {Yu}, {Li}, {Maccarone}  \& {Li}}{{Wu}
  et~al.}{2010}]{Wu+etal+2010+LMXB_period}
{Wu} Y.~X.,  {Yu} W.,  {Li} T.~P.,  {Maccarone} T.~J.,   {Li} X.~D.,  2010,
  \mn@doi [\apj] {10.1088/0004-637X/718/2/620}, \href
  {https://ui.adsabs.harvard.edu/abs/2010ApJ...718..620W} {718, 620}

\bibitem[\protect\citeauthoryear{{Wu}, {Xiong}  \& {Wang}}{{Wu}
  et~al.}{2022}]{Wu+etal+2022+CO+He}
{Wu} C.,  {Xiong} H.,   {Wang} X.,  2022, \mn@doi [\mnras]
  {10.1093/mnras/stac273}, \href
  {https://ui.adsabs.harvard.edu/abs/2022MNRAS.512.2972W} {512, 2972}

\bibitem[\protect\citeauthoryear{{Wu}, {Xiong}, {Lin}, {Guo}, {Wang}, {Han}  \&
  {Wang}}{{Wu} et~al.}{2023}]{Wu+etal+2023+ONe+CO}
{Wu} C.,  {Xiong} H.,  {Lin} J.,  {Guo} Y.,  {Wang} X.,  {Han} Z.,   {Wang} B.,
   2023, \mn@doi [\apjl] {10.3847/2041-8213/acb6f3}, \href
  {https://ui.adsabs.harvard.edu/abs/2023ApJ...944L..54W} {944, L54}

\bibitem[\protect\citeauthoryear{{Xia}, {Li}, {Chen}, {Guo}  \& {Gao}}{{Xia}
  et~al.}{2018}]{Xia+etal+2018+EW}
{Xia} Q.-Q.,  {Li} K.,  {Chen} X.,  {Guo} D.-F.,   {Gao} D.-Y.,  2018, \mn@doi
  [\pasj] {10.1093/pasj/psy103}, \href
  {https://ui.adsabs.harvard.edu/abs/2018PASJ...70..104X} {70, 104}

\bibitem[\protect\citeauthoryear{{Xia}, {Michel}, {Li}  \& {Higuera}}{{Xia}
  et~al.}{2021}]{Xia+etal+2021+contact}
{Xia} Q.,  {Michel} R.,  {Li} K.,   {Higuera} J.,  2021, \mn@doi [\pasp]
  {10.1088/1538-3873/abf32d}, \href
  {https://ui.adsabs.harvard.edu/abs/2021PASP..133e4202X} {133, 054202}

\bibitem[\protect\citeauthoryear{{Xiong} et~al.,}{{Xiong}
  et~al.}{2022}]{Xiong+etal+2022+SHeB}
{Xiong} H.,  et~al., 2022, arXiv e-prints, \href
  {https://ui.adsabs.harvard.edu/abs/2022arXiv221101564X} {p. arXiv:2211.01564}

\bibitem[\protect\citeauthoryear{{Yang} et~al.,}{{Yang}
  et~al.}{2020}]{Yang+etal+2020+LAMOST_binary}
{Yang} F.,  et~al., 2020, \mn@doi [\apjs] {10.3847/1538-4365/ab9b77}, \href
  {https://ui.adsabs.harvard.edu/abs/2020ApJS..249...31Y} {249, 31}

\bibitem[\protect\citeauthoryear{{Yi}, {Sun}  \& {Gu}}{{Yi}
  et~al.}{2019}]{Yi+etal+2019+search_BH}
{Yi} T.,  {Sun} M.,   {Gu} W.-M.,  2019, \mn@doi [\apj]
  {10.3847/1538-4357/ab4a75}, \href
  {https://ui.adsabs.harvard.edu/abs/2019ApJ...886...97Y} {886, 97}

\bibitem[\protect\citeauthoryear{{Yoon}, {Podsiadlowski}  \& {Rosswog}}{{Yoon}
  et~al.}{2007}]{Yoon+etal+2007+CO_merger}
{Yoon} S.~C.,  {Podsiadlowski} P.,   {Rosswog} S.,  2007, \mn@doi [\mnras]
  {10.1111/j.1365-2966.2007.12161.x}, \href
  {https://ui.adsabs.harvard.edu/abs/2007MNRAS.380..933Y} {380, 933}

\bibitem[\protect\citeauthoryear{{Yungelson}, {Lasota}, {Nelemans}, {Dubus},
  {van den Heuvel}, {Dewi}  \& {Portegies Zwart}}{{Yungelson}
  et~al.}{2006}]{Yungelson+etal+2006+BHUCXB_theory}
{Yungelson} L.~R.,  {Lasota} J.~P.,  {Nelemans} G.,  {Dubus} G.,  {van den
  Heuvel} E.~P.~J.,  {Dewi} J.,   {Portegies Zwart} S.,  2006, \mn@doi [\aap]
  {10.1051/0004-6361:20064984}, \href
  {https://ui.adsabs.harvard.edu/abs/2006A&A...454..559Y} {454, 559}

\bibitem[\protect\citeauthoryear{{Zellem}, {Hollon}, {Ballouz}, {Sion},
  {Godon}, {G{\"a}nsicke}  \& {Long}}{{Zellem}
  et~al.}{2009}]{Zellem+etal+2009+BKLyn}
{Zellem} R.,  {Hollon} N.,  {Ballouz} R.-L.,  {Sion} E.~M.,  {Godon} P.,
  {G{\"a}nsicke} B.~T.,   {Long} K.,  2009, \mn@doi [\pasp] {10.1086/605547},
  \href {https://ui.adsabs.harvard.edu/abs/2009PASP..121..942Z} {121, 942}

\bibitem[\protect\citeauthoryear{{Zhang} \& {Qian}}{{Zhang} \&
  {Qian}}{2020}]{Zhang+Qian+2020+cutoff}
{Zhang} X.-D.,  {Qian} S.-B.,  2020, \mn@doi [\mnras] {10.1093/mnras/staa2166},
  \href {https://ui.adsabs.harvard.edu/abs/2020MNRAS.497.3493Z} {497, 3493}

\bibitem[\protect\citeauthoryear{{Zhang} et~al.,}{{Zhang}
  et~al.}{2020}]{Zhang+etal+2020+tmts_performance}
{Zhang} J.-C.,  et~al., 2020, \mn@doi [\pasp] {10.1088/1538-3873/abbea2}, \href
  {https://ui.adsabs.harvard.edu/abs/2020PASP..132l5001Z} {132, 125001}

\bibitem[\protect\citeauthoryear{{Zhao}, {Zhao}, {Chu}, {Jing}  \&
  {Deng}}{{Zhao} et~al.}{2012}]{Zhao+etal+2012+LAMOST}
{Zhao} G.,  {Zhao} Y.-H.,  {Chu} Y.-Q.,  {Jing} Y.-P.,   {Deng} L.-C.,  2012,
  \mn@doi [Research in Astronomy and Astrophysics]
  {10.1088/1674-4527/12/7/002}, \href
  {https://ui.adsabs.harvard.edu/abs/2012RAA....12..723Z} {12, 723}

\bibitem[\protect\citeauthoryear{{Zhong} \& {Wang}}{{Zhong} \&
  {Wang}}{2011}]{Zhong+Wang+2011+0918}
{Zhong} J.,  {Wang} Z.,  2011, \mn@doi [\apj] {10.1088/0004-637X/729/1/8},
  \href {https://ui.adsabs.harvard.edu/abs/2011ApJ...729....8Z} {729, 8}

\bibitem[\protect\citeauthoryear{{Zhou}}{{Zhou}}{2010}]{Zhou+2010+oscillating_binary}
{Zhou} A.~Y.,  2010, arXiv e-prints, \href
  {https://ui.adsabs.harvard.edu/abs/2010arXiv1002.2729Z} {p. arXiv:1002.2729}

\bibitem[\protect\citeauthoryear{{Ziaali}, {Bedding}, {Murphy}, {Van Reeth}  \&
  {Hey}}{{Ziaali} et~al.}{2019}]{Ziaali+etal+2019+PL_relation}
{Ziaali} E.,  {Bedding} T.~R.,  {Murphy} S.~J.,  {Van Reeth} T.,   {Hey} D.~R.,
   2019, \mn@doi [\mnras] {10.1093/mnras/stz1110}, \href
  {https://ui.adsabs.harvard.edu/abs/2019MNRAS.486.4348Z} {486, 4348}

\bibitem[\protect\citeauthoryear{{Zickgraf}, {Engels}, {Hagen}, {Reimers}  \&
  {Voges}}{{Zickgraf} et~al.}{2003}]{Zickgraf+etal+2003+Rosat}
{Zickgraf} F.~J.,  {Engels} D.,  {Hagen} H.~J.,  {Reimers} D.,   {Voges} W.,
  2003, \mn@doi [\aap] {10.1051/0004-6361:20030679}, \href
  {https://ui.adsabs.harvard.edu/abs/2003A&A...406..535Z} {406, 535}

\bibitem[\protect\citeauthoryear{{Zucker}, {Mazeh}  \& {Alexander}}{{Zucker}
  et~al.}{2007}]{Zucker+etal+2007+beaming}
{Zucker} S.,  {Mazeh} T.,   {Alexander} T.,  2007, \mn@doi [\apj]
  {10.1086/521389}, \href
  {https://ui.adsabs.harvard.edu/abs/2007ApJ...670.1326Z} {670, 1326}

\bibitem[\protect\citeauthoryear{{de Miguel} et~al.,}{{de Miguel}
  et~al.}{2017}]{Miguel+etal+2017+superhump_ip}
{de Miguel} E.,  et~al., 2017, \mn@doi [\mnras] {10.1093/mnras/stx107}, \href
  {https://ui.adsabs.harvard.edu/abs/2017MNRAS.467..428D} {467, 428}

\bibitem[\protect\citeauthoryear{{in't Zand} \& {Weinberg}}{{in't Zand} \&
  {Weinberg}}{2010}]{Zand+weinberg+2010+superexpansion}
{in't Zand} J.~J.~M.,  {Weinberg} N.~N.,  2010, \mn@doi [\aap]
  {10.1051/0004-6361/200913952}, \href
  {https://ui.adsabs.harvard.edu/abs/2010A&A...520A..81I} {520, A81}

\bibitem[\protect\citeauthoryear{{van Roestel} et~al.,}{{van Roestel}
  et~al.}{2022}]{Roestel+etal+2022+AMCVn}
{van Roestel} J.,  et~al., 2022, \mn@doi [\mnras] {10.1093/mnras/stab2421},
  \href {https://ui.adsabs.harvard.edu/abs/2022MNRAS.512.5440V} {512, 5440}

\makeatother
\end{thebibliography}
\input{tmts_periodic.bbl}



\bsp	
\label{lastpage}
\end{document}